\documentclass[%
 reprint,
 amsmath,amssymb,
 aps,
 prc,nofootinbib,
 superscriptaddress
]{revtex4-2}

\usepackage{silence}
\usepackage{dcolumn}
\usepackage{bm}

\usepackage{float}
\usepackage{graphicx}
\usepackage{breqn}
\usepackage{mathtools}
\usepackage{listings}
\usepackage{comment}
\usepackage[hidelinks]{hyperref}
\usepackage{cleveref}

\usepackage{xcolor}

\usepackage{placeins}

\begin{document}

\crefname{figure}{Fig.}{Figs.}
\Crefname{figure}{Figure}{Figures}
\crefname{equation}{Eq.}{Eqs.}
\Crefname{equation}{Equation}{Equations}
\crefname{section}{Sec.}{Secs.}
\Crefname{section}{Section}{Sections}
\creflabelformat{equation}{#2\textup{#1}#3}
\renewcommand{\crefrangeconjunction}{--}

\newcommand{\reftexts}{Ref.~}
\newcommand{\lips}[1]{\tilde{d} #1 \;}
\newcommand{\vecIV}[1]{#1} %
\newcommand{\vecIII}[1]{\vec{#1}} %
\newcommand{\vecII}[1]{\boldsymbol{#1}} %

\makeatletter

\newcommand{\coll}[2]{%
	$\coll@process{#1} + \coll@process{#2}$%
}

\newcommand{\collThree}[3]{%
	$\coll@process{#1} / \coll@process{#2} + \coll@process{#3}$%
}

\newcommand{\collFour}[4]{%
  $\coll@process{#1} / \coll@process{#2} / \coll@process{#3} + \coll@process{#4}$%
}

\newcommand{\coleTwo}[1]{%
	{#1}%
}

\newcommand{\coleThree}[1]{%
	{#1}%
}

\newcommand{\coll@process}[1]{%
  \ifx#1A%
    #1%
  \else\ifx#1B%
    #1%
  \else\ifx#1x%
    #1%
  \else\ifx#1p%
    #1%
  \else\ifx#1d%
    #1%
  \else
    \coll@checkHeThree{#1}%
  \fi\fi\fi\fi\fi%
}

\newcommand{\coll@checkHeThree}[1]{%
  \ifnum\pdfstrcmp{#1}{He3}=0 %
    {}^3\mathrm{He}%
  \else
    \mathrm{#1}%
  \fi
}

\makeatother

\title{Collisional and radiative energy loss in small systems}

\author{Coleridge Faraday}
\email{frdcol002@myuct.ac.za}
\affiliation{Department of Physics\char`,{} University of Cape Town\char`,{} Private Bag X3\char`,{} Rondebosch 7701\char`,{} South Africa}

\author{W.\ A.\ Horowitz}
\email{wa.horowitz@uct.ac.za}
\affiliation{Department of Physics\char`,{} University of Cape Town\char`,{} Private Bag X3\char`,{} Rondebosch 7701\char`,{} South Africa}
\affiliation{Department of Physics\char`,{} New Mexico State University\char`,{} Las Cruces\char`,{} New Mexico\char`,{} 88003\char`,{} USA}

\date{\today}

\begin{abstract}

We present an energy loss model which includes small system size corrections to both the radiative and elastic energy loss. Our model is used to compute the nuclear modification factor $R_{AB}$ of light and heavy flavor hadrons, averaged over realistic collision geometries for central and peripheral \coll{A}{A} and central \collFour{p}{d}{He3}{A} collisions at LHC and RHIC. We find that the predicted suppression in small systems is almost entirely due to elastic energy loss. 
Our results are keenly sensitive to the crossover between elastic energy loss calculated with HTL propagators and vacuum propagators, respectively, which leads to a large theoretical uncertainty.
We find that the $R_{AB}$ is largely insensitive to the form of the elastic energy loss distribution---Gaussian or Poisson---surprisingly so in small systems where the central limit theorem is inapplicable.
We present an expansion of the $R_{AB}$ in terms of the moments of the energy loss probability distribution, which allows for a rigorous understanding of the dependence of the $R_{AB}$ on the underlying energy loss distribution.

\end{abstract}

\maketitle

\section{Introduction}

Due to their sensitivity to final state in-medium effects, hard probes, including jets and leading hadrons, are a crucial component of the heavy-ion program at the BNL Relativistic Heavy Ion Collider (RHIC) and the CERN Large Hadron Collider (LHC). One of the most useful hard probe observables is the nuclear modification factor $R_{AA}$, which captures the amount of in-medium energy loss suffered by high-$p_T$ partons. A measured $R_{AA} \sim 0.2$ for leading hadrons \cite{PHENIX:2001hpc, STAR:2002ggv} in central \coll{Au}{Au} collisions at RHIC---when compared to the photon $R_{AA} \simeq 1$ in the same collision \cite{PHENIX:2005yls}---is strong evidence of medium modifications to the hadron spectra. 
When one contrasts this large suppression with $R_{dA} \simeq 1$ of pions produced in minimum bias \coll{d}{Au} collisions \cite{PHENIX:2003qdw, STAR:2003pjh} and the independence of the $R_{AA}$ on the final state light hadron mass in \coll{Au}{Au} collisions \cite{PHENIX:2006mhb}, one is driven to conclude that the modification of the spectrum is due primarily to final state partonic energy loss. %
Various semi-classical perturbative Quantum Chromodynamics (pQCD) models, which make different sets of approximations, have been successful in making qualitative predictions for the energy loss of hard probes in \coll{A}{A} collisions \cite{Dainese:2004te, Schenke:2009gb, Wicks:2005gt, Horowitz:2012cf}.

The $p_T \sim 5\text{--}20$ GeV charged leading hadron $R_{AA}\sim 0.2$ in central \coll{Pb}{Pb} collisions at LHC \cite{ALICE:2010mlf} tells a similar story, when compared to null controls of the photon \cite{CMS:2012oiv} and $Z$ boson \cite{CMS:2011zfr} $R_{AA}\sim 1$, and charged leading hadron $R_{pA}\sim 1$ in minimum bias \coll{p}{Pb} collisions \cite{ALICE:2012mj}.
The LHC also allows for additional insight due to its high $\sqrt{s}$, which leads to a hotter medium and higher integrated luminosity, which allows for better statistics.
The high integrated luminosity additionally allows for the measurement of hadrons which fragment from heavy quarks \cite{ALICE:2012ab, ALICE:2012ab, ALICE:2015vxz}, produced in the earliest stages of the collision, making them an invaluable tomographic tool in probing the QGP. Heavy-quarks also probe the mass and color charge dependence predicted by energy loss models (see \cite{Andronic:2015wma} for a review).

Another vital set of insights from RHIC and LHC was the near-perfect fluidity of the strongly-coupled low momentum modes of the QGP formed in semi-central \coll{Au}{Au} \cite{STAR:2000ekf, PHENIX:2003qra} and \coll{Pb}{Pb} \cite{ALICE:2010suc} collisions, as inferred by relativistic, viscous, hydrodynamic model predictions \cite{Romatschke:2007mq, Song:2007ux, Schenke:2010nt}.

More recently, the very same collective signatures of QGP formation have been experimentally observed at both RHIC and LHC in high multiplicity \coll{p}{p} \cite{ATLAS:2015hzw, ALICE:2023ulm}, \coll{p}{Pb} \cite{ATLAS:2013jmi, ALICE:2014dwt, CMS:2015yux} and \collFour{p}{d}{He3}{Au} \cite{PHENIX:2013ktj, PHENIX:2014fnc, PHENIX:2015idk, PHENIX:2016cfs, PHENIX:2017xrm}
collisions, which are qualitatively consistent with predictions from hydrodynamic models \cite{Weller:2017tsr, Schenke:2020mbo}.
In addition to these collective signatures, other QGP signatures have been detected in these small systems, including quarkonium suppression \cite{ALICE:2016sdt} and strangeness enhancement \cite{ALICE:2015mpp, ALICE:2013wgn}, which appear to depend only on the final multiplicity and not the collision system.
This evidence suggests that small droplets of QGP are formed at high multiplicity in even the smallest collision systems.

	Operating under the premise that QGP forms in central small collision systems, it follows that final state energy loss suffered by high-$p_T$ partons moving through the medium should result in a nuclear modification factor less than 1 in these collisions. Experimental measurements of the nuclear modification factor $R_{AB}$ in small systems from RHIC and LHC provide an inconclusive suppression pattern, particularly as a function of centrality. ALICE \cite{ALICE:2016yta}  and ATLAS \cite{ATLAS:2022kqu} report leading hadron nuclear modification factors in central \coll{p}{Pb} collisions consistent with no suppression, and even a $20\%$ \emph{enhancement}. PHENIX \cite{PHENIX:2021dod} measures the nuclear modification factor in central \collFour{p}{d}{He3}{Au} as $R_{AB} \simeq 0.75$, with a pronounced enhancement in peripheral collisions.

However, measuring the centrality-dependent $R_{AB}$ in small systems presents experimental challenges, notably centrality bias \cite{ALICE:2014xsp, PHENIX:2013jxf, Kordell:2016njg, PHENIX:2023dxl, Bzdak:2014rca}. In the determination of the nuclear modification factor, the Glauber model \cite{Glauber:1970jm,Miller:2007ri} is typically used to map between the measured event activity, related to the number of soft particles produced, and the number of binary collisions, related to the number of hard particles produced. Centrality bias is a non-trivial correlation between the hard and soft particles, which may lead to an inaccurate estimate for the number of hard collisions, which are then used to normalize the nuclear modification factor.
A recent PHENIX study \cite{PHENIX:2023dxl} experimentally measures the number of binary collisions using the prompt photon $R_{AB}$, which they then use to determine the $R_{AB}$ for pions produced in central \coll{d}{Au} collisions, independently of the standard Glauber model mapping. They find $R_{AB} \simeq 0.75$ for central \coll{d}{Au} collisions, that the enhancement \cite{PHENIX:2021dod} in peripheral \coll{d}{Au} collisions disappears, and that the $0\text{--}100\%$ centrality $R_{AB}$ is consistent with the minimum bias value. These results are qualitatively consistent with a picture of QGP formation in central \coll{d}{Au} collisions wherein final state energy loss leads to $R_{AB} < 1$. Other theoretical and experimental work on this front includes measuring self-normalized observables which are not sensitive to this centrality bias \cite{ATLAS:2022iyq}, further understanding of high-$p_T$ $v_2$ measurements \cite{ATLAS:2019vcm}, minimum bias measurements in intermediate-sized systems like \coll{O}{O} \cite{Huss:2020whe, Huss:2020dwe}, and accounting for the correlations between the hard and soft modes \cite{JETSCAPE:2024dgu}.

Given these experimental challenges, theoretical input becomes a vital and complementary approach to determining the consistency of suppression measurements in small systems with a final state energy loss scenario.

However, applying existing theoretical models---which have seen success in broadly describing data in large systems---to small systems presents its own set of challenges. Many of the assumptions that underlie canonical semi-classical pQCD energy loss models are likely to be inapplicable in small collision systems \cite{Faraday:2023mmx}, making quantitative predictions in small systems difficult. An instance of an explicit assumption that the system size is large is in the dropping of terms exponentially suppressed according to the system size, which all radiative energy loss approaches based on the opacity expansion \cite{Gyulassy:2000er, Djordjevic:2003zk, Zakharov:1997uu, Baier:1996kr} do.  
A correction that includes these previously neglected terms has been derived \cite{Kolbe:2015rvk,Kolbe:2015suq}, and we discussed the phenomenological implications of this correction for small and large systems in our previous work \cite{Faraday:2023mmx}. 

BDMPS-Z based models \cite{Baier:1996kr, Baier:1996sk, Baier:1996vi, Baier:1998kq, Zakharov:1996fv, Zakharov:1997uu} rely on applying the central limit theorem \cite{Armesto:2011ht}, by assuming that the high-$p_T$ parton undergoes a large number of collisions. The $\mathcal{O}(5)$ scatters \cite{Armesto:2011ht} that are present in large collision systems does not well motivate this assumption in large collision systems, let alone in small collision systems with $\mathcal{O}(0\text{--}1)$ collisions \cite{Faraday:2023mmx}. %
A similar assumption is commonly utilized for the elastic energy loss, where it is assumed that incident high-$p_T$ partons undergo a large number of collisions, which allows one to model the elastic energy loss probability distribution as Gaussian \cite{Wicks:2005gt, Moore:2004tg, Zigic:2021rku, Faraday:2023mmx}. %

In our previous work \cite{Faraday:2023mmx}, we found that elastic energy loss comprises almost all energy loss in small systems, which we attributed to the breakdown of the assumption that the elastic energy loss distribution is Gaussian according to the central limit theorem. We reasoned that the small number of scatters present in small systems would result in a large probability weight associated with no energy loss, which cannot be captured by the Gaussian distribution, motivating a study of the validity of the Gaussian assumption.

In this work, we utilize an elastic energy loss kernel \cite{Wicks:2008zz}, which is derived in the hard thermal loop (HTL) \cite{Braaten:1989mz, Klimov:1982bv, Pisarski:1988vd, Weldon:1982aq, Weldon:1982bn} formalism but keeps the full kinematics of the hard exchanges (as opposed to a strict HTL calculation). With this approach, one may calculate the differential number of elastic scatters, allowing insight into the shape of the elastic energy loss distribution. This facilitates the treatment of the elastic and radiative energy loss in an identical manner by assuming that the number of scatters and number of radiated gluons emitted are separately independent and, therefore, that we may convolve the distributions according to a Poisson distribution \cite{Gyulassy:2001nm}.  
\
We also present results where the HTL elastic energy loss is modeled as a Gaussian distribution according to the central limit theorem \cite{Moore:2004tg}, which allows us to assess the validity of the Gaussian approximation. Finally, we compare the HTL-based and the Braaten and Thoma elastic energy loss models, which differ predominantly in the usage of vacuum and HTL propagators, to probe the theoretical uncertainty in the transition between HTL and vacuum propagators \cite{Romatschke:2004au,Gossiaux:2008jv,Wicks:2008zz}.	

	In this paper, we present results from our model for the nuclear modification factor of final state pions, $D$ mesons, and $B$ mesons produced in central, semi-central, and peripheral \coll{Au}{Au} and \coll{Pb}{Pb} collisions as well as central \coll{p}{Pb} and \collFour{p}{d}{He3}{Au} collisions. Our primary goals in this work are to introduce our energy loss model, which incorporates small system size corrections to both the radiative and elastic energy loss, and to investigate the impact of the Gaussian approximation and the transition between HTL and vacuum propagators in elastic energy loss calculations. We will show that while our results are largely insensitive to the Gaussian approximation, they are acutely sensitive to the uncertainty in the transition between HTL and vacuum propagators. This sensitivity leads to significantly different suppression predictions when using HTL versus BT elastic energy loss kernels. In principle, these differences could be absorbed into a change of the strong coupling constant to provide a more fair comparison with data.
As in our previous work \cite{Faraday:2023mmx}, we observe that the short pathlength correction becomes extremely large at high-$p_T$, which we previously concluded is likely unphysical due to the breakdown of the large formation time assumption in the model at high-$p_T$. Given the aforementioned goals of this work and the substantial theoretical uncertainties in our current approach, we do \emph{not} make comparison with experimental data from RHIC and the LHC.
Work in preparation \cite{Faraday:2024} will address the large formation time assumption breakdown by implementing a cut on the transverse radiated gluon momentum, which ensures that no contributions from regions where this assumption breaks down are included. Additionally, this future work will perform a global fit of the strong coupling to large system data from RHIC and the LHC. This global fit will enable us to draw quantitative conclusions about the consistency of suppression measurements in small and large systems at RHIC and LHC, as well as elucidate the qualitative and quantitative differences in predictions from HTL and BT energy loss calculations, which probes the uncertainty in the transition region between HTL and vacuum propagators. We believe that since a detailed analysis is underway, comparison with data in this paper would be premature. Moreover, given the quantity of content in this paper, incorporating such a comparison would obscure the focus of this discussion.

\section{Energy loss model}
\label{sec:model}

This work builds on the energy loss model described in detail in our previous work \cite{Faraday:2023mmx}, which is itself based on the Wicks-Horowitz-Djordjevic-Gyulassy (WHDG) energy loss model \cite{Wicks:2005gt}. In this work, we will focus on the effect of incorporating a more realistic elastic energy loss kernel and, as such, keep most other components of the model the same as in our previous work \cite{Faraday:2023mmx}. This section briefly reviews the various components of the energy loss model.

\subsection{Radiative energy loss}
\label{sec:radiative_energy_loss}

\subsubsection{DGLV radiative energy loss}
\label{sec:dglv_radiative_energy_loss}

The Djordjevic-Gyulassy-Levai-Vitev (DGLV) opacity expansion \cite{Djordjevic:2003zk, Gyulassy:1999zd} gives the inclusive differential distribution of medium-induced gluon radiation from a high-$p_T$ parent parton moving through a smooth brick of QGP. The expansion is in the expected number of scatterings or the \textit{opacity} $L / \lambda_g$, where $L$ is the length of the QGP brick and $\lambda_g$ is the mean free path of a gluon in the QGP.

The DGLV single inclusive gluon radiation spectrum is, to first order in opacity \cite{Gyulassy:2000er,Djordjevic:2003zk},
\begin{widetext}
	\begin{gather}
  \frac{\mathrm{d} N^g_{\text{DGLV}}}{\mathrm{d} x}=  \frac{C_R \alpha_s L}{\pi \lambda_g} \frac{1}{x} \int \frac{\mathrm{d}^2 \mathbf{q}_1}{\pi} \frac{\mu^2}{\left(\mu^2+\mathbf{q}_1^2\right)^2} \int \frac{\mathrm{d}^2 \mathbf{k}}{\pi} \int \mathrm{d} \Delta z \, \bar{\rho}(\Delta z) \nonumber\\
 -\frac{2\left\{1-\cos \left[\left(\omega_1+\tilde{\omega}_m\right) \Delta z\right]\right\}}{\left(\mathbf{k}-\mathbf{q}_1\right)^2+m_g^2+x^2 M^2}\left[\frac{\left(\mathbf{k}-\mathbf{q}_1\right) \cdot \mathbf{k}}{\mathbf{k}^2+m_g^2+x^2 M^2}-\frac{\left(\mathbf{k}-\mathbf{q}_1\right)^2}{\left(\mathbf{k}-\mathbf{q}_1\right)^2+m_g^2+x^2 M^2}\right].
 \label{eqn:DGLV_dndx}
\end{gather}
\end{widetext}
In \cref{eqn:DGLV_dndx} we have made use of the shorthand $\omega \equiv x E^+ / 2,~\omega_0 \equiv \mathbf{k}^2 / 2 \omega,~\omega_i \equiv (\mathbf{k} - \mathbf{q}_i)^2 / 2 \omega$, $\mu_i \equiv \sqrt{\mu^2 + \mathbf{q}_i^2}$, and $\tilde{\omega}_m \equiv (m_g^2 + M^2 x^2) / 2 \omega$ following \cite{Djordjevic:2003zk, Kolbe:2015rvk}. Additionally, $\mathbf{q}_i$ is the transverse momentum of the $i^{\mathrm{th}}$ gluon exchanged with the medium; $\mathbf{k}$ is the transverse momentum of the radiated gluon; $\Delta z$ is the distance between production of the hard parton, and scattering; $C_R$ ($C_A$) is the quadratic Casimir of the hard parton (adjoint) representation ($C_F = 4 / 3$ [quarks] and $C_A = 3$ [gluons]); $x$ is the fraction of parent parton's plus momentum carried away by the radiated gluon; $M$ is the mass of the incident parton; $m_g \approx \mu / \sqrt{2}$ is the mass of the gluon \cite{Wicks:2005gt}; and $\alpha_s$ is the strong coupling.

 The quantity $\bar{\rho}(\Delta z)$ is the \emph{distribution of scattering centers} in $\Delta z$ and is defined in terms of the density of scattering centers $\rho(\Delta z)$ in a static brick,
\begin{equation}
  \rho(\Delta z) = \frac{N}{A_{\perp}} \bar{\rho}(\Delta z),
  \label{eqn:density_scattering_centers}
\end{equation}
where $\Delta z$ is in the direction of propagation, $N$ is the number of scattering centers, $A_{\perp}$ is the perpendicular area of the brick, and $\int \mathrm{d}z \; \bar{\rho}(\Delta z) = 1$. The analysis of realistic collision geometries adds complexity to the scenario, as detailed in \cref{sec:geometry}.

\subsubsection{Short pathlength correction to DGLV radiative energy loss}
\label{sec:radiative_energy_loss_correction}
The derivation of the modification to the radiative energy loss in the DGLV \cite{Vitev:2002pf,Djordjevic:2003zk} opacity expansion approach with the relaxation of the large pathlength assumption $L \gg \mu^{-1}$ was considered in \cite{Kolbe:2015rvk,Kolbe:2015suq}. In the derivation of the short pathlength correction, all assumptions and approximations made in the original GLV and DGLV derivations were kept, except that the short pathlength approximation $L\gg\mu^{-1}$ was relaxed. The single inclusive radiative gluon distribution, including both the original DGLV contribution as well as the short pathlength correction, is
\begin{widetext}
	\begin{gather}
    \frac{\mathrm{d} N^g_{\text{DGLV+corr}}}{\mathrm{d} x}=  \frac{C_R \alpha_s L}{\pi \lambda_g} \frac{1}{x} \int \frac{\mathrm{d}^2 \mathbf{q}_1}{\pi} \frac{\mu^2}{\left(\mu^2+\mathbf{q}_1^2\right)^2} \int \frac{\mathrm{d}^2 \mathbf{k}}{\pi} \int \mathrm{d} \Delta z \, \bar{\rho}(\Delta z) \nonumber\\
   \times\left[-\frac{2\left\{1-\cos \left[\left(\omega_1+\tilde{\omega}_m\right) \Delta z\right]\right\}}{\left(\mathbf{k}-\mathbf{q}_1\right)^2+m_g^2+x^2 M^2}\left[\frac{\left(\mathbf{k}-\mathbf{q}_1\right) \cdot \mathbf{k}}{\mathbf{k}^2+m_g^2+x^2 M^2}-\frac{\left(\mathbf{k}-\mathbf{q}_1\right)^2}{\left(\mathbf{k}-\mathbf{q}_1\right)^2+m_g^2+x^2 M^2}\right] \right. \nonumber\\
   +\frac{1}{2} e^{-\mu_1 \Delta z}\left(\left(\frac{\mathbf{k}}{\mathbf{k}^2+m_g^2+x^2 M^2}\right)^2\left(1-\frac{2 C_R}{C_A}\right)\left\{1-\cos \left[\left(\omega_0+\tilde{\omega}_m\right) \Delta z\right]\right\}\right. \nonumber\\
   \left.\left.+\frac{\mathbf{k} \cdot\left(\mathbf{k}-\mathbf{q}_1\right)}{\left(\mathbf{k}^2+m_g^2+x^2 M^2\right)\left(\left(\mathbf{k}-\mathbf{q}_1\right)^2+m_g^2+x^2 M^2\right)}\left\{\cos \left[\left(\omega_0+\tilde{\omega}_m\right) \Delta z\right]-\cos \left[\left(\omega_0-\omega_1\right) \Delta z\right]\right\}\right)\right],
   \label{eqn:full_dndx}
\end{gather}
\end{widetext}
\noindent where the first two lines of the above equation are the original DGLV result \cite{Gyulassy:2000er,Djordjevic:2003zk}, \cref{eqn:DGLV_dndx}, while the last two lines are the short pathlength correction.
We emphasize that contributions from all diagrams which are not suppressed under the relevant assumptions are included. Of particular importance is the large formation time assumption, which allows one to systematically neglect a significant number of diagrams in both the original DGLV derivation \cite{Vitev:2002pf,Djordjevic:2003zk} and in the short pathlength correction \cite{Kolbe:2015rvk,Kolbe:2015suq}. For a detailed discussion of the self-consistency of the various approximations made in both the DGLV result and the DGLV result which receives a short pathlength correction, refer to our previous work \cite{Faraday:2023mmx}.

The short pathlength corrected DGLV radiative energy loss has unique qualitative features \cite{Kolbe:2015rvk, Kolbe:2015suq} including:
\begin{itemize}
	\item It grows \textit{linearly} in $L$ as opposed to the original DGLV result which grows as $L^2$.
	\item It breaks \textit{color triviality}, i.e.\ $dN^g / dx$ for incident gluons is not simply $C_A / C_F$ times that of incident quarks. We will see that this breaking of color triviality leads to the correction being anomalously large for gluons cf.\ quarks.
	\item It is \textit{nonzero for all lengths} because the $\Delta z$ dependence is integrated over, leading to contributions from small $\Delta z$ even at large lengths.
	\item It grows \textit{asymptotically faster in energy} than the uncorrected DGLV radiative energy loss ($\Delta E \sim E$ vs $\Delta E \sim \ln E$ for the corrected and uncorrected results, respectively).
	\item It \textit{reduces the radiative energy loss}, which can lead to less energy loss in the medium than in vacuum.
\end{itemize}
We will see these features numerically \coleTwo{at the level of $\Delta E / E$ in \cref{sec:numerical_elastic_radiative} and at the level of $R_{AA}$ in \cref{sec:results}}, and the asymptotics are discussed in depth in \cite{Faraday:2023mmx, Kolbe:2015rvk, Kolbe:2015suq}.

\subsection{Elastic energy loss}
\label{sec:elastic_energy_loss}

We will consider three different elastic energy loss \coleTwo{models} in this work, all of which utilize the Hard Thermal Loops (HTL) effective field theory \cite{Braaten:1989mz, Klimov:1982bv, Pisarski:1988vd, Weldon:1982aq, Weldon:1982bn}.

In \cref{sec:braaten_thoma_elastic}, we present a short review of the elastic energy loss derived by Braaten and Thoma (BT) \cite{Braaten:1991jj, Braaten:1991we}. The BT elastic energy loss kernel utilizes HTL propagators for soft momentum transfer and vacuum pQCD propagators for hard momentum transfer. The BT elastic energy loss kernel was used in our previous work \cite{Faraday:2023mmx} and in the WHDG energy loss model \cite{Wicks:2005gt}.

\Cref{sec:pure_htl_elastic_energy_loss} details the elastic energy loss kernel calculated purely with HTL propagators \cite{Wicks:2008zz}, which we will label \textit{Poisson HTL}. This kernel corresponds to the ``HTL-X1" prescription in \cite{Wicks:2008zz}, which calculates all energy loss with HTL propagators but includes the full kinematics of the hard scatters. We present a summary of the derivation of the HTL elastic energy loss kernel in \cref{sec:pure_htl_elastic_energy_loss}, due to a few typographical errors in the original work \cite{Wicks:2008zz}. 
Finally, we present results with the same average energy loss as the Poisson HTL results but with a Gaussian distribution, which we will label \textit{Gaussian HTL}. The Poisson and Gaussian distribution implementations are discussed in \cref{sec:probability_of_energy_loss_distributions}.

	This set of elastic energy loss distributions facilitates two important comparisons in this work. The comparison of the Gaussian BT and Gaussian HTL results shows the sensitivity of the $R_{AA}$ to the magnitude of the elastic energy loss, while disregarding the difference between Poisson and Gaussian distributions. Such a comparison is sensitive to the fundamental theoretical uncertainty in the transition region between the HTL and vacuum propagators. Comparing the Gaussian HTL and Poisson HTL results measures the sensitivity of the $R_{AA}$ to the distribution used. Through this comparison, we can assess the importance of more precise modeling of the elastic energy loss distribution distribution and evaluate the impact of the central limit theorem approximation used in our previous work \cite{Faraday:2023mmx} and the literature \cite{Wicks:2005gt, Horowitz:2011gd, Zigic:2021rku}.

\subsubsection{HTL vs vacuum propagators}
\label{sec:htl_vs_vacuum_propagators}

In our previous work \cite{Faraday:2023mmx}, we used the elastic energy loss calculated by Braaten and Thoma \cite{Braaten:1991jj, Braaten:1991we}. This formalism accounts for soft contributions to the elastic energy loss through HTL screened gluon propagators, which shield the infrared divergence. The hard contribution is calculated with vacuum propagators. The hard region is defined for momentum transfers $q \gtrsim T$, while the soft region is defined for $q \lesssim g T$, where $T$ is the temperature and $g$ is the strong coupling \cite{Braaten:1991jj}. 
In order to proceed, one must introduce an intermediary scale $q^*$ chosen such that $gT \ll q^* \ll T$, such that both the soft and hard regions are extrapolated outside of their domain of validity. This intermediary scale is necessary since neither the vacuum nor the HTL calculation is on a rigorous footing in the intermediary region. 
In the small coupling and high-temperature limit, the dependency on the transition scale $q^*$ falls out at the level of the fractional energy loss to leading logarithm accuracy \cite{Romatschke:2004au,Gossiaux:2008jv}. If one does not take the strict soft coupling and high-temperature limit, one finds a significantly increased sensitivity to the cutoff momentum chosen \cite{Romatschke:2004au, Wicks:2008zz}. 

One approach to treat this uncertainty \cite{Gossiaux:2008jv} is to use a ``semi-hard" gluon propagator $\sim 1 / (q^2 - \kappa \mu^2)$, where $\kappa$ is a dimensionless parameter, for momentum transfers $|q| > |q^*|$ in conjunction with the standard HTL propagator for soft momentum transfers $|q| < |q^*|$. 
The parameter $\kappa$ is then chosen such that the dependence on the intermediary scale $q^*$ is minimized.

In this current work, we will not make such a prescription but rather treat the uncertainty as a fundamental theoretical uncertainty by examining two extreme cases: the Braaten and Thoma and the HTL elastic energy loss. %
Future work could more rigorously capture this uncertainty by varying the intermediary scale $q^*$.

\subsubsection{Braaten and Thoma elastic energy loss}
\label{sec:braaten_thoma_elastic}

The BT elastic energy loss \cite{Braaten:1991jj, Braaten:1991we} of a quark is calculated in the regions $E \ll M^2 / T$ and $E \gg M^2 / T$, where $M$ is the mass of the incident quark, and $T$ is the temperature of the medium. For $E \ll M^2 / T$ the differential energy loss per unit distance is
\begin{multline}
\frac{\mathrm{d} E}{\mathrm{d} z} = 2 C_R \pi \alpha_s^2 T^2 \left[\frac{1}{v}- \frac{1-v^2}{2 v^2}\log \frac{1+v}{1-v}\right]\\
\times \log \left(2^{\frac{n_f}{6+n_f}} B(v) \frac{E T}{m_g M}\right)\left(1+\frac{n_f}{6}\right),
\label{eqn:elastic_energy_loss_low}
\end{multline}
where $B(v)$ is a smooth function satisfying constraints listed in \cite{Braaten:1991we}, $v$ is the velocity of the hard parton, and $n_f$ is the number of active quark flavors in the plasma (taken to be $n_f = 2$ throughout). For $E \gg M^2/T$ the differential energy loss per unit distance is
\begin{multline}
    \frac{\mathrm{d} E}{\mathrm{d} z} = 2 C_R \pi \alpha_s^2 T^2 \left(1 + \frac{n_f}{6}\right) \\
    \times \log \left(2^{\frac{n_f}{2(6+n_f)}} \, 0.92 \frac{\sqrt{E T}}{m_g}\right).
    \label{eqn:elastic_energy_loss_high}
\end{multline}
The energy loss at arbitrary incident energy is taken to be the connection of these two asymptotic results such that $\mathrm{d} E / \mathrm{d} z$ is continuous (determined numerically). 
The calculation was performed explicitly \cite{Braaten:1991jj, Braaten:1991we} for incident quarks with a Casimir of $C_R = C_F$, and incident gluons are taken into account by a Casimir change of $C_R = C_A$.

\subsubsection{Pure HTL elastic energy loss}
\label{sec:pure_htl_elastic_energy_loss}

In this section we present an outline of the steps involved in the HTL calculation of the elastic energy loss, following \cite{Wicks:2008zz, Braaten:1991jj, Braaten:1991we}. We duplicate the key steps in the calculation to highlight the differences in approaches in the literature and to fix some minor typographical errors in the equations in \cite{Wicks:2008zz}. 
The HTL elastic energy loss calculation differs from the BT result in a few key ways. First, the energy loss is computed for all momentum transfers (both hard and soft) with the HTL propagators (the ``HTL-X1" procedure in \cite{Wicks:2008zz}). Second, the full kinematics of the hard momentum transfers are kept, i.e.\ we do not make the following approximations, which are made in the BT \cite{Braaten:1991jj, Braaten:1991we} calculation:
\begin{enumerate}
	\item The expansion $1 + n_B(\omega) \approx \frac{T}{\omega}$,
	\item The partitioning of the calculation into regions where $E \ll M^2 / T$ and $E \gg M^2 / T$ respectively.
\end{enumerate}
This procedure is similar to the procedure used in \cite{Djordjevic:2006tw}, while the BT procedure is more akin to that used in \cite{Schenke:2009gb}. %

We begin by considering the $t$-channel matrix element $\mathcal{M}$ for the collision of two partons in a QGP, which is the dominant contribution to the full elastic scattering process in the eikonal limit. 
We will also only do the calculation for the scattering of a quark off of a quark and make the approximation that the gluon-quark and gluon-gluon scattering processes differ only by changing the relevant Casimirs. We can write the spin and color-averaged matrix element for an incident parton $q$ scattering off a medium parton $m$ as
\begin{equation}
	\left\langle \left| \mathcal{M}_{q m} \right|^{2}  \right\rangle  = \frac{1}{2 N_q} \frac{1}{2 N_m} \sum_{\text{spins, colors}} \left| \mathcal{M}_{qm} \right|^{2},
	\label{eqn:color-spin-average-matrix-squared}
\end{equation}
where $N_{q / m}$ is the number of colors of the parton $q / m$.

One may write down the interaction rate $dN / d z$, where $N$ is the number of elastic scatters and $z$ is the distance that the parton has traveled through the medium, for a hard parton $q$ as \cite{Wicks:2008zz} %
\begin{align}
\begin{split}
	\frac{dN}{dz} =& \frac{1}{2 E} \int \lips{p} \lips{k} \lips{k'} \times (2 \pi)^4 \delta^4(p + k - p' - k')\\
	&\times \sum_m n_m(k^0)(1 \pm_m n_m(k^{\prime 0}) \left\langle \left| \mathcal{M}_{q m} \right|^{2}  \right\rangle,
\end{split}
	\label{eqn:scattering_rate_first_step}
\end{align}
where $E$ is the energy of the incident parton $q$, the index $m$ sums over all of the medium partons ($N_c^2 - 1$ gluons and $4 N_c N_f$  for the active quark flavors), $\lips{p} \equiv d^3 \vecIII{p} / (2 \pi)^3 2 \vecIV{p}^0$  is the Lorentz invariant phase space, $n_m$ is the statistical distribution (Bose distribution for gluons and Fermi distribution for quarks), and $\pm_m$ is $+$ for bosons and $-$ for fermions.

The HTL quark propagator in the Coulomb gauge is given by \cite{Wicks:2008zz, Blaizot:2001nr, Bellac:2011kqa} %
\begin{gather}
	D_{\mu \nu}(Q) = Q_{\mu \nu}(Q) \Delta_L (Q)  + P_{\mu \nu}(Q) \Delta_T(Q),\label{eqn:HTL_propagator}\\
	\Delta_{L,T}(Q) = \frac{1}{Q^2 - \Pi_{L,T}(Q)}\label{eqn:delta_L_T}\\
	Q_{00}=\left(1-x^2\right),\quad P_{i j}=-\left(\delta_{i j}-\hat{q_i} \cdot \hat{q_j}\right)\\
	\Pi_L= \mu^2 (1 - x^2) \left(1 - \frac{x}{2} \ln \frac{x+1}{x-1}\right) \\
	\Pi_T = \frac{\mu^2}{2} \left( x^2 - \frac{x(1-x^2)}{2} \ln \frac{x+1}{x-1}\right) + \mu_M^2
\end{gather}
where $\hat{q_i} \equiv \vecIII{\hat{q_i}} / |\vecIII{\hat{q_i}}|$, $x \equiv \omega / q$, and all other components of $Q_{\mu \nu}$ and $P_{\mu \nu}$ are zero. 
Following \cite{Wicks:2008zz}, we include an \textit{ad hoc} magnetic mass $\mu_M$ in the transverse propagator $\Delta_T$ which is not present in standard HTL, although lattice calculations \cite{Nakamura:2003pu, Hart:2000ha} indicate that there is a nonzero magnetic mass in phenomenologically relevant QGPs. In this work we will take $\mu_M = \mu$ in all results for simplicity.

 Following \cite{Wicks:2008zz}, we perform the integral in \cref{eqn:scattering_rate_first_step} which collapses the energy-momentum conserving delta functions in $\left\langle |\mathcal{M}_{qm}|^2 \right\rangle$ and leads to
\begin{multline}
		\frac{dN}{d z} = \frac{4 C_R \alpha_s^2}{\pi} \int_q \frac{n_B(\omega)}{q} \left( 1 - \frac{\omega^2}{q^2} \right)^2 \times \\
		\sum_m\left[C_{L L}^{m}\left|\Delta_L\right|^2+2 C_{L T}^{m} \operatorname{Re}\left(\Delta_L \Delta_T\right)+C_{T T}^m\left|\Delta_T\right|^2\right],
	\label{eqn:scattering_rate_final}
\end{multline}
where
\begin{equation}
	\int_q \equiv \int \frac{d^3 \vecIII{q} d \omega}{2 \pi} \frac{E}{E^{\prime}} \delta\left(\omega+E-E^{\prime}\right).
	\label{eqn:integral_q}
\end{equation}
The transverse part of \cref{eqn:integral_q} can be performed trivially under the assumption that the medium is isotropic \cite{Wicks:2008zz}. One then performs the integral over $q=| \vecIII{q}|$ numerically, and the integral over $\omega$ will be changed to an integral over the fractional energy loss $\epsilon$ defined via $\omega = \epsilon E$, and left unevaluated since the quantity we want to calculate is $dN / d \epsilon$.

The coefficients $C_{L L}^{j m}$, $C_{L T}^{j m}$, and $C_{T T}^m$ are given in terms of various thermal integrals \cite{Wicks:2008zz}
\begin{equation}
	C^{m} \equiv \int_{\frac{1}{2}(\omega+q)}^{\infty} k^0 k d k\left(n_{m}\left(k^0-\omega\right)-n_{m}\left(k^0\right)\right) c^{\prime}
\end{equation}
where the coefficients $c^\prime$ are
\begin{align}
	c_{L L}^{\prime} \equiv  & \left(\left(1+\frac{\omega}{2 E}\right)^2-\frac{q^2}{4 E^2}\right)\left(\left(1-\frac{\omega}{2 k^0}\right)^2-\frac{q^2}{4\left(k^0\right)^2}\right) \\
	c_{L T}^{\prime} \equiv  & 0 \\
	c_{T T}^{\prime} \equiv  & \frac{1}{2}\left(\left(1+\frac{\omega}{2 E}\right)^2+\frac{q^2}{4 E^2}-\frac{1-v^2}{1-\frac{\omega^2}{q^2}}\right) \\
		& \left(\left(1+\frac{\omega}{2 k^0}\right)^2+\frac{q^2}{4\left(k^0\right)^2}-\frac{1-v_k^2}{1-\frac{\omega^2}{q^2}}\right).
\end{align}
In the above $v_k (v)$ is the velocity of the medium parton (incident parton), and the $c^\prime$ are independent of whether the medium parton is a quark or a gluon since we have assumed that the quarks and gluons differ only by the statistical distribution and the Casimir. One may then calculate the $C$ coefficients in terms of various thermal integrals, which one may perform analytically
\begin{equation}
	I_n^m \equiv \int_{\frac{1}{2} (\omega + q)}^\infty dk \; (n_m (k-\omega) - n_m(k)) k^n.
	\label{eqn:general_thermal_integral}
\end{equation}
We provide the results of these integrals for bosons ($+$) and fermions ($-$)
\begin{align*}
I_0^\pm &= \pm T \log\left(\frac{1 \mp e^{-\kappa_+}}{1 \mp e^{-\kappa_-}}\right) \\
I_1^\pm &= \pm T^2 \left[\operatorname{Li}_2\left(\pm e^{-\kappa-}\right) - \operatorname{Li}_2\left(\pm e^{-\kappa+}\right)\right] + \kappa_+ T I_0^\pm \\
I_2^\pm &= \pm 2 T^3 \left[\operatorname{Li}_3\left(\pm e^{-\kappa-}\right) - \operatorname{Li}_3\left(\pm e^{-\kappa+}\right)\right] \\
&\quad+ 2 \kappa_+ T I_1^\pm - \kappa_+^2 T^2 I_0^\pm,
\end{align*}
where $\kappa_{\pm} = (\omega \pm q)/2 T$ and $\operatorname{Li}_n$ is the polylogarithm function.

Finally, one may perform a change of variables from $\omega$ to $\epsilon$ where $\omega = \epsilon E$ to convert from the scattering rate $d N / d z$ to the single elastic scattering kernel $d N / d \epsilon$. Schematically, this change of variables can be written as
\begin{align}
	\frac{dN}{d \epsilon}  &= \frac{d\omega}{d \epsilon} \frac{d}{d \omega} \int d z \frac{dN}{d z} \\
	&\simeq L E \frac{dN}{d z \; d\omega},
	\label{eqn:dndx_pure_htl}
\end{align}
where practically $dN /d z d \omega$ is simply $dN / d z$ but with the integral over $\omega$ removed from $\int_q$ in \cref{eqn:integral_q}. In the last step, we have assumed that $d N /dz$ is independent of $z$, which is true in a static brick. In future work \cite{Bert:2024}, we will consider a more realistic model for the medium, which includes the $z$ dependence in the temperature $T(z)$.
The single elastic scattering kernel $dN / d \epsilon$ is the analog of the single emission kernel in the radiative energy loss \cite{Djordjevic:2003zk, Gyulassy:2000er, Gyulassy:2001nm} in \cref{sec:radiative_energy_loss}.

\subsection{Energy loss probability distributions}
\label{sec:probability_of_energy_loss_distributions}

\coleTwo{Energy loss calculations are typically divided into two steps: calculating the energy loss for 1) a single inclusive gluon emission or a single elastic collision and 2) modeling the effects of multiple gluons emission or multiple scattering. 
Generally, the average energy loss $\Delta E / E$ can be calculated analytically; however, higher order moments\footnote{\coleTwo{The effect of higher order moments will be discussed in detail in \cref{sec:distribution_dependence,sec:validity_of_power_law_approximation_to_r_aa}.}} of the probability distribution significantly impact the $R_{AB}$ \cite{Gyulassy:2001nm}, which introduces an additional modeling requirement.
\Cref{sec:radiative_energy_loss,sec:elastic_energy_loss} considered the analytic calculation of the energy loss for a single inclusive gluon emission or a single elastic collision, respectively. In this section, we will consider the impact of multiple radiated gluons and multiple elastic collisions on the energy loss probability distribution. }

\subsubsection{Gaussian approximation for elastic energy loss}
\label{sec:gaussian_distribution_approximation}

For the BT elastic energy loss and the Gaussian HTL elastic energy loss, we assume that there are enough elastic scatters such that the elastic energy loss follows a Gaussian distribution according to the central limit theorem. This assumption implies that the distribution of elastic energy loss is Gaussian with the mean provided by average elastic energy loss $\Delta E$, and width by the fluctuation dissipation theorem \cite{Moore:2004tg, Xu:2014ica}
\begin{equation}
    \sigma = \frac{2}{E} \int \mathrm{d}z \; \frac{\mathrm{d} E}{\mathrm{d}z} T(z),
    \label{eqn:sigma_elastic}
\end{equation}
where $z$ integrates along the path of the parton, and $T(z)$ is the temperature along the path. Therefore
\begin{multline}
  P_{\text{el}}(E_f | E_i,\,L,\,T) \\ \equiv \frac{1}{\sqrt{2\pi}\sigma}\exp\left[{-}\left( \frac{E_f-(E_i+\Delta E)}{\sqrt{2} \sigma} \right)^2 \right],
\end{multline}
where $\Delta E = \int dz \; \frac{dE}{dz}$. We can perform a change of variables from $E_f$ to $\epsilon$ where $E_f = (1-\epsilon) E_i$ resulting in the probability for a particle of initial energy $E_i$ losing a fraction $\epsilon$ of its energy

\begin{align}
	&P_{\text{el}}(\epsilon | E_i,\,L,\,T) \\ 
	&\equiv \frac{E_i}{\sqrt{2\pi}\sigma}\exp\left[{-}\left( \frac{(1-\epsilon)E_i-(E_i+\Delta E)}{\sqrt{2} \sigma} \right)^2 \right].\\
	&= \frac{1}{\sqrt{2\pi}\sigma'}\exp\left[{-}\left(\frac{\epsilon - \Delta E/E_i}{\sqrt{2} \sigma'} \right)^2 \right],
\end{align}
where $\sigma' = \sigma / E_i$.

\subsubsection{Poisson distribution for radiative and elastic energy loss}
\label{sec:multi_scatters}

The DGLV energy loss kernel (\ref{eqn:full_dndx}) gives the inclusive spectrum of medium-induced gluon emission. Thus, the expected number of radiated gluons may be different from one. In fact, one sees that for hard partons emerging from the center of a central heavy ion collision, the expected number of emitted gluons is $\sim 3$ \cite{Gyulassy:2001nm}. Similarly \cref{eqn:dndx_pure_htl} gives the differential number of scatters for the elastic energy loss, which is generically also different from one \cite{Wicks:2008zz}. We may treat these two situations identically, following the procedure outlined in \cite{Gyulassy:2001nm}. This treatment requires assuming that the gluon emissions and elastic scatters are both separately independent, which allows us to model gluon emission and elastic scattering with a Poisson distribution.

Explicitly, we can write 
\begin{equation}
  P(\epsilon | E)=\sum_{n=0}^{\infty} P_n(\epsilon | E),
  \label{eqn:poisson}
\end{equation}
where the $P_n$ are found via the convolution
\begin{align}
  P_{n+1}(\epsilon) & =\frac{1}{n+1} \int \mathrm{d} x_n \; \frac{\mathrm{d} N^{g}}{\mathrm{d} x} \; P_n(\epsilon-x_n)
    \label{eqn:pn}
\end{align}
and we have $P_0(\epsilon) \equiv e^{- \langle N^g \rangle} \delta(\epsilon)$. Here, and for the rest of the paper, we define $1-\epsilon$ as the fraction of initial momentum kept by the parton, such that in the eikonal limit the final energy of the parton in terms of the initial energy of the parton is $E_f\equiv(1-\epsilon)E_i$. The Poissonian form of \cref{eqn:poisson} guarantees the distribution is normalized to one and that the expected number of emitted gluons (number of elastic scatters) is $\sum_n\int \mathrm{d} \epsilon \: n \, P_n(\epsilon, E) = \langle N^g \rangle \; \left(\left\langle N_{\text{scatters}} \right\rangle\right)$. The support of $P(\epsilon)$ past $\epsilon = 1$ is unphysical and is interpreted as the probability for the parton to lose all of its energy before exiting the plasma. The fact that the support of $P(\epsilon)$ is not naturally constrained to be less than one stems from the fact that we do not update the parton's energy after each collision. Under this interpretation we include the excess weight $\int_1^\infty \mathrm{d} \epsilon \; P_{\text{rad}}(\epsilon)$ as the coefficient of a Dirac delta function at $\epsilon=1$.

\subsubsection{Total energy loss distribution}%
\label{sec:total_energy_loss}

As done in \cite{Wicks:2005gt}, we convolve the radiative and elastic energy loss probabilities to yield a total probability of energy loss,
\begin{align}
  P_{\text{tot}}(\epsilon) \equiv \int \mathrm{d} x \, P_{\text{el}}(x)P_{\text{rad}}(\epsilon-x).
\end{align}

Note that both the radiative distribution $P_{\text{rad.}}(\epsilon)$ and $P_{\text{el.}}(\epsilon)$ may contain Dirac delta functions at both $\epsilon = 0$ and $\epsilon = 1$, which correspond to the probability the parton losing no energy and the parton losing all of its energy before exiting the QGP, respectively. The presence of these delta functions implies that the total energy loss has a delta function at $\epsilon = 0$ with a weight of the product of the radiative and elastic delta functions at $\epsilon =0$. When one uses a Gaussian form of the elastic energy loss probability distribution, then the elastic energy loss does not contain a delta function at $\epsilon =0$, which in turn implies that there is no delta function at $\epsilon = 0$ in the total energy loss distribution.

\vspace{3em}

\subsection{Nuclear modification factor}
\label{sec:nuclear_modification_factor}

The observable that we will be computing is the nuclear modification factor $R^h_{AB}(p_T)$ for hadrons $h$ produced in a collision system \coll{A}{B}, defined by 
\begin{equation}
    R^h_{AB}(p_T) \equiv \frac{\mathrm{d} N^{AB \to h} / \mathrm{d} p_T}{\langle N_{\text{coll}} \rangle \mathrm{d} N^{pp \to h} / \mathrm{d} p_T},
    \label{eqn:nuclear_modification_factor}
\end{equation}
where $\mathrm{d} N^{AB / pp \to h} / \mathrm{d} p_T$ is the differential number of measured $h$ hadrons in \coll{A}{B} / \coll{p}{p} collisions, and $\langle N_{\text{coll}} \rangle$ is the expected number of binary collisions usually calculated according to the Glauber model \cite{Glauber:1970jm,Miller:2007ri}.

 We make several assumptions about the underlying collision and partons to access the $R_{AB}$ \emph{theoretically}. In the following, we will only refer to quarks, but all assumptions and formulae apply equally well to gluons. We first assume, following \cite{Wicks:2005gt,Horowitz:2010dm}, that the spectrum of produced quarks $q$ in the initial state of the plasma formed by the collisions \coll{A}{B} (before energy loss) is $\mathrm{d} N^{q}_{AB\text{; init.}} / \mathrm{d} p_i = N_{\text{coll}} \times \mathrm{d}N^q_{pp} / \mathrm{d}p_i$, where $\mathrm{d} N^q_{pp} / \mathrm{d} p_i$ is the quark production spectrum in \coll{p}{p} collisions. This assumption is equivalent to neglecting initial state effects, which is justified in heavy-ion collisions by the measured $R_{AA}$ consistent with unity for probes which do not interact strongly with the QGP \cite{CMS:2012oiv, CMS:2011zfr}.
In addition, we assume that the modification of the \coll{A}{B} production spectrum $\mathrm{d} N^{q}_{AB \text{; init.}}$ is due to energy loss,
\begin{equation}
d N^q_{AB; \text{ final}}\left(p_T\right)=\int  d \epsilon \; d N^q_{AB; \text{ init.}}\left(p_i\right) P_{\text{tot}}\left(\epsilon \mid p_i\right),
\end{equation}
where $P_{\text{tot}}(\epsilon | p_i)$ is the  probability of losing a fraction of transverse momentum $\epsilon$ given an initial transverse momentum $p_i$. Given these assumptions, the expression for the $R_{AB}$ in \cref{eqn:nuclear_modification_factor} simplifies to
\begin{equation}
    R_{AB} = \frac{1}{f\left(p_t\right)} \int \frac{d \epsilon}{1-\epsilon} f\left(\frac{p_t}{1-\epsilon}\right) P_{\text {tot}}\left(\epsilon \left| \frac{p_t}{1-\epsilon}\right.\right),
    \label{eqn:full_raa_spectrum_ratio}
\end{equation}
where the $1 / 1 - \epsilon$ factor is a Jacobian, and we have introduced the notation $f(p_T) \equiv dN^q / d p_T (p_T)$. In the literature \cite{Wicks:2005gt, Horowitz:2010dm} %
and in our previous work \cite{Faraday:2023mmx}, an assumption was made that the production spectra followed a slowly-varying power law $f(p_T) \simeq A p_T^{-n(p_T)}$, which resulted in further simplifications. In this work, we calculate all presented $R_{AB}$ results according to \cref{eqn:full_raa_spectrum_ratio}, and we discuss the validity of the slowly-varying power law assumption in \cref{sec:validity_of_power_law_approximation_to_r_aa}.

	We calculate heavy quark production spectra $f(p_T)$ using FONLL \cite{Cacciari:2001td}, and we compute gluons and light quark production spectra at leading order \cite{wang_private_communication}, as in \cite{Vitev:2002pf, Horowitz:2011gd}. 
	We note that for presented theoretical results on pion suppression, we used gluon and light quark production spectra $f(p_T)$ at $\sqrt{s_{NN}} = 5.5$ TeV and not $\sqrt{s_{NN}} = 5.02$ TeV, as this was conveniently available to us. The differing $\sqrt{s_{NN}}$ will lead to our results being slightly oversuppressed at $\sqrt{s_{NN}} = 5.02$ TeV.

There are uncertainties in both the fragmentation function fits and the generation of initial parton spectra; however, these uncertainties are negligible compared to others in this problem (running coupling, first order in opacity, etc.), and so we will not take these fitting uncertainties into account.
The spectrum $\mathrm{d} N^h / \mathrm{d} p_h$ for a hadron $h$ is related to the spectrum $\mathrm{d} N^q / \mathrm{d} p_q$ for a parton $q$ via \cite{Horowitz:2010dm}
\begin{equation}
        \frac{\mathrm{d} N^h}{\mathrm{d} p_h}\left(p_h\right) =\int \frac{\mathrm{d} N^q}{\mathrm{d} p_q}\left(\frac{p_h}{z}\right) \frac{1}{z} D^h_q(z, Q) \mathrm{d} z,
    \label{eqn:parton_to_hadron_spectrum}
\end{equation}
where $z \equiv p_h / p_q \in (0,1]$, $p_h$ is the observed hadron momentum, $D^h_q(z,Q)$ is the fragmentation function for the process $q \mapsto h$, and $Q$ is the hard scale of the problem taken to be $Q = p_q =  p_h / z$ \cite{Horowitz:2010dm}. The hadronic $R^h_{AB}$ is then found in terms of the partonic $R^q_{AB}$ (\cref{eqn:geometry_averaged_raa}) as \cite{Horowitz:2010dm}
\begin{align}
    R_{AB}^h\left(p_T\right)&=\frac{\sum_q \int d z \frac{1}{z} D_q^h(z) f\left(\frac{p_T}{z}\right) R_{AB}^q\left(\frac{p_T}{z}\right)}{\sum_q \int d z \frac{1}{z} D_q^h(z) f\left(\frac{p_T}{z}\right)}.
    \label{eqn:parton_to_hadron_raa}
\end{align}
For details of the derivation needed for \cref{eqn:parton_to_hadron_raa}, refer to Appendix B of \cite{Horowitz:2010dm}.

The fragmentation functions for $D$ and $B$ mesons are taken from \cite{Cacciari:2005uk}, and $\pi$ mesons from \cite{deFlorian:2007aj}. Note that the fragmentation functions for $\pi$ mesons were extrapolated outside their domain in $Q^2$, as we found that the extrapolation was smooth. %

\subsection{Numerical implementation}
\label{sec:numerical_elastic_radiative}

We neglect the running of the strong coupling constant for all numerical calculations and use $\alpha_s = 0.3$, consistent with \cite{Kolbe:2015rvk, Wicks:2005gt, Djordjevic:2003zk}. Additionally, we use charm and bottom quark masses of $m_c = 1.2~\mathrm{GeV}/c^2$ and $m_b = 4.75~\mathrm{GeV}/c^2$, respectively. We set the effective light quark and gluon masses to the asymptotic one-loop medium-induced thermal masses, $m_{\text{light}} = \mu / 2$ and $m_g = \mu / \sqrt{2}$, respectively \cite{Djordjevic:2003be}. The upper bounds on the $|\mathbf{k}|$ and $|\mathbf{q}|$ integrals are given by $k_{\text{max}}=2x (1-x) E$ and $q_{\text{max}}=\sqrt{3 E \mu}$, following \cite{Wicks:2005gt}. This choice of $k_{ \text{max}}$ guarantees that the momentum of the radiated gluon and the initial and final momenta of the parent parton are all collinear. %
In our previous work \cite{Faraday:2023mmx}, we motivated an alternative upper bound for the transverse radiated gluon momentum of $\operatorname{Min}\left(\sqrt{2 x E \mu_1}, 2 x(1-x) E\right)$ which guarantees that the energy loss kernel receives no contributions from regions of phase space where either the large formation time assumption or the collinear assumption are invalid. We showed \cite{Faraday:2023uay} that this upper bound on $k_{\text{max}}$ resulted in a slight decrease of the standard DGLV radiative energy loss, and a dramatic reduction of the short pathlength correction, particularly for gluons. Future work \cite{Faraday:2024,Bert:2024} will explore the impact of utilizing such a bound at the level of the $R_{AB}$; however, in this work, we utilize the standard collinear upper bound for consistency with previous work \cite{Faraday:2023mmx, Wicks:2005gt, Wicks:2008zz}.

We assume the distribution of scattering centers is exponential: $\rho_{\text{exp.}} (z) \equiv \frac{2}{L} \exp[ - 2 z / L]$, consistent with the WHDG model \cite{Wicks:2005gt}. The exponential distribution makes the integral in $\Delta z$ analytically simple. The physical motivation for the exponential form of the distribution of scattering centers is that an exponentially decaying distribution of scattering centers captures the decreasing density of the QGP due to Bjorken expansion \cite{Bjorken:1982qr}. It is likely that using an exponential rather than, say, a power law decay distribution is an overestimate of the effect of the expansion since Bjorken expansion obeys a power law decay along the incident partons path, not exponential. It turns out that in DGLV, the distribution of scattering centers $\bar{\rho}(\Delta z)$ affects the characteristic shape of $\mathrm{d} N^g / \mathrm{d} x$; however the radiative energy loss $\int dx x \frac{dN^g}{dx}$ is largely insensitive to the distribution of scattering centers \cite{Armesto:2011ht, Faraday:2023mmx}. Once one includes the short pathlength correction, however, the energy loss becomes far more sensitive to the distribution of scattering centers, particularly at small $\Delta z$ \cite{Kolbe:2015rvk, Faraday:2023mmx}.

\begin{figure}[!htbp]
	\includegraphics[width=\linewidth]{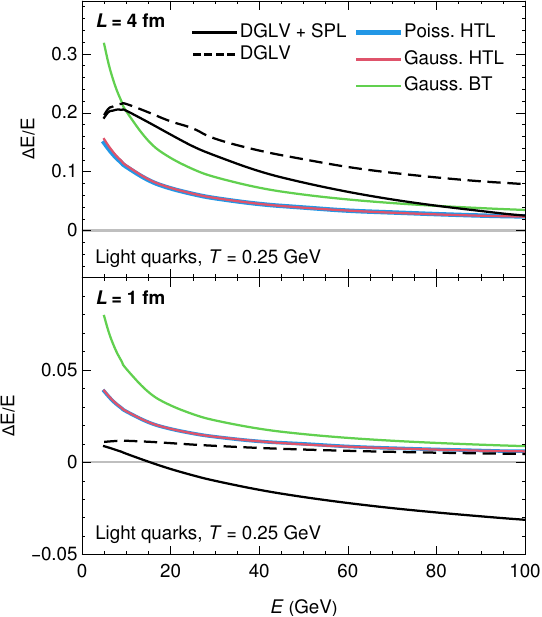}
	\caption{Plot of the fractional energy loss $\Delta E / E$ of light quarks as a function of incident energy $E$. The energy loss curves shown are calculated from elastic energy loss distributions of Gaussian Braaten and Thoma \cite{Braaten:1991jj, Braaten:1991we}, Gaussian HTL, and Poisson HTL \cite{Wicks:2008zz}, as well as radiative energy loss distributions of DGLV \cite{Djordjevic:2003zk} and DGLV with the short pathlength correction \cite{Kolbe:2015rvk}. Top pane shows results for a large pathlength $L = 4$~fm and bottom pane for a small pathlength $L = 1$~fm. The temperature is kept constant at $T = 0.25$~GeV and the strong coupling is also kept constant at $\alpha_s = 0.3$.}
	\label{fig:deltaEoverE_small_large}
\end{figure}

\begin{figure}[!htbp]
	\includegraphics[width=\linewidth]{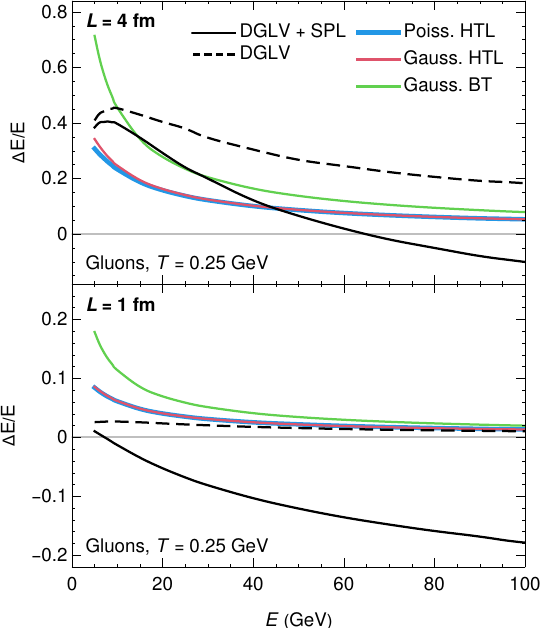}
	\caption{Plot of the fractional energy loss $\Delta E / E$ of gluons as a function of incident energy $E$. The energy loss curves shown are calculated from elastic energy loss distributions of Gaussian Braaten and Thoma \cite{Braaten:1991jj, Braaten:1991we}, Gaussian HTL, and Poisson HTL \cite{Wicks:2008zz}, as well as radiative energy loss distributions of DGLV \cite{Djordjevic:2003zk} and DGLV with the short pathlength correction \cite{Kolbe:2015rvk}. Top pane shows results for a large pathlength $L = 4$~fm and bottom pane for a small pathlength $L = 1$~fm. The temperature is kept constant at $T = 0.25$~GeV and the strong coupling is also kept constant at $\alpha_s = 0.3$.}
	\label{fig:deltaEoverE_small_large_gluon}
\end{figure}

\Cref{fig:deltaEoverE_small_large} shows the fractional energy loss $\Delta E / E$ of light quarks for the various radiative and elastic energy loss kernels discussed in \cref{sec:radiative_energy_loss,sec:elastic_energy_loss}, for both a large pathlength of $L = 4$~fm (top pane) and a small pathlength of $L = 1$~fm (bottom pane). Similarly, we plot the fractional energy loss of gluons in \cref{fig:deltaEoverE_small_large_gluon}. The radiative energy loss curves shown are for the DGLV radiative energy loss \cite{Djordjevic:2003zk} and the DGLV radiative energy loss which receives a short pathlength correction \cite{Kolbe:2015rvk, Kolbe:2015suq}, both of which we model with a Poisson distribution \cite{Gyulassy:2001nm}. The elastic energy loss curves shown are for the HTL elastic energy \cite{Wicks:2008zz} loss with both a Poisson distribution and Gaussian distribution, and the Braaten and Thoma (BT) elastic energy loss \cite{Braaten:1991jj, Braaten:1991we} with a Gaussian distribution. 

From \cref{fig:deltaEoverE_small_large,fig:deltaEoverE_small_large_gluon} we see the various effects of the short pathlength correction to the radiative energy loss which were discussed in \cref{sec:radiative_energy_loss_correction}. The short pathlength correction is significantly larger for gluons in comparison to light quarks, which is because of the breaking of color triviality in \cref{eqn:full_dndx}; the short pathlength correction can lead to negative energy loss; the short pathlength correction grows faster in $E$ than standard DGLV; and the short pathlength correction is proportionally larger at small lengths.

Comparing the different elastic fractional energy loss kernels in \cref{fig:deltaEoverE_small_large,fig:deltaEoverE_small_large_gluon} we see that the BT elastic energy loss is significantly larger than the HTL elastic energy loss at low--moderate $p_T \lesssim \mathcal{O}(50)~\text{GeV}$. We see that the HTL and BT fractional energy loss results approach each other at asymptotically high momenta; however, the convergence is slow. We see also that the effect of using a Gaussian compared to a Poisson distribution is negligible at the level of the fractional energy loss. 
The similarity in average energy loss between the Poisson and Gaussian elastic energy loss distributions is expected, as both are constrained to have an identical average energy loss if the kinematic bounds on the $\epsilon$ integral are ignored. 

Therefore, any minor observed differences in the Gaussian and Poisson average energy loss plots arise from the portion of the distribution beyond the kinematic bound at $\epsilon = 1$.

Comparing the relative size of the elastic and radiative energy loss results, we see that for large pathlengths, elastic energy loss is of similar importance to radiative energy loss at low momenta, while at large momenta, the radiative energy loss is $\sim\!\! 2\text{--}4$ times larger than the elastic. For small $L=1$~fm pathlengths the elastic energy loss is $\sim\!\! 2\text{--}8$ times as large as the radiative energy loss for $5~\text{GeV} \leq p_T  \leq 25~\text{GeV}$, and becomes relatively less important at large momenta. The variation in the magnitude of elastic vs radiative energy loss, as a function of pathlength and energy, highlights the importance of including both forms of energy loss in order to make phenomenological predictions over a wide range of system sizes and momenta.

\begin{figure}[!htbp]
	\includegraphics[width=\linewidth]{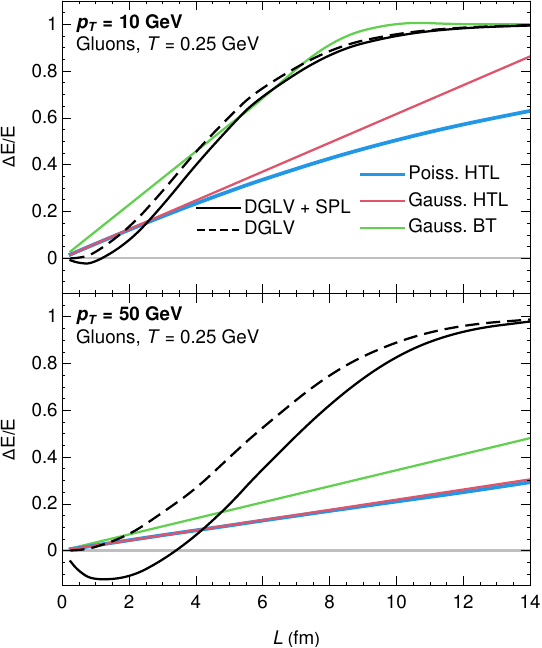}
	\caption{Plot of the fractional energy loss $\Delta E / E$ of gluons as a function of the pathlength $L$. The energy loss curves shown are calculated from elastic energy loss distributions of Gaussian Braaten and Thoma \cite{Braaten:1991jj, Braaten:1991we}, Gaussian HTL, and Poisson HTL \cite{Wicks:2008zz}, as well as radiative energy loss distributions of DGLV \cite{Djordjevic:2003zk} and DGLV with the short pathlength correction \cite{Kolbe:2015rvk}. Top pane shows results for $p_T = 10$ GeV and bottom pane for $p_T = 50$ GeV. The temperature is kept constant at $T = 0.25$~GeV and the strong coupling is also kept constant at $\alpha_s = 0.3$.}
	\label{fig:deltaEoverE_vs_L_gluon}
\end{figure}

\Cref{fig:deltaEoverE_vs_L_gluon} shows the fractional energy loss $\Delta E / E$ as a function of the pathlength $L$ for set $p_T = 10$~GeV (top pane) and $p_T = 50$~GeV bottom pane, at constant temperature $T = 0.25$~GeV. In this figure, we show that the short pathlength correction approaches zero at asymptotically large lengths, while at small lengths, the short pathlength corrected DGLV radiative energy loss decreases linearly with $L$.
The DGLV radiative energy loss grows like $L^2$ at small lengths $L$ due to the LPM effect, while the elastic energy loss is linear in $L$. These different $L$ dependencies lead to the elastic energy loss dominating at small pathlengths compared to the radiative energy loss.

\coleTwo{We briefly examine the effect of hadronization on the small system size corrections. \Cref{fig:pp_spectrum} plots the hadron $h$ production spectrum $d \sigma^{p + p \to h + X} / d p_T$ as a function of the final hadron momentum $p_T$ for the process $p + p \to h + X$. The spectra produced by our model are shown for pions at $\sqrt{s_{NN}} = 200 ~\mathrm{GeV}$ and $\sqrt{s_{NN}} = 5.5 ~\mathrm{TeV}$, as well as $D$ and $B$ mesons at $\sqrt{s_{NN}} = 5.02 ~\mathrm{TeV}$. Fragmentation is performed following \cite{Cacciari:2005uk} for $D$ and $B$ mesons and \cite{deFlorian:2007aj} for pions, with partonic production spectra from \cite{Cacciari:2001td} and \cite{wang_private_communication} respectively. The fragmentation procedure is described in \cref{sec:nuclear_modification_factor}. 
	We also show the $p + p \to h + X$ data for charged hadrons \cite{ATLAS:2022kqu} and charged pions \cite{ALICE:2019hno} at $\sqrt{s_{NN}} = 5.02 ~\mathrm{TeV}$, charged hadrons at $\sqrt{s_{NN}} = 200 ~\mathrm{GeV}$ \cite{PHENIX:2007kqm}, charged $B$ mesons at $\sqrt{s_{NN}} = 5.02 ~\mathrm{TeV}$ \cite{CMS:2017uoy}, and neutral $D$ mesons at $\sqrt{s_{NN}} = 5.02 ~\mathrm{TeV}$ \cite{ALICE:2016yta}.
	The theoretical results are fit to the \coll{p}{p} data by multiplying by a constant $K$-factor for each data set, and both the theoretical curves and experimental data points are shifted vertically for visual clarity. Since $R_{AB}$ is insensitive to multiplicative differences in the spectra, these changes have no impact on our conclusions.
	We observe that all of the measured \coll{p}{p} spectra are well-described in our model. 
Notably, the approximation that charged hadron spectra share similar $p_T$ dependence with $\pi^0$ meson spectra, and likewise for $D^0$ and $D$ mesons and $B^\pm$ and $B$ mesons, is justified by the agreement with \coll{p}{p} data. The small difference between $\sqrt{s_{NN}} = 5.02~\mathrm{TeV}$ for the measured charged hadron and pion data and $\sqrt{s_{NN}} = 5.5~\mathrm{TeV}$ for the theoretical neutral pion cross section has a negligible effect, as seen by the agreement between the experimental data and theoretical calculation.

\begin{figure}[!htbp]
	\centering
	\includegraphics[width=\linewidth]{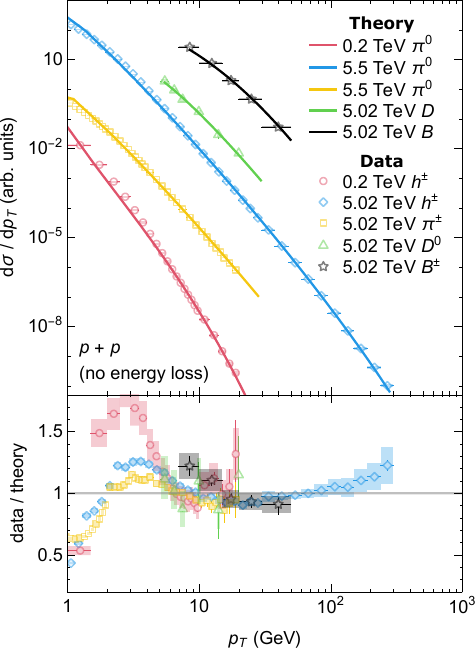}
	\caption{The top panel shows the $p_T$-differential production cross section $d \sigma / d p_T$ in \coll{p}{p} collisions for final state hadrons as a function of final hadron $p_T$. Experimental data is shown for charged hadrons \cite{ATLAS:2022kqu} and charged pions \cite{ALICE:2019hno} at $\sqrt{s_{NN}} = 5.02 ~\mathrm{TeV}$, charged hadrons at $\sqrt{s_{NN}} = 200 ~\mathrm{GeV}$ \cite{PHENIX:2007kqm}, charged $B$ mesons at $\sqrt{s_{NN}} = 5.02 ~\mathrm{TeV}$ \cite{CMS:2017uoy}, and neutral $D$ mesons at $\sqrt{s_{NN}} = 5.02 ~\mathrm{TeV}$ \cite{ALICE:2016yta}. Theoretical calculations are shown for pions at $\sqrt{s_{NN}} = 200 ~\mathrm{GeV}$ and $\sqrt{s_{NN}} = 5.5 ~\mathrm{TeV}$, as well as $D$ and $B$ mesons at $\sqrt{s_{NN}} = 5.02 ~\mathrm{TeV}$. 
		Fragmentation is performed following \cite{Cacciari:2005uk} for $D$ and $B$ mesons and \cite{deFlorian:2007aj} for pions, with partonic spectra from \cite{Cacciari:2001td} and \cite{wang_private_communication} respectively. The theoretical calculations are multiplied by a $K$-factor to best fit the \coll{p}{p} data, and both experimental data and theoretical curves are shifted vertically for visual clarity. \coleThree{The bottom panel shows the ratio of the same experimental data to the theoretical predictions as a function of $p_T$.}
Statistical uncertainties are represented by bars while systematic uncertainties are represented by shaded boxes.}
	\label{fig:pp_spectrum}
\end{figure}

We note that the pion hadronization procedure we used in our previous work \cite{Faraday:2023mmx} had a bug which effectively led to our code neglecting light quark contributions to the pion energy loss. We will see in \cref{sec:results}: that the corrected pion $R_{AA}$ shows a faster rise in $p_T$ compared to our previous results due to the smaller fraction of gluons compared to light quarks at higher momenta. The agreement between the corrected pion spectra and \coll{p}{p} data indicates that no further errors are present in the hadronization procedure.

	\Cref{fig:ratio_gluon_to_lq} plots the ratio of pions which fragment from gluons to pions which fragment from either gluons or light quarks. Of particular interest is the interaction between the short pathlength correction and the amount of gluons compared to light quarks which form pions because the short pathlength correction breaks color triviality, and is very large for gluons compared to light quarks (as demonstrated by \cref{fig:deltaEoverE_small_large_gluon,fig:deltaEoverE_vs_L_gluon,fig:deltaEoverE_small_large}). We observe from the figure that far fewer gluons fragment to pions at RHIC compared to LHC, implying that we should expect the effects of the short pathlength correction to be diminished at RHIC in comparison to LHC. Additionally, the fraction of gluons which fragment to pions decreases as a function of $p_T$, which will compete with the growth of the short pathlength correction in $p_T$. We will see in \cref{sec:results}, that the growth in $p_T$ of the short pathlength correction wins this battle at small to intermediate $p_T$, but at large $p_T$ the fraction of gluons becomes small enough that the effect of the short pathlength correction on $R_{AA}$ is dramatically reduced.

\begin{figure}[!htbp]
	\includegraphics[width=\linewidth]{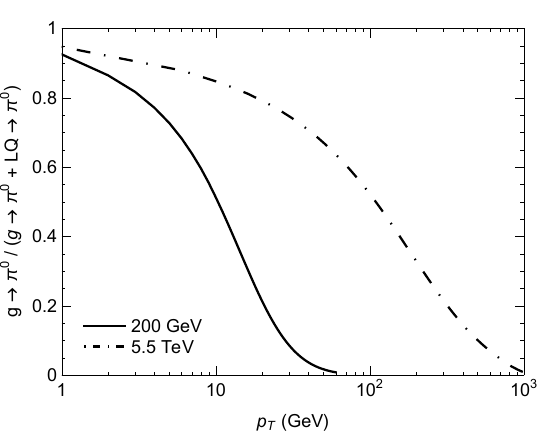}
	\caption{Ratio of the production spectrum for pions which fragment from gluons to the production spectrum for pions which fragment from either gluons or light quarks, as a function of final pion $p_T$. Hadronization is performed following \cite{deFlorian:2007aj} with LO pQCD production spectra from \cite{wang_private_communication} at $\sqrt{s_{NN}} = 5.5 ~\mathrm{TeV}$ (dot-dashed) and $\sqrt{s_{NN}} = 200 ~\mathrm{GeV}$ (solid). No energy loss is used in calculating this ratio.}
	\label{fig:ratio_gluon_to_lq}
\end{figure}

}

\subsection{Geometry}%
\label{sec:geometry}

For large systems, using the Glauber model for the collision geometry is standard, with Wood-Saxon distributions for the nucleon density inside the heavy ions \cite{Glauber:1970jm,Miller:2007ri}. For small \collFour{p}{d}{He3}{A} collisions, the Glauber model cannot be applied in its most simple form since one expects subnucleonic features of the proton to be important \cite{Schenke:2020mbo}. Additionally, one may treat the subsequent evolution of the medium in a more sophisticated way than simply assuming a Bjorken expansion of the initial Glauber model geometry as was done, e.g., in \cite{Wicks:2005gt}.
 In this work, we will use collision profiles generated with \cite{Schenke:2020mbo} and sourced from \cite{shen_private_communication}.
In these calculations, initial conditions are given by the IP-Glasma model \cite{Schenke:2012hg, Schenke:2012wb}, which are then evolved with the \texttt{MUSIC} \cite{Schenke:2010rr,Schenke:2011bn, Schenke:2010nt} viscous relativistic (2+1)D hydrodynamics code, followed by UrQMD microscopic hadronic transport \cite{Bass:1998ca, Bleicher:1999xi}.

Consistent with our previous work \cite{Faraday:2023mmx} we use only the initial temperature profile $T(\tau = \tau_0 = 0.4~\mathrm{fm})$ for our collision geometry, where $\tau_0$ is the turn-on time for hydrodynamics. The collision geometry used for all results presented in this work is, therefore, effectively IP-Glasma initial conditions \cite{Schenke:2012hg, Schenke:2012wb} coupled with Bjorken expansion time dependence, unless otherwise stated. Future work \cite{Bert:2024} will incorporate the full hydrodynamic evolution, where results for small systems may require (3+1D) hydrodynamics, as suggested by its significance for flow observables in small systems \cite{Zhao:2022ayk, Zhao:2022ugy}.

From HTL perturbation theory \cite{Blaizot:2001nr} one may derive the leading order expression for the Debye mass 
\begin{equation}
	\mu  = g T \sqrt{\frac{2 N_c  + n_f}{6}} \text{ where } g = \sqrt{4 \pi \alpha_s}.
	\label{eqn:debye_mass}
\end{equation}

The QGP is treated as an ultrarelativistic mixture of a Fermi and Bose gas with zero chemical potential, following \cite{Wicks:2005gt, Horowitz:2010dm, Faraday:2023mmx, Wicks:2008zz}. Then
\begin{subequations}
    \label{eqn:thermodynamic_quantities}
		\begin{alignat}{2}
			\sigma_{gg}  =& \frac{C_A^2 \pi \alpha_s^2}{2 \mu^2} \quad \text{and} \quad \sigma_{qg}  = \frac{C_F}{C_A} \sigma_{gg}&\\
			\rho_g =& 2 (N_c^2 - 1) \frac{\zeta(3)}{\pi^2} T^3 \label{eqn:rho_thermal_g}&\\
			\rho_q =& 3 N_c n_f \frac{\zeta(3)}{\pi^2} T^3 \label{eqn:rho_thermal_q}&\\
			\rho =& \rho_g + \frac{\sigma_{q g}}{\sigma_{gg}} \rho_{q}& \nonumber\\
			=&\frac{\zeta(3) (N_c^2 - 1)}{\pi^2} T^3 \left( 2 + \frac{3n_f}{2N_c}\right)&\\
			\lambda_g^{-1}  =& \rho_g \sigma_{g g}+\rho_q \sigma_{q g} = \sigma_{gg} \rho,& \label{eqn:mean_free_path}
    \end{alignat}
\end{subequations}
where $\zeta$ is the Riemann zeta function, $T$ is the temperature, $\rho_q$ ($\rho_g$) is the density of quarks (gluons), $\sigma_{qg}$ ($\sigma_{gg}$) is the gluon-gluon (quark-gluon) elastic cross section, $n_f$ is the number of active quark flavors (taken to be $n_f = 2$ throughout), and $N_c=3$ is the number of colors. We denote the cross section weighted density as $\rho$, which we will subsequently refer to as the density for simplicity. Due to color triviality, one need only calculate these results for an incident gluon, and the change for an incident quark is simply a change of Casimir in the relevant energy loss kernels in \cref{eqn:DGLV_dndx,eqn:full_dndx,eqn:elastic_energy_loss_low,eqn:elastic_energy_loss_high,eqn:dndx_pure_htl}.

The radiative [\cref{eqn:DGLV_dndx,eqn:full_dndx}] and elastic [\cref{eqn:elastic_energy_loss_low,eqn:elastic_energy_loss_high,eqn:dndx_pure_htl}] energy loss results were derived using a ``brick" model, which represents a medium with a fixed length $L$ and constant temperature $T$. In order to capture fluctuations in temperature and density, we need a mapping from the path that a parton takes through the plasma, to a brick with an effective length $L_{\text{eff}}$ and effective temperature $T_{\text{eff}}$.

We follow WHDG \cite{Wicks:2005gt} and define the effective pathlength as
\begin{equation}
	L_{\text{eff}} (\mathbf{x}_i, \boldsymbol{\hat{\phi}}) = \frac{1}{\rho_{\text{eff}}} \int_{0}^\infty \mathrm{d}z \; \rho(\mathbf{x}_i + z \boldsymbol{\hat{\phi}}, \tau_0),
    \label{eqn:effective_length}
\end{equation}
and the effective density as
\begin{equation}
  \rho_{\text{eff}} \equiv \frac{\int \mathrm{d}^2 \mathbf{x} \; \rho^2(\mathbf{x}, \tau_0)}{\int \mathrm{d}^2 \mathbf{x} \; \rho(\mathbf{x}, \tau_0)} \iff T_{\text{eff}}^{3} \equiv \frac{\int \mathrm{d}^2 \mathbf{x} \; T^6(\mathbf{x}, \tau_0)}{\int \mathrm{d}^2 \mathbf{x} \; T^3(\mathbf{x}, \tau_0)}.
  \label{eqn:effective_density}
\end{equation}
Here, the effective pathlength $L_{\text{eff}}$ includes all $(\mathbf{x}_i, \phi)$ dependence, and $\rho_{\text{eff}}$ is a constant for all paths that a parton takes through the plasma for a fixed centrality class. In principle, one can allow both the effective density and effective pathlength to depend on the specific path taken through the plasma. However, such a numerically intensive model is beyond the scope and objective of this work.

In WHDG \cite{Wicks:2005gt}, the prescription $\rho \equiv \rho_{\text{part}}$ was made, where $\rho_{\text{part}}$ is the participant density---the density of nucleons which participate in at least one binary collision. This prescription is not necessary in our case, since we have access to the temperature profile \cite{Schenke:2020mbo, shen_private_communication}. We extract the temperature profile $T(\mathbf{x}, \tau)$ from the hydrodynamics output \cite{Schenke:2020mbo,shen_private_communication}, and for $L_{\mathrm{eff}}$ one evaluates the temperature at the initial time set by the hydrodynamics simulation, $\tau_0=0.4~\mathrm{fm}$. 
There is no unique mapping from realistic collision geometries to simple brick geometries and more options are explored in \cite{Wicks:2008zz} and our previous work \cite{Faraday:2023mmx}. Future work will perform a careful analysis of the geometrical mapping procedure between realistic media and brick geometries wherein theoretical calculations take place \cite{Bert:2024}.

Bjorken expansion \cite{Bjorken:1982qr} is then taken into account by approximating
\begin{equation}
T_{\text{eff}}(\tau) \approx T_{\text{eff}}(\tau_0) \left( \frac{\tau_0}{\tau} \right)^{1/3} \approx T_{\text{eff}}(\tau_0) \left( \frac{2 \tau_0}{L_{\text{eff}}} \right)^{1/3}
  \label{eqn:bjorken_expansion}
\end{equation}
where in the last step we have evaluated $T( \mathbf{x}, \tau)$ at the average time $\tau=L / 2$, following what was done in \cite{Wicks:2005gt, Djordjevic:2005db, Djordjevic:2004nq}. In \cite{Wicks:2005gt}, this average time approximation was found to be a good approximation to the full integration through the Bjorken expanding medium. For a given collision system, we can then calculate the distribution of effective pathlengths that a hard part will travel in the plasma. We assume, as is standard and consistent with WHDG \cite{Wicks:2005gt} and our previous work \cite{Faraday:2023mmx}, that the hard partons have starting positions weighted by the density of binary nucleon-nucleon collisions, provided by IP-Glasma \cite{shen_private_communication}. 

\begin{figure}[!htbp]
    \centering
    \includegraphics[width=\linewidth]{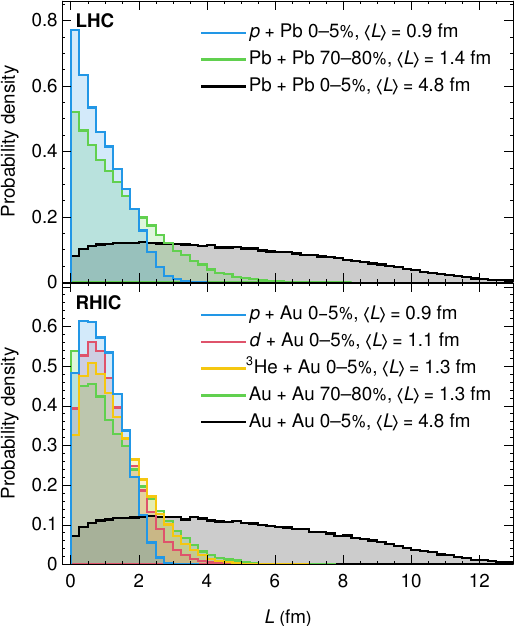}
		\caption{
			(Top) Distribution of the effective pathlengths in $0\text{--}5\%$ central \coll{p}{Pb}, $0\text{--}5\%$ central \coll{Pb}{Pb}, and $70\text{--}80\%$ peripheral \coll{Pb}{Pb} collision systems at $\sqrt{s_{NN}} = 5.02$ TeV. (Bottom) Distribution of the effective pathlengths in $0\text{--}5\%$ central \coll{p}{Au}, \coll{d}{Au}, \coll{He3}{Au}, and \coll{Au}{Au} as well as $70\text{--}80\%$ peripheral \coll{Au}{Au} collisions at $\sqrt{s_{NN}} = 200$ GeV. The average pathlength $\langle L \rangle$ is shown in the legend for each collision.
		}
    \label{fig:path_length_distribution}
\end{figure}

\begin{figure}[!htbp]
	\includegraphics[width=\linewidth]{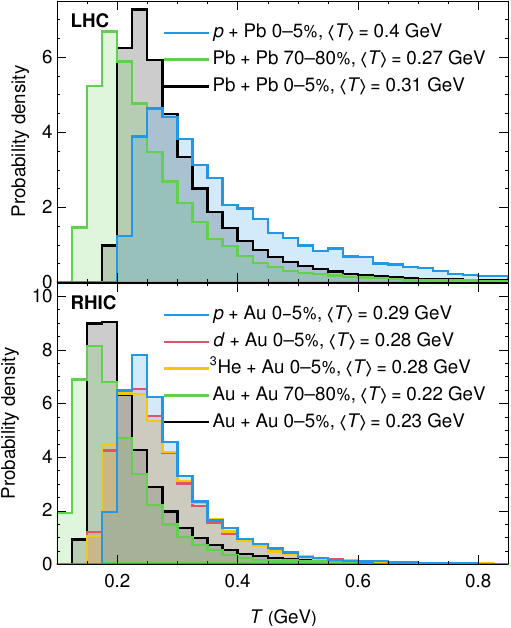}
	\caption{(Top) Distribution of the effective temperatures $T$ in $0\text{--}5\%$ central \coll{p}{Pb}, $0\text{--}5\%$ central \coll{Pb}{Pb}, and $70\text{--}80\%$ peripheral \coll{Pb}{Pb} collision systems at $\sqrt{s_{NN}} = 5.02$ TeV. (Bottom) Distribution of the effective temperatures $T$ in $0\text{--}5\%$ central \coll{p}{Au}, \coll{d}{Au}, \coll{He3}{Au}, and \coll{Au}{Au} as well as $70\text{--}80\%$ peripheral \coll{Au}{Au} collisions at $\sqrt{s_{NN}} = 200$ GeV. The average temperature $\langle T \rangle$ is shown in the legend for each collision.}
	\label{fig:temperature_distribution}
\end{figure}

\Cref{fig:path_length_distribution} shows the probability distribution of effective pathlengths for $0\text{--}5\%$ centrality \coll{p}{Pb} and \coll{Pb}{Pb} collisions, as well as for $70\text{--}80\%$ peripheral \coll{Pb}{Pb} collisions in the top panel. The bottom panel shows the probability distribution for $0\text{--}5\%$ centrality \coll{p}{Au}, \coll{d}{Au}, \coll{He3}{Au}, and \coll{Au}{Au} collisions, as well as for $70\text{--}80\%$ peripheral \coll{Au}{Au} collisions. 
In the legend, we indicate the average pathlength in each system. 
Note especially that the effective pathlength distribution for $0\text{--}5\%$ central \coll{p}{Pb} and \collFour{p}{d}{He3}{Au} collisions are nearly indistinguishable from peripheral $70\text{--}80\%$ \coll{Pb}{Pb} and \coll{Au}{Au} collisions. \Cref{fig:temperature_distribution} plots the distribution of temperatures for the same set of collision systems as shown in \cref{fig:path_length_distribution}, where the distributional shape of the figure is simply a mapping of the length distribution in \cref{fig:path_length_distribution} according to \cref{eqn:bjorken_expansion}.

The average pathlength in $0\text{--}5\%$ centrality \collFour{p}{d}{He3}{A} collisions is $L \sim 1~\mathrm{fm}$, with an average temperature of $T \approx 0.3\text{--}0.4~\mathrm{GeV}$, corresponding to a mean free path $\lambda_g \approx 0.7\text{--}1~\mathrm{fm}$ and an inverse Debye mass $\mu^{-1} \approx 0.2\text{--}0.3~\text{fm}$.
We see from this simple analysis that in small \collFour{p}{d}{He3}{A} collision systems $L/\lambda_g\sim1$ for most of the distribution of effective pathlengths, which implies that approaches that assume many soft scatterings are inapplicable. Similarly, the small amount of expected scatters does not provide a good motivation for modeling the elastic energy loss as a Gaussian distribution according to the central limit theorem \cite{Moore:2004tg}.
Moreover, in small \collFour{p}{d}{He3}{A} collisions, the large pathlength assumption $\lambda \ll L \iff 0.2\text{--}0.3 ~\mathrm{fm} \ll 1 ~\mathrm{fm}$ is not particularly well founded, as discussed in our previous work \cite{Faraday:2023mmx}.

 The average pathlength of $\langle L \rangle \sim1$~fm in small systems, while substantially smaller than the $\langle L \rangle \sim5$~fm in large systems, is similar to the $\langle L \rangle \sim 1.4$~fm in $70\text{--}80\%$ peripheral \coll{Pb}{Pb} and \coll{Au}{Au} collisions. The average temperatures of these small collision systems are however notably larger: while $0\text{--}5\%$ \collFour{p}{d}{He3}{A} collisions and $70\text{--}80\%$ \coll{A}{A} collisions have comparable average lengths, the temperature in central \collFour{p}{d}{He3}{A} collisions is $\mathcal{O}(30\text{--}50\%)$ larger than that of peripheral \coll{A}{A} collisions. 

\Cref{fig:temperature_distribution} shows how the small average pathlengths in small \collFour{p}{d}{He3}{A} collisions result in a temperature distribution with a larger average temperature since the parton spends a significantly larger fraction of its lifetime in the high-temperature region of the medium.

The temperature distributions between the various collision systems are more similar than the corresponding length distributions. This similarity is due to the $T \sim L^{- 1 /3}$ dependency from \cref{eqn:bjorken_expansion}, which results in the temperature distributions having significantly less variation than the corresponding length distributions.

\subsubsection{Energy loss outside the QGP}
\label{sec:prethermalization_time_energy_loss}

As an incident high-$p_T$ parton propagates through the QGP formed during a collision, it first traverses through a non-thermalized medium, then a thermalized QGP, and finally a hadronic medium of some sort. The first and last of these three phases are not rigorously modeled by pQCD energy loss, even though they must contribute to the energy loss, which means we must make a phenomenological choice on how to perform energy loss during this period of the collision.

There are a few common phenomenological choices on how to treat energy loss before the medium has thermalized, including \cite{Xu:2014ica}

\begin{enumerate}
	\item \textit{Free streaming}. The temperature distribution takes the form $T(\vecII{x}, \tau) \equiv \Theta(\tau - \tau_0) T(\vecII{x}) \left(\frac{\tau_0}{\tau}\right)^{1 / 3}$. This prescription is equivalent to assuming no energy loss occurs for the first $\tau_0$ of propagation distance through the plasma.
	\item \textit{Linear thermalization}. The temperature distribution takes the form $T(\vecII{x}, \tau) \equiv T(\vecII{x})\left[ \Theta(\tau - \tau_0) \left(\frac{\tau_0}{\tau}\right)^{1 / 3}  + \Theta(\tau_0 - \tau) \left(\frac{\tau}{\tau_0}\right)^{1 / 3}\right] $.  This prescription assumes that the temperature starts at $0$ GeV and grows to the temperature at the thermalization time, after which the temperature decays according to Bjorken expansion.
	\item \textit{Instant thermalization}. The temperature distribution takes the form $T(\vecII{x}, \tau) \equiv T(\vecII{x}) \left(\frac{\tau_0}{\tau}\right)^{1 / 3}$. The pre-thermalization time behaviour is extrapolated from the post-thermalization time behavior of Bjorken expansion. 
\end{enumerate}

The prescriptions outlined above have implications for both the effective temperature in \cref{eqn:bjorken_expansion} and the domain of the integrals in \cref{eqn:effective_length,eqn:effective_density}. In this work, all calculations presented are performed according to the instant thermalization prescription for simplicity, meaning that the integral in \cref{eqn:effective_length} is evaluated from $\tau = 0$ to $\tau = \infty$. The instant thermalization prescription is the most theoretically consistent prescription for the exponential distribution of scattering centers, which is used for the radiative energy loss \cref{eqn:density_scattering_centers}. It was found \cite{Xu:2014ica} that while $R_{AB}$ results were sensitive to the choice of pre-thermalization time energy loss, this sensitivity could largely be absorbed by a change in the strong coupling. Small systems likely have a greatly increased sensitivity to the choice of pre-thermalization time energy loss, and therefore the system size dependence of the $R_{AB}$ may not be able to be absorbed into changes in the strong coupling. Future work should examine the different pre-thermalization time energy loss scenarios, particularly in relation to the dependence on system size.

For energy loss at temperatures below the freezeout temperature of $T_{\text{f.o.}} \simeq 0.155$ GeV \cite{Borsanyi:2011sw}, one may choose to turn off energy loss, extrapolate the pQCD energy loss, or perform fragmentation and switch to a hadronic energy loss model. High-$p_T$ particles can interact with matter in the hadronic phase, and so a good compromise between simplicity and realism seems to be to extrapolate the pQCD energy loss calculation into the hadronic phase. We note that although the hydrodynamics temperature profiles have a turn-off at $T_{\text{f.o.}} = 0.155$ GeV \cite{Schenke:2020mbo}, the Bjorken expansion formula for the temperature has no such turn-off. Additionally, since all of our temperatures and lengths are calculated from the initial distribution, we are, to a good approximation, ignoring the freezeout temperature and extrapolating our energy loss into the hadronic phase.

\subsubsection{Geometry and event averaged nuclear modification factor}
\label{sec:geometry_averaged_nuclear_modification_factor}

To incorporate our model for the collision geometry, we expand \cref{eqn:full_raa_spectrum_ratio} to average over the effective pathlengths according to the length distribution described in \cref{sec:geometry}. Additionally, one must perform an average over the various collision events due to statistical fluctuations in the IP-Glasma initial conditions \cite{Schenke:2020mbo}. Then

\begin{align}
  R&_{A B}^q\left(p_T\right) \nonumber\\
  & =  \langle R_{A B}^{q}(p_T, L, T) \rangle_{\text{geometry, events}}	\label{eqn:geometry_averaged_raa}\\
	& = \left\langle \int \mathrm{d} L\; P_L(L) \; R_{AB}^q\left(p_T, L, T_{\text{eff.}} (L / 2)\right) \right\rangle_{\text{events}}\nonumber\\
	&\approx \int \mathrm{d} L\; \left\langle P_L \right\rangle_{\text{events}}(L) \; R_{AB}^q\left(p_T, L, \left\langle  T_{\text{eff.}}\right\rangle_{\text{events}} (L / 2)\right),\nonumber
\end{align}
where $P_L(L_{\mathrm{eff}})$ is the normalized distribution of effective pathlengths weighted by the binary collision density (\cref{fig:path_length_distribution}), and $T(L_{\text{eff}} / 2)$ is the effective temperature (\cref{fig:temperature_distribution}) determined according to \cref{eqn:bjorken_expansion}. 
In the last line of the above, we have approximated the event average of the $R_{AB}$ by an event average of the distribution of effective lengths $\left\langle P_L \right\rangle_{\text{events}}$ and effective temperatures $\left\langle T_{\text{eff}} \right\rangle_{\text{events}}$.
This procedure is strictly correct in its application to $\left\langle P_L \right\rangle_{\text{events}}$, because the sum over the events and the integral over $L$ are linear operators and therefore commute. The approximation occurs in the fact that we average over the $T_{\text{eff.}}$ functions, which appear inside the non-linear $R_{AB}$ function. We make this approximation for computational simplicity, and because including the variation at the level of the initial temperatures would lead to an overestimate in the variation of the temperatures at later times \cite{Faraday:2023mmx}. This overestimation occurs because the Bjorken expansion \cite{Bjorken:1982qr} does not smooth out the initial temperature distribution, while a more realistic hydrodynamic evolution \cite{Schenke:2020mbo} would \cite{Faraday:2023mmx}.

\vspace{3em}

\section{Results}
\label{sec:results}

In this section, we present nuclear modification factor results from our model for final-state pions, $D$ mesons, and $B$ mesons produced in \coll{p}{Pb} and \coll{Pb}{Pb} collisions at the LHC, as well as for final-state pions produced in \collFour{p}{d}{He3}{Au} collisions, and pions and $D$ mesons produced in \coll{Au}{Au} collisions at RHIC.
\coleTwo{These hadron species and collision systems were chosen as a representative sample of those reported by experiments at RHIC and LHC. While small system suppression is of particular interest for our work, large systems serve as a crucial consistency check \cite{Faraday:2023mmx,Faraday:2023uay}.} In this work, our objective is to \emph{qualitatively} understand the influence of small system size corrections to the elastic and radiative energy loss on the nuclear modification factor in different collision systems. For this reason, we produce all theoretical curves with a constant coupling of $\alpha_s = 0.3$ and make no comparison with data; however, the choice of displayed collision systems \emph{is} motivated by available RHIC and LHC data. Future work in preparation \cite{Faraday:2024} will examine the extent to which both small and large system suppression data can be \emph{quantitatively and simultaneously} described by pQCD final state energy loss models, through a one-parameter fit to data of the strong coupling $\alpha_s$.

There are six sets of results presented for each collision system, which are constructed by varying both the radiative and elastic energy loss kernels. The radiative energy loss kernel is varied between the DGLV radiative energy loss kernel (\textit{DGLV}) \cite{Djordjevic:2003zk} and the DGLV radiative energy loss kernel which receives a short pathlength correction (\textit{DGLV + SPL}) \cite{Kolbe:2015rvk, Kolbe:2015suq}. The elastic energy loss kernel is varied between the Braaten and Thoma elastic energy loss with a Gaussian distribution (\textit{Gaussian BT}), the HTL result from Wicks \cite{Wicks:2008zz} with a Poisson distribution (\textit{Poisson HTL}), and the HTL result from Wicks \cite{Wicks:2008zz} with a Gaussian distribution (\textit{Gaussian HTL}).
We presented the model implementation details for each theoretical curve in \cref{sec:model}.

\coleTwo{%
These theoretical curves were chosen to enable three key comparative analyses. First, examining the DGLV and DGLV + SPL curves shows the impact of the short pathlength correction as a function of system size, $p_T$, and hadron species. Second, contrasting the Gaussian BT and Gaussian HTL models provides insight into the fundamental uncertainties associated with the transition between HTL and vacuum propagators in elastic energy loss. Third, the comparison between Gaussian HTL and Poisson HTL results quantifies the impact of the commonly used Gaussian approximation for the convolution of multiple elastic scatters on the $R_{AA}$.}

\subsection{Large system suppression at LHC}
\label{sec:large_system_suppression_LHC}

\begin{figure}[!htpb]
	\includegraphics[width=\linewidth]{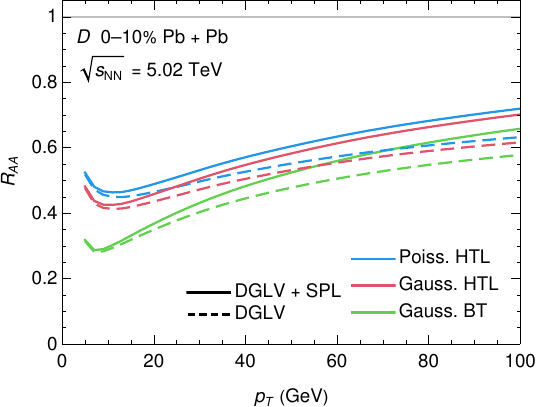}
	\caption{ Nuclear modification factor $R_{AA}$ as a function of $p_T$ for $D$ mesons produced in $0\text{--}10\%$ centrality \coll{Pb}{Pb} collisions at $\sqrt{s_{NN}} = 5.02$ TeV. We produce theoretical results through our convolved radiative and elastic energy loss model by varying the elastic model between Gaussian BT \cite{Braaten:1991jj, Braaten:1991we}, Gaussian HTL, and Poisson HTL \cite{Wicks:2008zz}, and the radiative model between DGLV \cite{Djordjevic:2003zk} and DGLV + SPL \cite{Kolbe:2015rvk, Kolbe:2015suq}. }
	\label{fig:raa_pbpb_D0010}
\end{figure}

\begin{figure}[!htpb]
	\includegraphics[width=\linewidth]{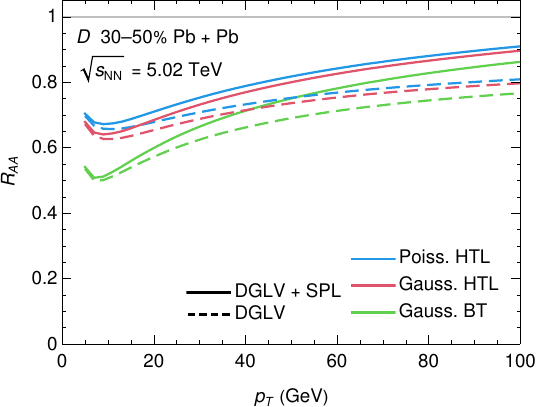}
	\caption{Nuclear modification factor $R_{AA}$ as a function of $p_T$ for $D$ mesons produced in $30\text{--}50\%$ centrality \coll{Pb}{Pb} collisions at $\sqrt{s_{NN}} = 5.02$ TeV. Theoretical results are produced through our convolved radiative and elastic energy loss model by varying the elastic model between Gaussian BT \cite{Braaten:1991jj, Braaten:1991we}, Gaussian HTL, and Poisson HTL \cite{Wicks:2008zz}, and the radiative model between DGLV \cite{Djordjevic:2003zk} and DGLV + SPL \cite{Kolbe:2015rvk, Kolbe:2015suq}.}
	\label{fig:raa_pbpb_D3050}
\end{figure}

\begin{figure}[!htpb]
	\includegraphics[width=\linewidth]{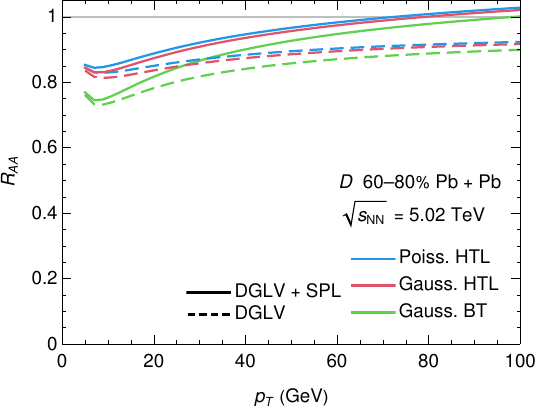}
	\caption{Nuclear modification factor $R_{AA}$ as a function of $p_T$ for $D$ mesons produced in $60\text{--}80\%$ centrality \coll{Pb}{Pb} collisions at $\sqrt{s_{NN}} = 5.02$ TeV. Theoretical results are produced through our convolved radiative and elastic energy loss model by varying the elastic model between Gaussian BT \cite{Braaten:1991jj, Braaten:1991we}, Gaussian HTL, and Poisson HTL \cite{Wicks:2008zz}, and the radiative model between DGLV \cite{Djordjevic:2003zk} and DGLV + SPL \cite{Kolbe:2015rvk, Kolbe:2015suq}.}
	\label{fig:raa_pbpb_D6080}
\end{figure}

\begin{figure}[!htpb]
	\includegraphics[width=\linewidth]{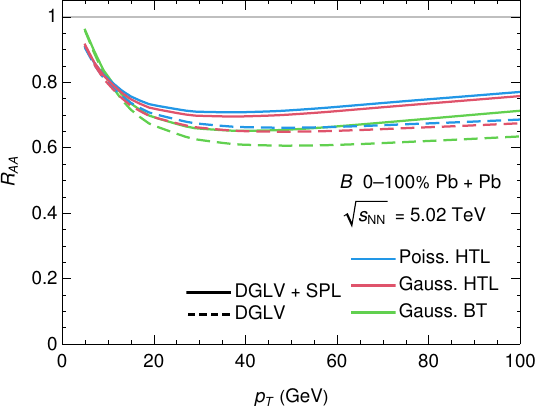}
	\caption{Nuclear modification factor $R_{AA}$ as a function of $p_T$ for $B$ mesons produced in $0\text{--}100\%$ centrality \coll{Pb}{Pb} collisions at $\sqrt{s_{NN}} = 5.02$ TeV. Theoretical results are produced through our convolved radiative and elastic energy loss model by varying the elastic model between Gaussian BT \cite{Braaten:1991jj, Braaten:1991we}, Gaussian HTL, and Poisson HTL \cite{Wicks:2008zz}, and the radiative model between DGLV \cite{Djordjevic:2003zk} and DGLV + SPL \cite{Kolbe:2015rvk, Kolbe:2015suq}.}
	\label{fig:raa_pbpb_B}
\end{figure}

\Cref{fig:raa_pbpb_D0010,fig:raa_pbpb_D3050,fig:raa_pbpb_D6080} show the nuclear modification factor $R_{AA}$ of $D$ mesons produced in $0\text{--}10\%$, $30\text{--}50\%$, and $60\text{--}80\%$ centrality \coll{Pb}{Pb} collisions at $\sqrt{s_{NN}} = 5.02$ TeV from our convolved elastic and radiative energy loss model. \Cref{fig:raa_pbpb_B} shows the nuclear modification factor $R_{AA}$ of $B$ mesons produced in $0\text{--}100\%$ centrality \coll{Pb}{Pb} collisions at $\sqrt{s_{NN}} = 5.02$ TeV from our convolved elastic and radiative energy loss model. We obtain theoretical model results by varying both the elastic and radiative energy loss kernels used in the calculation, as previously described.

Comparing the results calculated with the Poisson HTL elastic energy loss kernel to those calculated with the Gaussian HTL elastic energy loss kernel, shown in \cref{fig:raa_pbpb_D0010,fig:raa_pbpb_D3050,fig:raa_pbpb_D6080,fig:raa_pbpb_B}, we see that the heavy-quark $R_{AA}$ is largely insensitive to the shape of the elastic energy loss distribution. We conclude that approximating the elastic energy loss distribution as a Gaussian distribution is not important phenomenologically for heavy flavor $D$ and $B$ mesons.

At low momenta, a comparison between the Gaussian BT results and the Gaussian HTL results is sensitive to the uncertainty in the crossover region between HTL and vacuum propagators. 
We observe in \cref{fig:raa_pbpb_D0010,fig:raa_pbpb_D3050,fig:raa_pbpb_D6080} that the $D$ meson $R_{AA}$ is acutely sensitive to this choice for all centrality classes. 
This sensitivity is significantly reduced in the $B$ meson $R_{AA}$ shown in \cref{fig:raa_pbpb_B}, because the energy loss goes to zero as $p_T$ goes to zero. 
We conclude that quantitative suppression predictions that include heavy-flavor predictions, must consider the theoretical uncertainty due to the crossover between HTL and vacuum propagators in the elastic energy loss sector.
In lieu of a theoretical framework which can resolve this uncertainty, future phenomenological models may use the different dependencies of elastic vs radiative energy loss on system size and $p_T$ to phenomenologically extract the changeover between the HTL and vacuum propagators.

Comparing the DGLV and DGLV + SPL curves in \cref{fig:raa_pbpb_D0010,fig:raa_pbpb_D3050,fig:raa_pbpb_D6080,fig:raa_pbpb_B}, we see that the short pathlength correction has a negligible impact on the heavy flavor $R_{AA}$ in central collisions, but the size of the correction grows in $p_T$ in agreement with our previous results \cite{Faraday:2023mmx}. 
The relative size of the short pathlength correction grows as a function of centrality simply because of the smaller average pathlength in less central collisions.

\begin{figure}[!htpb]
	\includegraphics[width=\linewidth]{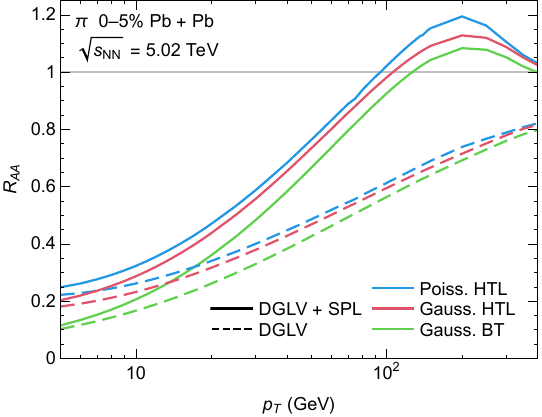}
	\caption{Nuclear modification factor $R_{AA}$ as a function of $p_T$ for pions produced in $0\text{--}5\%$ centrality \coll{Pb}{Pb} collisions at $\sqrt{s_{NN}} = 5.02$ TeV. Theoretical results are produced for pions through our convolved radiative and elastic energy loss model by varying the elastic model between Gaussian BT \cite{Braaten:1991jj, Braaten:1991we}, Gaussian HTL, and Poisson HTL \cite{Wicks:2008zz}, and the radiative model between DGLV \cite{Djordjevic:2003zk} and DGLV + SPL \cite{Kolbe:2015rvk, Kolbe:2015suq}.}
	\label{fig:raa_pbpb_pions}
\end{figure}

\begin{figure}[!htpb]
	\includegraphics[width=\linewidth]{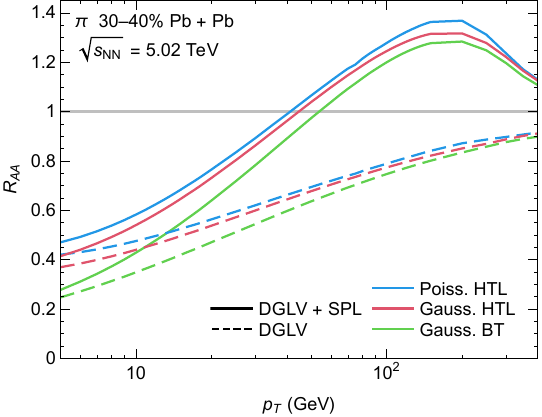}
	\caption{Nuclear modification factor $R_{AA}$ as a function of $p_T$ for pions produced in $30\text{--}40\%$ centrality \coll{Pb}{Pb} collisions at $\sqrt{s_{NN}} = 5.02$ TeV. Theoretical results are produced for pions through our convolved radiative and elastic energy loss model by varying the elastic model between Gaussian BT \cite{Braaten:1991jj, Braaten:1991we}, Gaussian HTL, and Poisson HTL \cite{Wicks:2008zz}, and the radiative model between DGLV \cite{Djordjevic:2003zk} and DGLV + SPL \cite{Kolbe:2015rvk, Kolbe:2015suq}.}
	\label{fig:raa_pbpb_pions_3040}
\end{figure}

\begin{figure}[!htpb]
	\includegraphics[width=\linewidth]{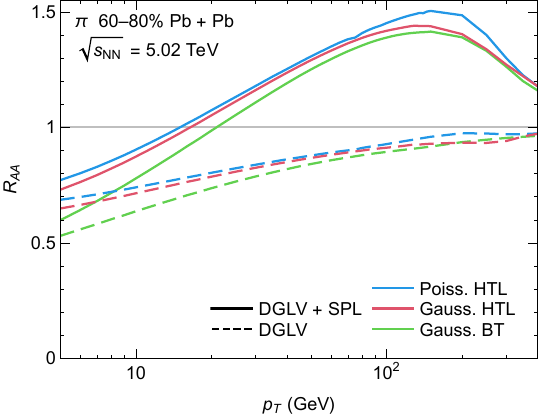}
	\caption{Nuclear modification factor $R_{AA}$ as a function of $p_T$ for pions produced in $60\text{--}80\%$ centrality \coll{Pb}{Pb} collisions at $\sqrt{s_{NN}} = 5.02$ TeV. Theoretical results are produced for pions through our convolved radiative and elastic energy loss model by varying the elastic model between Gaussian BT \cite{Braaten:1991jj, Braaten:1991we}, Gaussian HTL, and Poisson HTL \cite{Wicks:2008zz}, and the radiative model between DGLV \cite{Djordjevic:2003zk} and DGLV + SPL \cite{Kolbe:2015rvk, Kolbe:2015suq}.}
	\label{fig:raa_pbpb_pions_6080}
\end{figure}

	In \cref{fig:raa_pbpb_pions,fig:raa_pbpb_pions_3040,fig:raa_pbpb_pions_6080} we plot the nuclear modification factor $R_{AA}$ as a function of $p_T$ for pions produced in $0\text{--}5\%$, $30\text{--}40\%$, and $60\text{--}80\%$ centrality \coll{Pb}{Pb} collisions at $\sqrt{s_{NN}} = 5.02$ TeV for all aforementioned radiative and elastic energy loss kernels. While LHC typically measures charged hadrons and RHIC typically measures neutral pions, we neglect this distinction in this current work. All results labeled $\pi$ mesons or pions are calculated with DSS \cite{deFlorian:2007aj} fragmentation functions corresponding to neutral pions.

We observe in \cref{fig:raa_pbpb_pions} that $R_{AA}$ results calculated with Poisson HTL and Gaussian HTL elastic energy loss kernels differ negligibly, and this difference is reduced as a function of $p_T$, similar to the heavy-flavor results. For small momenta $p_T \lesssim 20$ GeV, there is a large $\sim 50\text{--}100\%$ difference between the $R_{AA}$ calculated with the Gaussian BT elastic energy loss and that calculated with the Gaussian HTL elastic energy loss. The sensitivity to the choice of elastic energy loss kernel reduces as a function of $p_T$.

We note that at higher momenta the pion $R_{AA}$ results are not consistent with our previous work \cite{Faraday:2023mmx}. This difference is due to an error in our pion hadronization code which effectively led to our code neglecting light quark contributions to the pion energy loss. 
The corrected results presented in \cref{fig:raa_pbpb_pions} show a faster rise of the $R_{AA}$ in $p_T$, now correctly capturing the changeover from pions fragmenting primarily from gluons to pions fragmenting primarily from light quarks \cite{Horowitz:2011gd}. We note the interesting phenomenological effect of a steep rise in the $R_{AA}$, caused by including the short pathlength correction to the radiative energy loss. 

\coleTwo{The effect of the short pathlength correction is much larger for pions compared to heavy-flavor $D$ and $B$ mesons, which is due to the breaking of color triviality by the short pathlength correction \cite{Kolbe:2015suq,Kolbe:2015rvk,Faraday:2023mmx}. This was shown numerically for the $\Delta E / E$ in \cref{sec:numerical_elastic_radiative}, where we saw that the short pathlength correction leads to a large energy \emph{gain} at high energies. }
Interestingly, the crossover from gluon-dominated pion production to light quark-dominated pion production leads to a reduction of the effect of the short pathlength correction, which is especially large for gluons, causing the $R_{AA}$ to decrease for $p_T \gtrsim 200$ GeV. \coleTwo{Referring to \cref{fig:ratio_gluon_to_lq}, we see that this sharp reduction of the short pathlength correction is consistent with the sharp reduction in the number of gluons which fragment to pions, also at around $p_T \simeq 200 ~\mathrm{GeV}$. While the short pathlength correction leads to a faster rise in the $R_{AA}$ as a function of $p_T$, it is not as large of an effect as one might naively expect from the computation of the fractional energy loss in \cref{sec:numerical_elastic_radiative}. The reduction of the effects of the short pathlength correction stems from the presence of \emph{both} elastic and radiative energy loss in the full $R_{AA}$ calculation, the decreasing fraction of gluons which fragment to pions compared to light quarks (see \cref{fig:ratio_gluon_to_lq}), and contributions from higher-order moments of the energy loss distribution than the average fractional energy loss (discussed further in \cref{sec:distribution_dependence}).
}

Contrasting \cref{fig:raa_pbpb_pions,fig:raa_pbpb_pions_3040,fig:raa_pbpb_pions_6080} we see that the short pathlength correction increases dramatically for more peripheral collisions. One may understand the increase in the short pathlength correction in peripheral systems as a function of the shorter pathlengths involved in these smaller collision systems.
The relaxation of the large formation time assumption---an assumption which we found was not satisfied self-consistently at large momenta within the DGLV formalism \cite{Faraday:2023mmx}---will likely significantly reduce the size of the short pathlength correction \cite{Faraday:2023uay}. 
Future phenomenological work \cite{Faraday:2024} will examine the effect of placing a cut on the radiated transverse momentum, which ensures that the large formation time approximation is never explicitly violated \cite{Faraday:2023mmx}.

\subsection{Large system suppression at RHIC}
\label{sec:large_system_predictions_rhic}

\begin{figure}[!htpb]
	\includegraphics[width=\linewidth]{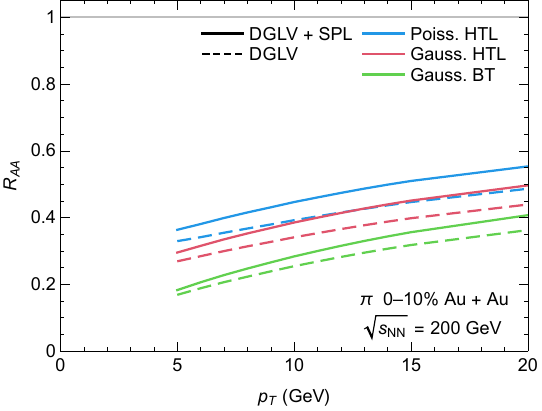}
	\caption{Nuclear modification factor $R_{AA}$ as a function of $p_T$ for pions produced in $0\text{--}10\%$ centrality \coll{Au}{Au} collisions at $\sqrt{s_{NN}} = 200$ GeV. Theoretical results are produced for pions through our convolved radiative and elastic energy loss model by varying the elastic model between Gaussian BT \cite{Braaten:1991jj, Braaten:1991we}, Gaussian HTL, and Poisson HTL \cite{Wicks:2008zz}, and the radiative model between DGLV \cite{Djordjevic:2003zk} and DGLV + SPL \cite{Kolbe:2015rvk, Kolbe:2015suq}.}
	\label{fig:raa_auau_pions}
\end{figure}

\begin{figure}[!htpb]
	\includegraphics[width=\linewidth]{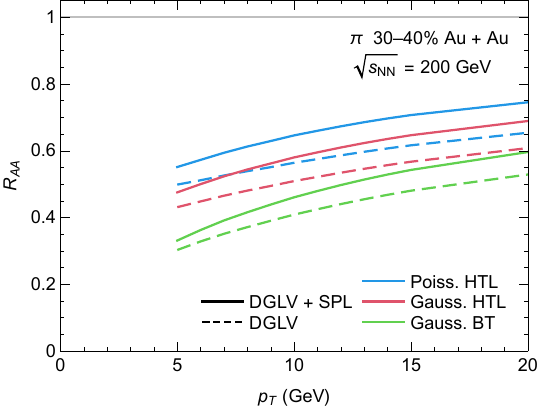}
	\caption{Nuclear modification factor $R_{AA}$ as a function of $p_T$ for pions produced in $30\text{--}40\%$ centrality \coll{Au}{Au} collisions at $\sqrt{s_{NN}} = 200$ GeV. Theoretical results are produced for pions through our convolved radiative and elastic energy loss model by varying the elastic model between Gaussian BT \cite{Braaten:1991jj, Braaten:1991we}, Gaussian HTL, and Poisson HTL \cite{Wicks:2008zz}, and the radiative model between DGLV \cite{Djordjevic:2003zk} and DGLV + SPL \cite{Kolbe:2015rvk, Kolbe:2015suq}.}
	\label{fig:raa-auau-pion-3040}
\end{figure}

\begin{figure}[!htpb]
	\includegraphics[width=\linewidth]{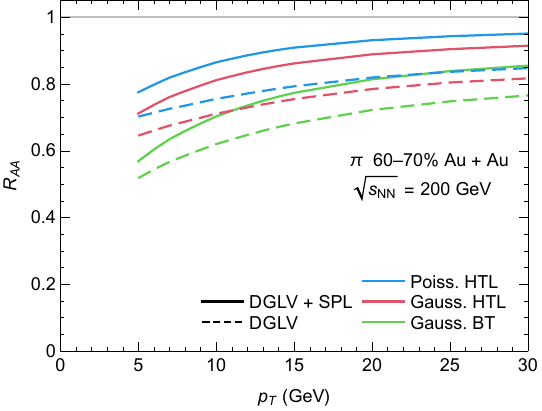}
	\caption{Nuclear modification factor $R_{AA}$ as a function of $p_T$ for pions produced in $60\text{--}70\%$ centrality \coll{Au}{Au} collisions at $\sqrt{s_{NN}} = 200$ GeV. Theoretical results are produced for pions through our convolved radiative and elastic energy loss model by varying the elastic model between Gaussian BT \cite{Braaten:1991jj, Braaten:1991we}, Gaussian HTL, and Poisson HTL \cite{Wicks:2008zz}, and the radiative model between DGLV \cite{Djordjevic:2003zk} and DGLV + SPL \cite{Kolbe:2015rvk, Kolbe:2015suq}.}
	\label{fig:raa-auau-pion-6070}
\end{figure}

\begin{figure}[!htpb]
	\includegraphics[width=\linewidth]{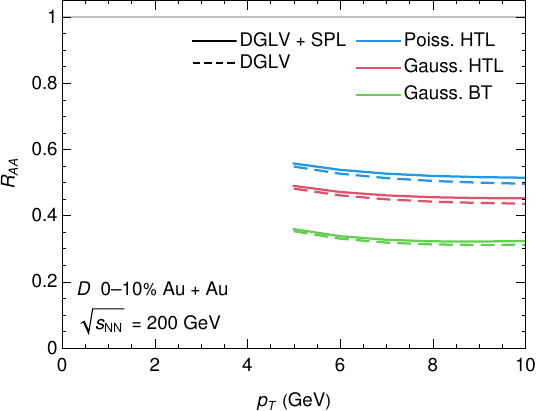}
	\caption{Nuclear modification factor $R_{AA}$ as a function of $p_T$ for $D$ mesons produced in $0\text{--}10\%$ centrality \coll{Au}{Au} collisions at $\sqrt{s_{NN}} = 200$ GeV. Theoretical results are produced for $D$ mesons through our convolved radiative and elastic energy loss model by varying the elastic model between Gaussian BT \cite{Braaten:1991jj, Braaten:1991we}, Gaussian HTL, and Poisson HTL \cite{Wicks:2008zz}, and the radiative model between DGLV \cite{Djordjevic:2003zk} and DGLV + SPL \cite{Kolbe:2015rvk, Kolbe:2015suq}.}
	\label{fig:raa_auau_dmeson}
\end{figure}

\Cref{fig:raa_auau_pions,fig:raa-auau-pion-3040,fig:raa-auau-pion-6070} show the nuclear modification factor $R_{AA}$ for pions produced in $0\text{--}10\%$, $30\text{--}40\%$, and $60\text{--}80\%$ centrality \coll{Au}{Au} collisions at $\sqrt{s_{NN}} = 200$ GeV based on our theoretical model. \Cref{fig:raa_auau_dmeson} shows the nuclear modification factor $R_{AA}$ for $D$ mesons produced in $0\text{--}10\%$ centrality \coll{Au}{Au} collisions at $\sqrt{s_{NN}} = 200$ GeV based on our theoretical model.

Both the pion and $D$ meson $R_{AA}$ exhibit a similar dependence on the choice of elastic energy loss distribution as observed for the same final states in \coll{Pb}{Pb} collisions at LHC.
There is an $\mathcal{O}(10\text{--}25\%)$ relative difference between the $R_{AA}$ results using the Poisson HTL and Gaussian HTL elastic energy loss distributions, and this difference decreases as a function of $p_T$.
This relative difference is significantly larger than the equivalent relative difference of the $R_{AA}$ calculated with the same two elastic energy loss distributions in central \coll{Pb}{Pb} collisions at LHC, as shown in \cref{fig:raa_pbpb_pions}.
The difference between the $R_{AA}$ calculated with the Gaussian HTL and Gaussian BT elastic energy loss is $\mathcal{O}(20\text{--}50\%)$, and it decreases as a function of $p_T$. This difference is similar to that in \coll{Pb}{Pb} at LHC, shown in \cref{fig:raa_pbpb_D0010,fig:raa_pbpb_pions,fig:raa_pbpb_pions_3040,fig:raa_pbpb_pions_6080,fig:raa_pbpb_D3050,fig:raa_pbpb_D6080}.

The impact of the short pathlength correction is small for both pions and $D$ mesons in central \coll{Au}{Au} collisions; however, it is smaller for $D$ mesons than for pions, as expected due to the breaking of color triviality in the short pathlength correction. The correction is significantly smaller for pions in central \coll{Au}{Au} collisions than for pions in central \coll{Pb}{Pb} collisions (compare \cref{fig:raa_pbpb_pions,fig:raa_auau_pions}) due to the much lower maximum momentum shown in the figure, as well as the smaller ratio of gluons to light quarks produced at RHIC compared to LHC \cite{Horowitz:2011gd}.
At low momenta, the short pathlength correction is small because it grows faster in $p_T$ than the uncorrected DGLV radiative energy loss \cite{Kolbe:2015suq, Kolbe:2015rvk}; see \cref{fig:deltaEoverE_small_large,fig:deltaEoverE_vs_L_gluon}. Additionally, because fewer gluons fragment to pions at RHIC compared to LHC, and because the short pathlength correction is significantly larger for gluons compared to quarks, the short pathlength correction is reduced at RHIC compared to LHC.

Comparing \cref{fig:raa_auau_pions,fig:raa-auau-pion-3040,fig:raa-auau-pion-6070}, we see that the relative effect of the short pathlength correction grows for more peripheral collisions, consistent with the behavior observed in \coll{Pb}{Pb} collisions in \cref{fig:raa_pbpb_pions,fig:raa_pbpb_pions_3040,fig:raa_pbpb_pions_6080}. The effect is not nearly as dramatic as it is for pions produced at LHC, simply because the short pathlength correction is significantly smaller at RHIC compared to LHC.

\subsection{Small system suppression at LHC}
\label{sec:small_system_predictions_LHC}

\begin{figure}[!htpb]
	\includegraphics[width=\linewidth]{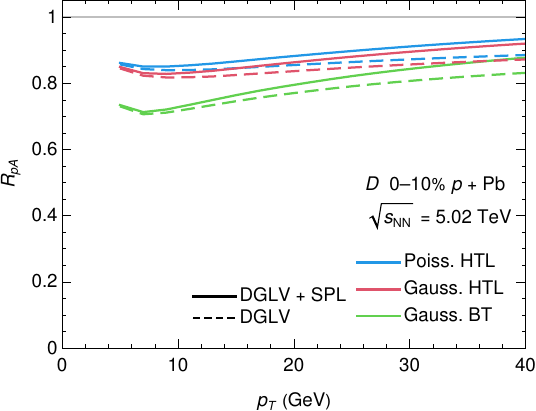}
	\caption{Nuclear modification factor $R_{AB}$ as a function of $p_T$ for $D$ mesons produced in $0\text{--}10\%$ centrality \coll{p}{Pb} collisions at $\sqrt{s_{NN}} = 5.02$ TeV. Theoretical results are produced for $D$ mesons through our convolved radiative and elastic energy loss model by varying the elastic model between Gaussian BT \cite{Braaten:1991jj, Braaten:1991we}, Gaussian HTL, and Poisson HTL \cite{Wicks:2008zz}, and the radiative model between DGLV \cite{Djordjevic:2003zk} and DGLV + SPL \cite{Kolbe:2015rvk, Kolbe:2015suq}.}
	\label{fig:raa_ppb_D}
\end{figure}

\begin{figure}[!htpb]
	\includegraphics[width=\linewidth]{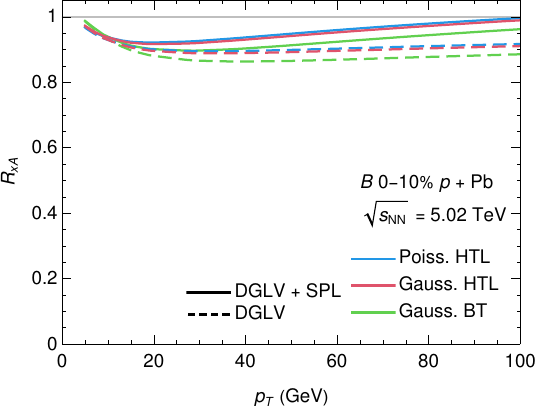}
	\caption{Nuclear modification factor $R_{AB}$ as a function of $p_T$ for $B$ mesons produced in $0\text{--}10\%$ centrality \coll{p}{Pb} collisions at $\sqrt{s_{NN}} = 5.02$ TeV. Theoretical results are produced for $B$ mesons through our convolved radiative and elastic energy loss model by varying the elastic model between Gaussian BT \cite{Braaten:1991jj, Braaten:1991we}, Gaussian HTL, and Poisson HTL \cite{Wicks:2008zz}, and the radiative model between DGLV \cite{Djordjevic:2003zk} and DGLV + SPL \cite{Kolbe:2015rvk, Kolbe:2015suq}.}
	\label{fig:raa_ppb_B}
\end{figure}

\Cref{fig:raa_ppb_D} shows the nuclear modification factor $R_{pA}$ as a function of $p_T$ for $D$ mesons produced in central \coll{p}{Pb} collisions at $\sqrt{s_{NN}} = 5.02$ TeV based on our theoretical energy loss model. \Cref{fig:raa_ppb_B} shows the same results for $B$ mesons.

We see from \cref{fig:raa_ppb_D} that for $D$ mesons produced in central \coll{p}{Pb} collisions, there is a negligible difference between the $R_{pA}$ calculated with the Poisson HTL elastic energy loss kernel, and the $R_{pA}$ calculated with the Gaussian HTL elastic energy loss kernel. This negligible difference is surprising because the $\mathcal{O}(0\text{--}1)$ elastic scatters in small collision systems is significantly fewer than the $\sim \!\! 80$ scatters required for the Poisson distribution to converge to a Gaussian distribution \cite{Wicks:2008zz}.
We discuss this point further in \cref{sec:surprising_similarity_gaussian_poisson}. 

The difference between $R_{pA}$ results with and without the short pathlength correction to the radiative energy loss is negligible, even in this small collision system where we expect that the impact of the correction is largest (see \cref{fig:deltaEoverE_small_large,fig:deltaEoverE_small_large_gluon,fig:deltaEoverE_vs_L_gluon}).
The small effect of the short pathlength correction arises because most energy loss in small collision systems is due to elastic energy loss and not radiative energy loss \cite{Faraday:2023mmx}, as evidenced by the plot of $\Delta E / E (L)$ shown in \cref{fig:deltaEoverE_vs_L_gluon}. 
One can understand the dominance of the elastic over the radiative energy loss by noting that the elastic energy loss scales like $L$, while the radiative energy loss scales like $L^2$ due to LPM interference effects. \Cref{sec:relative_elastic_radiative} provides a detailed discussion of the relative contribution of radiative vs elastic energy loss as a function of system size.

We additionally see in \cref{fig:raa_ppb_D} that the $D$ meson $R_{pA}$ at low-$p_T$ shows a large $\mathcal{O}(20\%)$ difference between results calculated with the HTL elastic energy loss and those calculated with the BT elastic energy loss. This considerable sensitivity is even more pronounced when considering the degree of suppression: the quantity $1 - R_{pA}$ displays a $\mathcal{O}(70\%)$ relative difference between results computed with these two elastic energy loss kernels---much larger than the equivalent relative difference in \coll{Pb}{Pb} collisions, shown in \cref{fig:raa_pbpb_D0010}. As in \coll{Pb}{Pb}, the relative difference between results calculated with these two elastic energy loss distributions decreases as a function of $p_T$ and displays reduced sensitivity to the choice of elastic energy loss kernel for $B$ mesons, shown in \cref{fig:raa_ppb_B}, compared to $D$ mesons, shown in \cref{fig:raa_ppb_D}.

\begin{figure}[!htpb]
	\includegraphics[width=\linewidth]{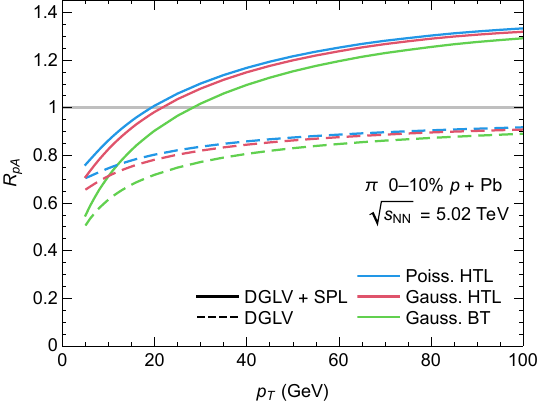}
	\caption{Nuclear modification factor $R_{AB}$ as a function of $p_T$ for pions produced in $0\text{--}10\%$ centrality \coll{p}{Pb} collisions at $\sqrt{s_{NN}} = 5.02$ TeV. Theoretical results are produced for pions through our convolved radiative and elastic energy loss model by varying the elastic model between Gaussian BT \cite{Braaten:1991jj, Braaten:1991we}, Gaussian HTL, and Poisson HTL \cite{Wicks:2008zz}, and the radiative model between DGLV \cite{Djordjevic:2003zk} and DGLV + SPL \cite{Kolbe:2015rvk, Kolbe:2015suq}.}
	\label{fig:raa_ppb_pions}
\end{figure}

\Cref{fig:raa_ppb_pions} shows the nuclear modification factor $R_{pA}$ of pions produced in central \coll{p}{Pb} collisions at $\sqrt{s_{NN}} = 5.02$ TeV based on our model. The difference between results calculated with the Poisson HTL and the Gaussian HTL elastic energy loss distributions is negligible, as it was for $D$ mesons in \coll{p}{Pb} collisions. 
We observe a similarly large sensitivity to the choice between Gaussian HTL and Gaussian BT elastic energy loss distributions for pions in \coll{p}{Pb} as for $D$ mesons in \coll{p}{Pb}.

Including the short pathlength correction produces a nuclear modification factor $R_{pA} \sim 1.2$ for $p_T \gtrsim 40$ GeV with slow $p_T$ dependence. The flattening of the $R_{pA}$ can be understood as a crossover from gluon-dominated energy loss to light quark-dominated energy loss, similar to what was observed for pions produced in \coll{Pb}{Pb} collisions.

\subsection{Small system suppression at RHIC}
\label{sec:small_system_predictions_rhic}

\begin{figure}[!htbp]
	\includegraphics[width=\linewidth]{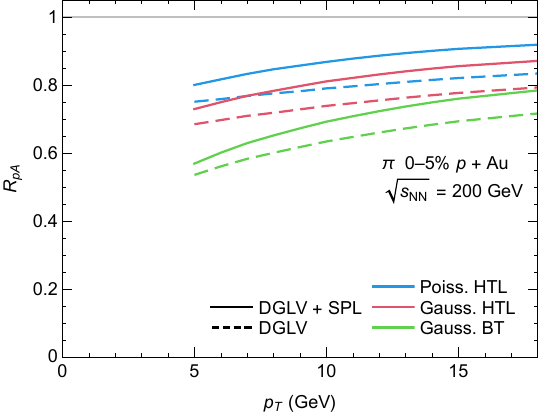}
	\caption{Nuclear modification factor $R_{pA}$ as a function of $p_T$ for pions produced in $0\text{--}5\%$ centrality \coll{p}{Au} collisions at $\sqrt{s_{NN}} = 200$ GeV. Theoretical results are produced for pions through our convolved radiative and elastic energy loss model by varying the elastic model between Gaussian BT \cite{Braaten:1991jj, Braaten:1991we}, Gaussian HTL, and Poisson HTL \cite{Wicks:2008zz}, and the radiative model between DGLV \cite{Djordjevic:2003zk} and DGLV + SPL \cite{Kolbe:2015rvk, Kolbe:2015suq}.}
	\label{fig:raa-pion-pau-0005}
\end{figure}

\begin{figure}[!htpb]
	\includegraphics[width=\linewidth]{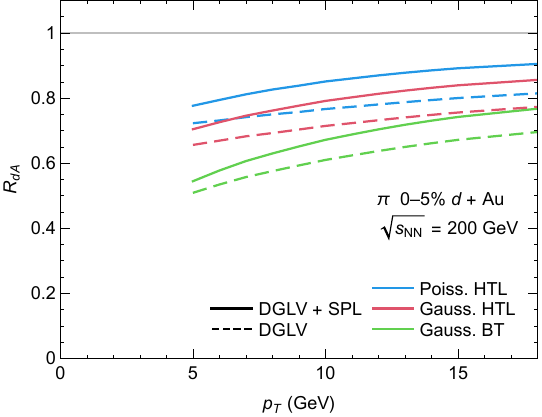}
	\caption{Nuclear modification factor $R_{dA}$ as a function of $p_T$ for pions produced in $0\text{--}5\%$ centrality \coll{d}{Au} collisions at $\sqrt{s_{NN}} = 200$ GeV. Theoretical results are produced for pions through our convolved radiative and elastic energy loss model by varying the elastic model between Gaussian BT \cite{Braaten:1991jj, Braaten:1991we}, Gaussian HTL, and Poisson HTL \cite{Wicks:2008zz}, and the radiative model between DGLV \cite{Djordjevic:2003zk} and DGLV + SPL \cite{Kolbe:2015rvk, Kolbe:2015suq}.}
	\label{fig:raa_dau_pions}
\end{figure}

\begin{figure}[!htpb]
	\includegraphics[width=\linewidth]{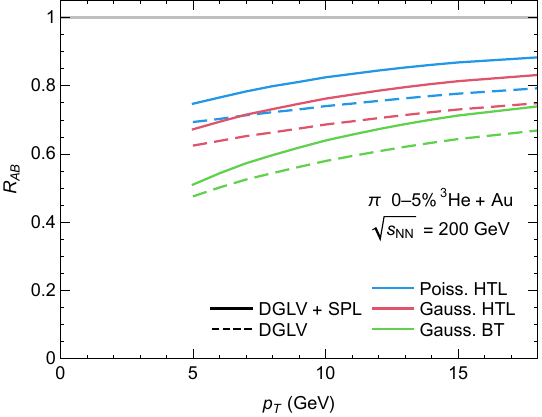}
	\caption{Nuclear modification factor $R_{{}^3\mathrm{He}A}$ as a function of $p_T$ for pions produced in $0\text{--}5\%$ centrality \coll{He3}{Au} collisions at $\sqrt{s_{NN}} = 200$ GeV. Theoretical results are produced for pions through our convolved radiative and elastic energy loss model by varying the elastic model between Gaussian BT \cite{Braaten:1991jj, Braaten:1991we}, Gaussian HTL, and Poisson HTL \cite{Wicks:2008zz}, and the radiative model between DGLV \cite{Djordjevic:2003zk} and DGLV + SPL \cite{Kolbe:2015rvk, Kolbe:2015suq}.}
	\label{fig:raa_he3au_pions}
\end{figure}

\Cref{fig:raa-pion-pau-0005,fig:raa_dau_pions,fig:raa_he3au_pions} show the nuclear modification factor $R_{AB}$ as a function of $p_T$ for pions produced in central \coll{p}{Au}, \coll{d}{Au}, and \coll{He3}{Au} collisions at $\sqrt{s_{NN}} = 200$ GeV based on our theoretical energy loss model.

We see that the nuclear modification factor $R_{AB}$ model results for \coll{p}{Au}, \coll{d}{Au}, and \coll{He3}{Au} collisions are almost identical. This can be understood by the similarity of the collisions, as evidenced by the length and temperature distributions shown in \cref{fig:path_length_distribution,fig:temperature_distribution}. 
The absolute difference between the Poisson HTL and Gaussian HTL pion $R_{AB}$ in \collFour{p}{d}{He3}{Au} collisions is similar to the same difference in the pion $R_{AB}$ in \coll{Au}{Au} collisions, as shown in \cref{fig:raa_auau_pions}.
The largest difference between the theoretical curves is the choice of elastic energy loss kernel (HTL vs BT), accounting for the $\sim \! 50\%$ difference in the predicted $R_{AB}$. The short pathlength correction to the radiative energy loss has a relatively small effect in \collFour{p}{d}{He3}{Au} collisions compared to its impact in \coll{p}{Pb} collisions, as shown in \cref{fig:raa_ppb_pions}. The small impact of the short pathlength correction is due to the same reasons as the \coll{Pb}{Pb} vs \coll{Au}{Au} case: the lower maximum $p_T$ and smaller fraction of gluons relative to light quarks at RHIC compared to LHC leads to a significantly smaller short pathlength correction.

\section{Surprising independence of the \texorpdfstring{$R_{AB}$}{R_AB} on the elastic energy loss distribution}
\label{sec:surprising_similarity_gaussian_poisson}

\begin{figure}[!htpb]
	\includegraphics[width=\linewidth]{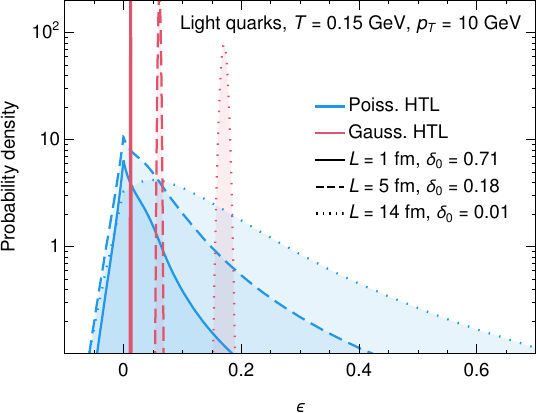}
	\caption{Comparison of realistic Poisson distribution to Gaussian approximation for various pathlengths $L$. The Gaussian distribution is constrained to have the same expectation value $\langle \epsilon \rangle$ as the Poisson distribution, and its variance is given by the fluctuation dissipation theorem \cite{Moore:2004tg}. The probability weight attached to the delta function at $\epsilon = 0$ for the Poisson distribution is given by the value of $\delta_0$ in the legend.} 
	\label{fig:comparison_gaussian_poisson}
\end{figure}

In this section, we aim to understand the surprisingly small difference between the $R_{AB}$ calculated with the Poisson HTL elastic energy loss kernel and the Gaussian HTL elastic energy loss kernel (see \cref{sec:elastic_energy_loss} for a comparison). \Cref{fig:comparison_gaussian_poisson} plots the Poisson HTL and Gaussian HTL elastic energy loss probability distributions for pathlengths of $L = 1$ fm, $L = 5$ fm, and $L = 14$ fm for incident light quarks at constant temperature $T = 0.15$ GeV and momentum $p_T = 10$ GeV. 
Note that we choose a relatively low temperature for illustrative purposes.

According to \cref{fig:path_length_distribution} we may interpret the $L = 1$ fm results as typical of small system results in central \coll{p}{Pb} and \collFour{p}{d}{He3}{Au}, the $L = 5$ fm results as typical of large system results in central \coll{Pb}{Pb} and \coll{Au}{Au}, and the $L=14$ fm results as being from an unrealistically large system. \Cref{fig:comparison_gaussian_poisson} shows that for all system sizes, the Gaussian approximation does not visually appear to be a reasonable approximation to the full Poisson distribution. One may conclude that the small difference observed between the Gaussian HTL and Poisson HTL results in all systems and final states shown in \crefrange{fig:raa_pbpb_D0010}{fig:raa_dau_pions} is not attributable to convergence of the Poisson distribution to the Gaussian distribution according to the central limit theorem.

\begin{figure}[!htpb]
	\includegraphics[width=\linewidth]{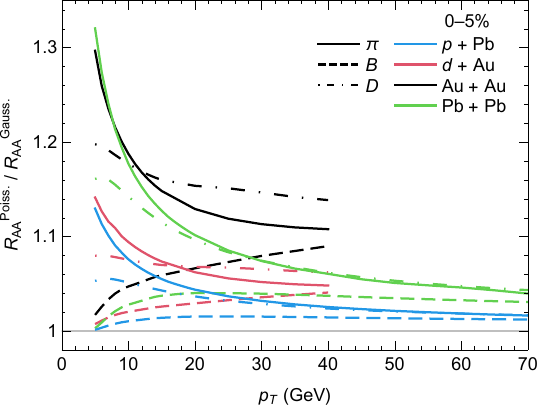}
	\caption{Plot of the double ratio $R_{\text{PG}} \equiv R_{AB}^{\text{Pois.}} / R_{AB}^{\text{Gauss.}}$, where the $R_{AB}$ is calculated with the DGLV radiative energy loss \cite{Djordjevic:2003zk} convolved with the HTL Poisson elastic energy loss for $R^{\text{Pois.}}_{\text{AB}}$, and HTL Gaussian elastic energy loss for $R^{\text{Gauss.}}_{AB}$. The ratio $R^{\text{PG}}_{AB}$ is plotted as a function of $p_T$ for pions, $D$ mesons, and $B$ mesons produced in $0\text{--}5\%$ centrality \coll{d}{Au}, \coll{Au}{Au}, \coll{p}{Pb}, and \coll{Pb}{Pb} collisions.
} 
	\label{fig:raa_poisson_over_gaussian_vs_pt}
\end{figure}

As further evidence that the approximation of the Poisson distribution to a Gaussian distribution does not make sense on the grounds of convergence according to the central limit theorem, we compute the double ratio $R_{\text{PG}} \equiv R_{AB}^{\text{Pois.}} / R_{AB}^{\mathrm{Gauss.}}$, where $R_{AB}^{\text{Pois.}}$ is the $R_{AB}$ calculated with Poisson HTL elastic energy loss convolved with the DGLV radiative energy loss and $R_{AB}^{\text{Gauss.}}$ is the $R_{AB}$ calculated with the is the Gaussian HTL elastic energy loss convolved with the DGLV radiative energy loss. The ratio $R_{\text{PG}}$ is a measure of the relative difference between the $R_{AB}$ calculated with Poisson HTL and Gaussian HTL elastic energy loss kernels, respectively. \Cref{fig:raa_poisson_over_gaussian_vs_pt} plots $R_{\text{PG}}$ as a function of $p_T$ for pions, $B$ mesons, and $D$ mesons produced in central \coll{p}{Pb}, \coll{d}{Au}, \coll{Au}{Au}, and \coll{Pb}{Pb} collisions. 

	\Cref{fig:raa_poisson_over_gaussian_vs_pt} displays several qualitative features that warrant explanation, especially in light of the inadequacy of the appeal to the central limit theorem according to \cref{fig:comparison_gaussian_poisson}.
We explain these features in order in \cref{sec:distribution_dependence,sec:relative_elastic_radiative}. We see that the ratio $R_{\text{PG}}$
\begin{itemize}
	\item \emph{Is close to one} for all final states and collision systems of interest.
	\item \emph{Increases as a function of system size} (comparing systems at the same $\sqrt{s_{NN}}$). This system size dependence is opposite to that which the central limit theorem predicts.
	\item \emph{Decreases as a function of $p_T$}.
	\item Exhibits a clear \emph{mass ordering} at low-$p_T$ of $R_{\text{PG}}^{B}< R_{\text{PG}}^{D}< R_{\text{PG}}^{\pi}$, based on the final state hadron.
	\item \emph{Decreases as a function of $\sqrt{s_{NN}}$}, when considering systems of similar geometrical size (refer to \cref{fig:path_length_distribution,fig:temperature_distribution})---\coll{d}{Au} vs \coll{p}{Pb} and \coll{Pb}{Pb} vs \coll{Au}{Au}---at RHIC and LHC energies.
\end{itemize}

Two primary mechanisms contribute to the observed effects listed above. First, the relative contributions of elastic and radiative energy loss helps to explain why $R_{\text{PG}} \sim 1$, and the system size and  $p_T$ dependence of $R_{\text{PG}}$. The effect of the relative contribution of radiative vs elastic energy loss is discussed in \cref{sec:relative_elastic_radiative}. Second, the dependence of the $R_{AB}$ on the various moments of the elastic energy loss distribution helps to explains why $R_{\text{PG}} \sim 1$ as well as the system size, $\sqrt{s_{NN}}$, and final state dependence. We discuss the effect of the various moments of the energy loss distributions in \cref{sec:distribution_dependence}. These two mechanisms comprehensively account for all the aforementioned features of the $R_{\text{PG}}$ ratio.

\subsection{Relative contribution of elastic and radiative energy loss to the \texorpdfstring{$R_{AB}$}{RAB}}
\label{sec:relative_elastic_radiative}

While asymptotically the $p_T$ dependence of the radiative and elastic energy loss is the same, $\Delta E \sim \ln E$, %
we saw numerically in \cref{sec:numerical_elastic_radiative} that at non-asymptotic energies the $p_T$ dependence differs dramatically. We saw explicitly in \cref{fig:deltaEoverE_small_large,fig:deltaEoverE_small_large_gluon} that while the final asymptotic ratio of elastic vs radiative energy loss depends strongly on both the pathlength $L$ and the energy loss model used for both the elastic and radiative energy loss, the trend in $p_T$ is clear: as $p_T$ increases the ratio of elastic energy loss to radiative energy loss $\Delta E_{\text{el.}} / \Delta E_{\text{rad.}}$ decreases. This trend in $p_T$ accounts for the $p_T$ dependence of \cref{fig:raa_poisson_over_gaussian_vs_pt}.

To quantify this effect, we define
\begin{equation}
	\left(\frac{\Delta E^{\text{el.}}}{\Delta E^{\text{rad.}}}\right)_{\text{eff.}} \equiv \frac{1-R_{AB}^{\mathrm{el.}}}{1-R_{AB}^{\mathrm{rad.}}},
	\label{eqn:effective_deltaE_elastic_over_radiative}
\end{equation}
where $R_{AB}^{\text{el.}}$ is the nuclear modification factor calculated with the radiative energy loss turned off, and similarly $R_{AB}^{\text{rad.}}$ has the elastic energy loss turned off.

We consider the ratio of $1-R_{AB}$ in the above because of the following expansion\footnote{For a detailed discussion of such an expansion in the lost fractional energy $\epsilon$ refer to \cref{sec:validity_of_power_law_approximation_to_r_aa}} in the lost fractional energy $\epsilon$. From \cref{eqn:full_raa_spectrum_ratio} we have
\begin{align}
	R_{AB}^{\text{tot.}} =& \frac{1}{f\left(p_t\right)} \int \frac{d \epsilon}{1-\epsilon} f\left(\frac{p_t}{1-\epsilon}\right) P_{\text {tot.}}\left(\epsilon \left | \frac{p_t}{1-\epsilon}\right. \right)\nonumber\\
R_{AB}^{\text{tot.}}=& \int d \epsilon \left[1 + (1 + p_T f'(p_T)) \epsilon + \mathcal{O}(\epsilon^2) \right] \times \nonumber\\
& \int dx \; P_{\text{rad.}}(x) P_{\text{el.}}(\epsilon - x),\\
\end{align}
which implies that
\begin{align}
1 - R_{AB}^{\text{tot.}} =& \int d \epsilon' \left[\left(1 + p_T f'(p_T) \right) (\epsilon' + x) + \mathcal{O}(x + \epsilon')^2\right]   \nonumber\\
& \times \int dx \; P_{\text{rad.}}(x) P_{\text{el.}}(\epsilon') \nonumber\\
=& \left(1 + p_T \frac{f'(p_T)}{f(p_T)} \right) \left[ \frac{\Delta E^{\text{rad.}}}{E} + \frac{\Delta E ^{\text{el.}}}{E} \right]\nonumber\\
&+ \mathcal{O}(\langle \epsilon^2 \rangle_{\text{tot.}})\nonumber\\
	1-R_{AB}^{\text{tot.}} =& (1-R_{AB}^{\text{rad.}}) + (1-R_{AB}^{\text{el.}}) + \mathcal{O}(\langle \epsilon^2 \rangle_{\text{tot.}}),
	\label{eqn:one_minus_RAB}
\end{align}
which facilitates the interpretation of \cref{eqn:effective_deltaE_elastic_over_radiative} as the fraction of elastic energy loss vs radiative energy loss which contributes to the $R_{AB}$.  In \cref{eqn:one_minus_RAB} we see that to leading order in $\left\langle  \epsilon \right\rangle_{\text{tot.}}$ in both the numerator and denominator, $(\Delta E^{\text{el.}}/\Delta E^{\text{rad.}})_{\text{eff.}} = \Delta E^{\text{rad.}}/\Delta E^{\text{el.}}$. The ratio $(\Delta E^{\text{el.}}/\Delta E^{\text{rad.}})_{\text{eff.}}$ is normalized such that it is equal to one when radiative and elastic energy loss contribute equally to suppression, below one when radiative energy loss is the dominant suppression mechanism, and above one when elastic energy loss is the dominant suppression mechanism. Higher order corrections to \cref{eqn:one_minus_RAB} indicate an interplay between the radiative and elastic energy loss distributions, and one cannot separate these corrections into purely radiative or purely elastic contributions.

\begin{figure}[!htpb]
	\includegraphics[width=\linewidth]{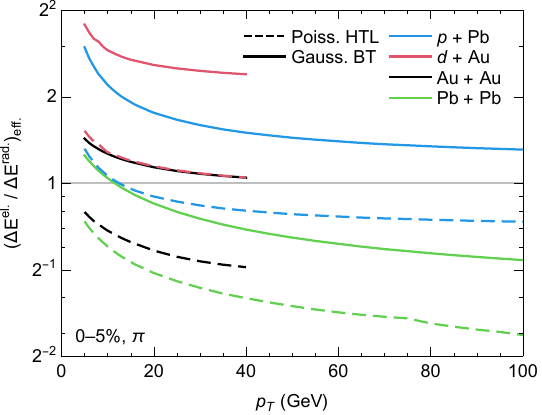}
	\caption{Plot of the ratio $\left(\Delta E^{\text{el.}} / \Delta E^{\text{rad.}}\right)_{\text{eff.}} \equiv (1-R_{AB}^{\mathrm{el.}})/(1-R_{AB}^{\mathrm{rad.}})$, where $R_{AB}^{\text{el.}}$ is the nuclear modification factor calculated with the radiative energy loss turned off, and similarly $R_{AB}^{\text{rad.}}$ has the elastic energy loss turned off. The ratio $(\Delta E^{\text{el.}}/\Delta E^{\text{rad.}})_{\text{eff.}}$ is plotted as a function of $p_T$ $0\text{--}5\%$ centrality \coll{d}{Au}, \coll{Au}{Au}, \coll{p}{Pb}, and \coll{Pb}{Pb} collisions. We calculate the ratio for both Poisson HTL \cite{Wicks:2008zz} and Gaussian BT \cite{Braaten:1991jj,Braaten:1991we} elastic energy loss distributions convolved with the Poisson DGLV radiative energy loss distribution \cite{Djordjevic:2003zk}. Note the $\log_2$ scale used for the vertical axis in the figure, with a linear scale for the minor ticks.
	}
	\label{fig:raa_elastic_over_radiative_vs_pt}
\end{figure}

\Cref{fig:raa_elastic_over_radiative_vs_pt} plots the ratio $(\Delta E^{\text{el.}}/\Delta E^{\text{rad.}})_{\text{eff.}}$ as a function of $p_T$ for $0\text{--}5\%$ centrality \coll{d}{Au}, \coll{Au}{Au}, \coll{p}{Pb}, and \coll{Pb}{Pb} collisions. We use the DGLV radiative energy loss kernel and the Gaussian BT elastic energy loss kernel for all curves and vary the elastic energy loss kernel between the Poisson HTL and Gaussian BT elastic energy loss kernels. We use a $\log_2$ vertical scale in the figure, which gives equal visual weight to scenarios where radiative energy loss is twice elastic energy loss and vice versa.

From \cref{fig:raa_elastic_over_radiative_vs_pt}, the elastic energy loss is $\sim 1\text{--}3$  times more important in small systems than large collision systems. Additionally, using the Poisson HTL vs  Gaussian BT elastic energy loss distributions dramatically changes the relative importance of the elastic energy loss. We conclude that further theoretical control is needed in the elastic energy loss sector to make quantitative predictions in small systems. Indeed, even \textit{qualitative} questions, such as whether elastic or radiative energy loss is of primary importance in small and large systems, are sensitive to this uncertainty.

The extensive range of relative contributions of the elastic vs radiative energy loss as a function of system size and $\sqrt{s_{NN}}$ leads us to propose that a system size scan will help to disentangle the relative contribution of radiative and elastic energy loss.

\subsection{Distribution dependence of the nuclear modification factor}
\label{sec:distribution_dependence}

One approach to understanding how the shape of energy loss distributions affect the nuclear modification factor is by expanding the integrand of the $R_{AB}$ in powers of the fractional energy loss $\epsilon$. This approach is motivated by the fact that the Gaussian and Poisson distributions are constrained to have identical\footnote{In practice, they may differ slightly due to the kinematic cut at $\epsilon =1$. See \cref{fig:deltaEoverE_small_large_gluon,fig:deltaEoverE_small_large,fig:deltaEoverE_vs_L_gluon} for a comparison of the first moment ($\Delta E / E$) between the Gaussian and Poisson distributions.} zeroth moments $\left\langle \epsilon^0 \right\rangle \equiv \int \mathrm{d} \epsilon P(\epsilon) = 1$ (probability conservation) and identical first moments $\left\langle \epsilon^1 \right\rangle \equiv \int \mathrm{d} \epsilon P(\epsilon) = \Delta E / E$. Starting with the expression for $R_{AB}$ from \cref{eqn:full_raa_spectrum_ratio} one may schematically expand the integrand in powers of $\epsilon$ to obtain 
\begin{align}
	R_{AB}(p_T) =& \frac{1}{f\left(p_t\right)} \int \frac{d \varepsilon}{1-\varepsilon} f\left(\frac{p_t}{1-\varepsilon}\right) P_{\text {tot.}}\left(\varepsilon \left | \frac{p_t}{1-\varepsilon}\right. \right)\nonumber\\
	R_{AB}(p_T) =& \sum_n c_n(p_T) \int d \epsilon \; \epsilon^n P_{\text{tot.}}(\epsilon | p_T)\nonumber\\
	=& \sum_n c_n(p_T) \left\langle \epsilon^n(p_T) \right\rangle_{\text{tot.}}
	\label{eqn:schematic_moment_expansion}
\end{align}
where $\left\langle \epsilon^n(p_T) \right\rangle_{\text{tot.}} \equiv \int d \epsilon \; \epsilon^n P_{\text{tot.}}(\epsilon | p_T)$ is the $n^{\text{th}}$ raw moment of the $P_{\text{tot.}}$ distribution and $c_n(p_T)$ are the coefficients of the expansion. Importantly, this form of the $R_{AB}$ partitions the effect of the production spectrum into the $c_n(p_T)$ and the effects of energy loss into the $\left\langle \epsilon^n(p_T) \right\rangle_{\text{tot.}}$. The $c_n$ do not have a simple analytic form; however, we list the first few:

\begin{align}
	c_0(p_T) =& 1 \nonumber\\
	c_1(p_T) =& 1 + p_T \frac{f'(p_T)}{f(p_T)} \nonumber\\
	c_2(p_T) =& 1 + p_T \frac{f'(p_T)}{f(p_T)} + \frac{1}{2} p_T^2 \frac{f''(p_T)}{f(p_T)}\\
	c_3(p_T) =& 1 + 3 p_T \frac{f'(p_T)}{f(p_T)} + \frac{3}{2} p_T^2 \frac{f''(p_T)}{f(p_T)} + \frac{1}{6} p_T^3 \frac{f'''(p_T)}{f(p_T)}.\nonumber
	\label{eqn:cn}
\end{align}

Since the total energy loss distribution is the convolution of the radiative and elastic energy loss distributions, the moments $\left\langle \epsilon^n(p_T) \right\rangle_{\text{tot.}}$ of the total probability distribution $P_{\text{tot.}}$ are related to the moments $\left\langle \epsilon^n(p_T) \right\rangle_{\text{rad.}}$ and $\left\langle \epsilon^n(p_T) \right\rangle_{\text{el.}}$ of the radiative $P_{\text{rad.}}$ and elastic $P_{\text{el.}}$ distributions respectively through a binomial expansion
\begin{align}
	\left\langle \epsilon^n \right\rangle_{\text{tot.}} \equiv& \int d \epsilon \; \epsilon^n P_{\text{tot.}}(\epsilon | p_T)\\
	=& \int d \epsilon \; \epsilon^n \int d x \; P_{\text{el.}}(\epsilon - x | p_T) P_{\text{rad.}}(x | p_T)\\
	=& \int d \epsilon' \; (\epsilon' + x)^n \int d x \; P_{\text{el.}}(\epsilon' | p_T) P_{\text{rad.}}(x | p_T)\\
  =&\sum_k \binom{n}{k} \left\langle \epsilon^k \right\rangle_{\text{rad.}} \left\langle \epsilon^{n-k} \right\rangle_{\text{el.}}.
	\label{eqn:moment_binomial_radiative_elastic}
\end{align}

\begin{figure}[!htbp]
	\includegraphics[width=\linewidth]{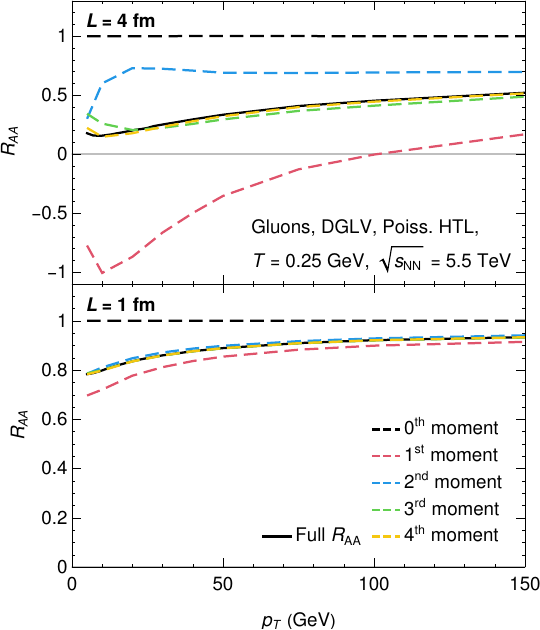}
	\caption{\coleTwo{Order-by-order approximation of the $R_{AB}$ in terms of moments $\langle \epsilon^i \rangle \equiv \int d \epsilon \epsilon^i P_{\text{tot.}}(\epsilon)$ of the total energy loss probability distribution, as a function of $p_T$. Results are presented for a gluon at constant temperature $T=0.25$~GeV, with the top pane corresponding to a constant pathlength $L = 4 ~\mathrm{fm}$ and the bottom pane to $L = 1 ~\mathrm{fm}$. The solid curve shows the $R_{AB}$ calculated according to \cref{eqn:full_raa_spectrum_ratio} while the dashed curves shows the cumulative contributions up to the $i^{\text{th}}$ moment (indicated in the legend). All curves are computed with DGLV radiative energy loss \cite{Djordjevic:2003zk} and Poisson HTL elastic energy loss \cite{Wicks:2008zz} and at $\sqrt{s_{NN}} = 5.5$ TeV.}}
	\label{fig:raa_moment_breakdown}
\end{figure}

\Cref{fig:raa_moment_breakdown} shows the order-by-order contributions to $R_{AA}$ as a function of $p_T$ for the expansion described by \cref{eqn:schematic_moment_expansion}. This calculation is performed for a gluon moving through a constant brick of length $L = 4$~fm (top pane) and $L = 1$~fm (bottom pane), and constant temperature $T = 0.25$~GeV. \coleTwo{The DGLV radiative energy loss and Poisson HTL elastic energy loss are used for this calculation.} From the figure, we see that when there is a large amount of suppression of $R_{AA} \sim 0.2$ (top pane), we need up to the fourth moment to correctly converge to the full $R_{AA}$. In contrast, for smaller suppression of $R_{AA} \sim 0.75$ (bottom pane), only the first one or two moments are needed to correctly reproduce the full $R_{AA}$.  

\begin{figure}[!htbp]
	\includegraphics[width=\linewidth]{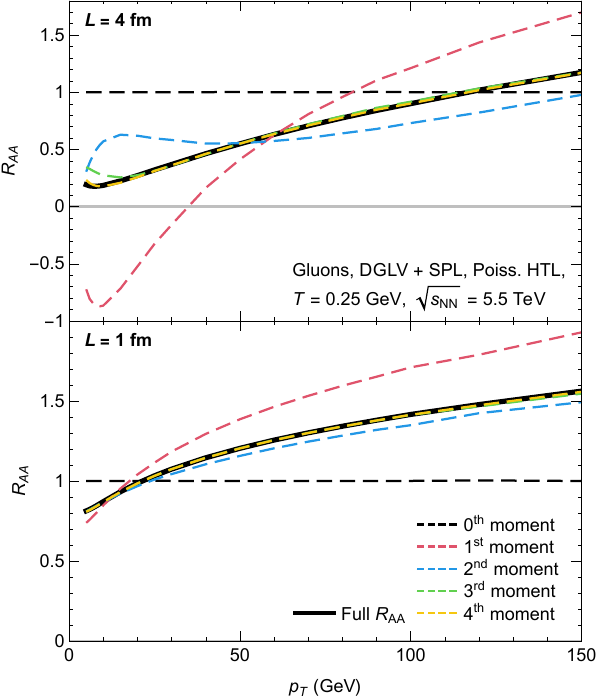}
	\caption{\coleTwo{Order-by-order approximation of the $R_{AB}$ in terms of moments $\langle \epsilon^i \rangle \equiv \int d \epsilon \epsilon^i P_{\text{tot.}}(\epsilon)$ of the total energy loss probability distribution, as a function of $p_T$. Results are presented for a gluon at constant temperature $T=0.25$~GeV, with the top pane corresponding to a constant pathlength $L = 4 ~\mathrm{fm}$ and the bottom pane to $L = 1 ~\mathrm{fm}$. The solid curve shows the $R_{AB}$ calculated according to \cref{eqn:full_raa_spectrum_ratio} while the dashed curves shows the cumulative contributions up to the $i^{\text{th}}$ moment (indicated in the legend). All curves are computed with DGLV + SPL radiative energy loss \cite{Kolbe:2015rvk, Kolbe:2015suq} and Poisson HTL elastic energy loss \cite{Wicks:2008zz} and at $\sqrt{s_{NN}} = 5.5$ TeV.}}
	\label{fig:raa_moment_breakdown_SPL}
\end{figure}

\coleTwo{\Cref{fig:raa_moment_breakdown_SPL} shows the same order-by-order contributions to the $R_{AA}$ as a function of $p_T$; however, the DGLV + SPL radiative energy loss is used with the Poisson HTL elastic energy loss. We observe, as in \cref{fig:raa_moment_breakdown}, that the first four moments are needed to converge to the full $R_{AA}$ for large systems; however, only the first two moments are required in the small system case. In the figure, we observe that for positive energy loss ($R_{AA} < 1$), higher order moments increase $R_{AA}$, while for negative energy loss ($R_{AA} > 1$) they decrease $R_{AA}$. We note that for a more realistic radiative energy loss probability distribution, such as the Skellam distribution \cite{skellam:1946,Faraday:2023mmx}, the latter effect may be reversed. In such a distribution the moments are likely to alternate due to the significant probability weight for negative fractional energy loss, resulting in $\langle \epsilon \rangle < 0, \langle \epsilon^2 \rangle > 0,$ and so on. Our simpler treatment of the probability distribution follows the canonical approach in the literature \cite{Gyulassy:2001nm} which convolves the single emission kernel $dN^g / dx$. For negative energy loss this kernel is less than zero which leads to $\langle \epsilon \rangle <0, \langle \epsilon^2 \rangle < 0$ and, consequently, an alternating series of contributions to the $R_{AA}$ due to the power law spectrum (discussed further in \cref{sec:validity_of_power_law_approximation_to_r_aa}). Further studies are required to understand the implications of a more realistic distribution on the $R_{AA}$.
}

From these illustrative examples, one may already understand the system size dependence of $R_{\text{PG}}$ in \cref{fig:raa_poisson_over_gaussian_vs_pt}. In small systems, there is a small amount of suppression, and so only the $0^{th}$ and $1^{st}$ moments are necessary to describe the $R_{AB}$, for which both the Gaussian and Poisson distributions produce identical values. In larger systems, more moments are required for a satisfactory fit; however, the elastic energy loss is only a small contribution to suppression compared to the radiative energy loss (see \cref{sec:relative_elastic_radiative}).

In order to illustrate this mechanism for a more extensive range of collision systems and final states, we summarize this effect by introducing the \textit{average important moment}

\begin{equation}
\left\langle n \right\rangle \equiv \frac{\sum_n n \; \left| c_n  \left\langle \epsilon^n \right\rangle \right|}{\sum_n \left| c_n \left\langle \epsilon^n \right\rangle \right|},
	\label{eqn:average_important_moment}
\end{equation}
as a measure of how much each moment of the expansion in \cref{eqn:schematic_moment_expansion} contributes to the $R_{AB}$. Note that the reason we consider the absolute value of the summed terms is that the expansion in \cref{eqn:schematic_moment_expansion} is generically oscillatory\footnote{The oscillation in the expansion is due to the power-law behavior of the spectrum meaning that successive derivatives of the spectrum $f(p_T)$ have alternating signs.}, as evidenced in \cref{fig:raa_moment_breakdown}, and so to capture the ``impact" of a particular moment we consider the absolute value of that moment's contribution.
Since the zeroth and first moments of both the Poisson and Gaussian elastic energy loss distributions are identical, we expect that smaller values of $\left\langle n \right\rangle$ correspond to a more similar $R_{AB}$ for the two distributions. 

\begin{figure}[!htbp]
	\includegraphics[width=\linewidth]{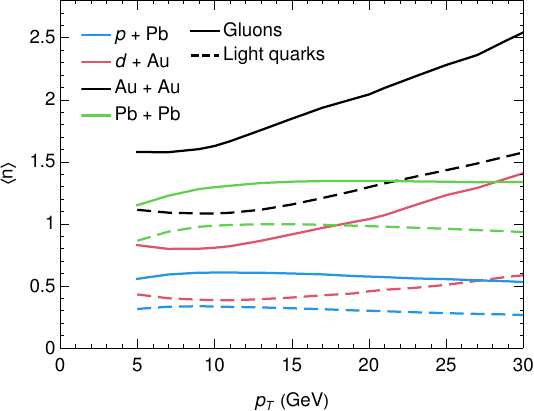}
	\caption{Plot of the average important moment $\left\langle n \right\rangle$ as a function of $p_T$ for gluons and light quarks produced in $0\text{--}5\%$ centrality \coll{p}{Pb}, \coll{d}{Au}, \coll{Au}{Au}, and \coll{Pb}{Pb} collisions. All curves are computed with Poisson HTL elastic energy loss and DGLV radiative energy loss.}
	\label{fig:average_important_moment}
\end{figure}

\Cref{fig:average_important_moment} plots $\left\langle n \right\rangle$ as a function of $p_T$ for gluons and light quarks produced in central \coll{p}{Pb}, \coll{d}{Au}, \coll{Au}{Au}, and \coll{Pb}{Pb} collisions. In the figure, we observe $\left\langle n \right\rangle$ is larger in \coll{Au}{Au} and \coll{Pb}{Pb} collisions than it is in \coll{d}{Au} and \coll{p}{Pb} collisions. 
This system size dependence of $\left\langle n \right\rangle$ explains the system size dependence of the ratio $R_{\text{PG}}$ in \cref{fig:raa_poisson_over_gaussian_vs_pt}---in large \coll{Au}{Au} and \coll{Pb}{Pb} collision systems there is more energy loss and hence the higher order moments are more important in comparison to the smaller collision systems of \coll{p}{Pb} and \coll{d}{Au}. Since the zeroth and first moment contributions to the $R_{AB}$ are constrained to be identical, the smaller $\left\langle n \right\rangle$ means that the $R_{AB}$ computed with Gaussian and Poisson elastic energy loss distributions respectively will be more similar.

The $p_T$ dependence of $\left\langle n \right\rangle$ shown in \cref{fig:average_important_moment} is weak at LHC but moderate at RHIC. On may understand this $p_T$ dependence as a competition between the $p_T$ dependence of the production spectrum and the $p_T$ dependence of the energy loss. Less energy loss corresponds to a lower $\left\langle n \right\rangle$ because $\left\langle \epsilon^n \right\rangle \sim (\Delta E / E)^n \sim ( \ln E / E)^n$, if one neglects the effects of the shape of the distribution. The $p_T$ dependence of the spectra affects the $p_T$ dependence of the $R_{AB}$ through its effect on the $c_n(p_T)$ coefficients of \cref{eqn:schematic_moment_expansion}, listed in \cref{eqn:cn}. From \cref{eqn:cn}, we see that a faster-changing spectra leads to a larger $f'(p_T)$ and a larger magnitude for the various $c_n$. We conclude that at LHC, the $p_T$ dependence of the energy loss is more important to understand the $p_T$ dependence of $\left\langle n \right\rangle$, while at RHIC, the $p_T$ dependence of the spectra is more important.

\Cref{fig:average_important_moment} shows a strong $\sqrt{s_{NN}}$ dependence in $\left\langle n \right\rangle$; the \coll{Au}{Au} and \coll{d}{Au} collisions at $\sqrt{s_{NN}} = 200$ GeV have a larger $\left\langle n \right\rangle$ than the corresponding \coll{Pb}{Pb} and \coll{p}{Pb} collisions at LHC at $\sqrt{s_{NN}} = 5.02$ TeV. One may understand this $\sqrt{s_{NN}}$ dependence from the steeper spectra at lower $\sqrt{s_{NN}}$ leading to larger values for the $c_n$ coefficients, thereby making the average important moment $\left\langle n \right\rangle$ larger. The $\sqrt{s_{NN}}$ dependence explains why in \cref{fig:raa_poisson_over_gaussian_vs_pt} the $R_{AB}$ results calculated with Poisson and Gaussian elastic energy loss distributions respectively, are more similar in \coll{p}{Pb} and \coll{Pb}{Pb} collisions than in the corresponding similarly sized systems of \coll{d}{Au} and \coll{Au}{Au} collisions respectively.

The flavor dependence of the $R_{\text{PG}}$ ratio in \cref{fig:raa_poisson_over_gaussian_vs_pt} can be understood simply as a result of the flavor dependence of the suppression. Since $\Delta E^{B} < \Delta E^{D} < \Delta E^{\pi}$ we have that $\left\langle n \right\rangle^{B} < \left\langle n \right\rangle^{D} < \left\langle n \right\rangle^{\pi}$ from which it follows that $R_{\text{PG}}^{B} < R_{\text{PG}}^{D} < R_{\text{PG}}^{\pi}$.

We conclude that the sensitivity of the $R_{AB}$ to the shape of the underlying energy loss probability distribution decreases as a function of $\sqrt{s_{NN}}$. Additionally, when our model predicts a small amount of energy loss, as in small systems, then the $R_{AB}$ is largely insensitive to the shape of the underlying elastic energy loss distribution.

\section{Power law approximation to nuclear modification factor}
\label{sec:validity_of_power_law_approximation_to_r_aa}

In our previous work \cite{Faraday:2023mmx}, we calculated the $R_{AB}$ under the assumption that the production spectra followed a power law
\begin{equation}
d N^q_{pp} / d p_T \equiv f(p_T) \approx A p_T^{-n(p_T)},
\label{eqn:eqn-power-law}
\end{equation}
where $A$ is a normalization constant. This power law approximation followed previous work \cite{Wicks:2005gt, Wicks:2008zz, Horowitz:2011gd} and is similar to other power law approximations in the literature \cite{Arleo:2017ntr, Baier:2002tc}.
This assumption allows one to simplify the expression for the $R_{AB}$ in \cref{eqn:full_raa_spectrum_ratio} as follows \cite{Faraday:2023mmx, Horowitz:2010dm}
\begin{align}
	R_{A B}^q\left(p_T\right) =& \frac{1}{f(p_T)} \int d \epsilon  \frac{1}{1 - \epsilon} f\left( \frac{p_T}{1-\epsilon}\right) P_{\text{tot.}}\left(\epsilon \left| \frac{p_T}{1-\epsilon}\right) \right.\nonumber\\
  =& \frac{\int \frac{\mathrm{d} \epsilon}{1-\epsilon} \frac{A}{\left(p_T / 1-\epsilon\right)^{n\left(p_T / [1-\epsilon]\right)}} P\left(\epsilon \mid \frac{p_T}{1-\epsilon}\right)}{A \; p_T^{- n\left(p_T\right)}}\label{eqn:eqn-raa-power-law-simplifications}\\
	=& \int d \epsilon (1-\epsilon)^{n(p_T / 1-\epsilon) - 1}  \nonumber\\
	&\times \frac{p_T^{n(p_T / 1 - \epsilon)}}{p_T^{n(p_T)}} P\left(\epsilon | \frac{p_T}{1 - \epsilon}\right).\nonumber
\end{align}
If one then assumes that $n(p_T)$ is slowly varying and that $\epsilon \ll 1$, then $n(p_T / 1 - \epsilon) \simeq n(p_T)$ and \cref{eqn:eqn-raa-power-law-simplifications} simplifies to \cite{Horowitz:2010dm}
\begin{equation}
	R_{AB} = \int d \epsilon (1 - \epsilon)^{n(p_T) - 1} P(\epsilon | p_T).
	\label{eqn:power-law-raa}
\end{equation}

The practical method in which the function $n(p_T)$ was calculated in our previous work \cite{Faraday:2023mmx} and the literature \cite{Wicks:2005gt, Horowitz:2010dm} is to differentiate \cref{eqn:eqn-power-law} resulting in
\begin{equation}
	\frac{f^{\prime}(p_T)}{f(p_T)} = -n^{\prime}(p_T) \log p_T -\frac{n(p_T)}{p_T}.
\end{equation}
If one further assumes that $n'(p_T) \log p_T$ is small according to the assumption that $n(p_T)$ varies slowly we obtain
\begin{equation}
	n(p_T) = - p_T \frac{f^{\prime}(p_T)}{f(p_T)}.
\label{eqn:eqn-npt-derivative}
\end{equation}

\begin{figure}[!htpb]
	\includegraphics[width=\linewidth]{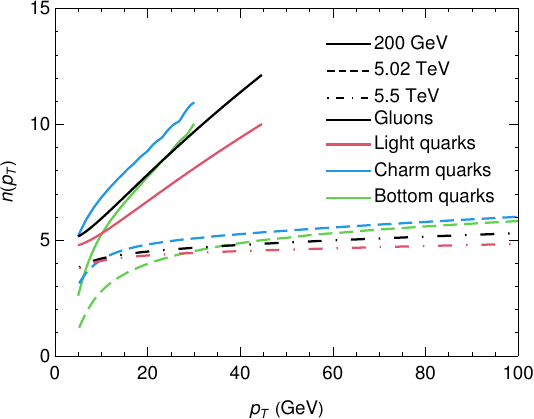}
	\caption{Production spectrum power $n(p_T)$ as a function of $p_T$ for gluons, light quarks, charm quarks, and bottom quarks produced at RHIC and LHC.}
	\label{fig:npt}
\end{figure}

\Cref{fig:npt} shows the $n(p_T)$ calculated according to \cref{eqn:eqn-npt-derivative} for various partons at RHIC and LHC energies. We see in the figure that lower $\sqrt{s_{NN}}$ has a larger $n(p_T)$, which corresponds to a steeper spectrum. 
	Also evident in the figure is that $n(p_T)$ is not always slowly varying. At RHIC energies $\sqrt{s_{NN}} = 200$ GeV, $n(p_T)$ grows linearly from $\mathcal{O}(5)\text{--}\mathcal{O}(10)$ over the $p_T$ range $\mathcal{O}(5)\text{--}\mathcal{O}(40)$ GeV, and for bottom and charm quarks at LHC energies $\sqrt{s_{NN}} = 5.02$ TeV the charm and bottom quark $n(p_T)$ grows from $\mathcal{O}(1)\text{--}\mathcal{O}(5)$ over the $p_T$ range $\mathcal{O}(5)\text{--}\mathcal{O}(40)$ GeV. Based on these apparent inconsistencies, it is reasonable to question the validity of the above set of approximations.

The utility of the slowly-varying power law $R_{AB}$ expression in \cref{eqn:power-law-raa} is largely in its explanatory power. The slowly-varying power law $R_{AB}$ cleanly separates the effect of the production spectra from the effect of energy loss, which is not as obvious in the original expression in \cref{eqn:full_raa_spectrum_ratio}. This clean separation of the $p_T$ dependencies in the power law $R_{AB}$ allows for a good qualitative description of, for instance, the $p_T$ dependence of the $R_{AB}$. To first order in $\epsilon$ we have that \cref{eqn:power-law-raa} is
\begin{equation}
	R^q_{AB} \sim 1 - (n(p_T) - 1) \frac{\Delta E}{E} (p_T)
\end{equation}
One may then understand the slow $p_T$ dependence of the $R^{\pi}_{AA}$ at RHIC in \cref{fig:raa_auau_pions,fig:raa-auau-pion-3040,fig:raa-auau-pion-6070} as a partial cancellation of the growth in $p_T$ of $n(p_T)$ (see \cref{fig:npt}) with the decrease in $p_T$ of $\Delta E / E (p_T)$ (see \cref{fig:deltaEoverE_small_large,fig:deltaEoverE_small_large_gluon}). 

\begin{figure}[!htpb]
	\includegraphics[width=\linewidth]{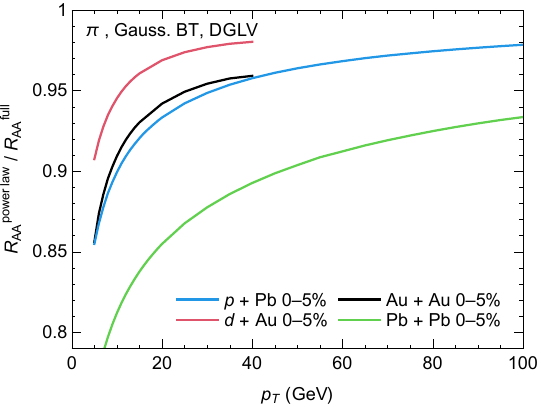}
	\caption{Plot of the ratio of the full $R_{AB}$ to the slowly-varying power law approximation to the full $R_{AB}$, $R_{AB}^{\text{power law}} / R_{AB}^{\text{full}}$ as a function of $p_T$. Curves are produced for $0\text{--}5\%$ centrality \coll{p}{Pb}, \coll{d}{Au}, \coll{Au}{Au}, and \coll{Pb}{Pb} collisions.}
	\label{fig:spectrum-ratio-vs-full-raa}
\end{figure}

To examine the validity of the slowly-varying power law approximation we plot in \cref{fig:spectrum-ratio-vs-full-raa} the ratio of the slowly-varying power law nuclear modification factor $R_{AB}^{\text{power law}}$ (\cref{eqn:power-law-raa}) to the full $R_{AB}$ (\cref{eqn:full_raa_spectrum_ratio}) as a function of $p_T$ for pions produced in central \coll{p}{Pb}, \coll{d}{Au}, \coll{Au}{Au}, and \coll{Pb}{Pb} collisions. We see that the difference is maximal in \coll{Pb}{Pb} collisions at low momenta, reaching around $20\%$, while the difference is $\lesssim 10\%$ in all other colliding systems. 
We note that \cref{fig:spectrum-ratio-vs-full-raa} is generated only for the elastic energy loss kernel set to Gaussian BT and the radiative energy loss kernel set to DGLV, as the dependence on the specific energy loss kernel used was minimal.

\begin{figure}[!htpb]
	\includegraphics[width=\linewidth]{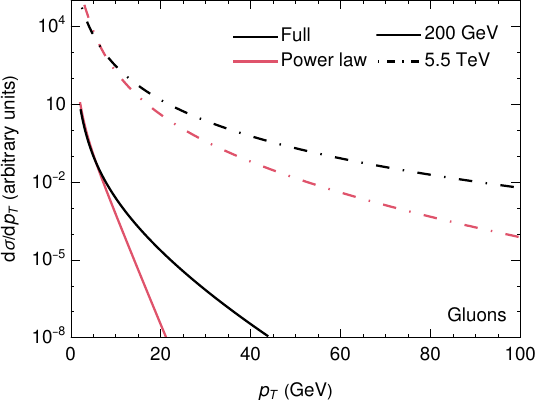}
	\caption{Comparison of the production spectra $d \sigma / d p_T$ to the slowly-varying power law approximation in \cref{eqn:eqn-power-law} as a function of $p_T$ for gluons produced at $\sqrt{s_{NN}} = 200$ GeV and $\sqrt{s_{NN}} = 5.5$ TeV.}
	\label{fig:spectra-comparison}
\end{figure}

The relatively good agreement between the full $R_{AB}$ and the slowly-varying power law approximation to the $R_{AB}$ apparently validates the set of approximations used to derive \cref{eqn:power-law-raa} from \cref{eqn:full_raa_spectrum_ratio}. \Cref{fig:spectra-comparison} plots the production spectra $d \sigma / d p_T$ and the slowly-varying power law approximation to the production spectra calculated according to \cref{eqn:eqn-power-law}, where we found the parameter $A$ by fitting the power law spectra to the original spectra. 
The figure demonstrates that the slowly varying power law spectra deviate from the original spectra by many orders of magnitude. This substantial disagreement makes the agreement at the level of the $R_{AB}$ shown in \cref{fig:spectrum-ratio-vs-full-raa} astounding and calls into question the validity of the steps taken to arrive at the result in \cref{eqn:power-law-raa}.

In order to understand why the slowly-varying power law $R_{AB}$ in \cref{eqn:power-law-raa} is a good approximation to the full $R_{AB}$ for most phenomenologically relevant systems, we consider an expansion of the integrand of the full $R_{AB}$ in \cref{eqn:full_raa_spectrum_ratio} in powers of the lost fractional energy $\epsilon$. We will perform the expansion presented in \cref{eqn:schematic_moment_expansion} explicitly for the full spectrum $R_{AB}$ in \cref{eqn:full_raa_spectrum_ratio} and the slowly-varying power law approximation to the $R_{AB}$ in \cref{eqn:power-law-raa}. For simplicity, we neglect the Jacobian $1 / 1 - \epsilon$ since it is common to both expressions and simply expand the factor $f(p_T / 1-\epsilon) / f(p_T)$ in powers of $\epsilon$. We obtain
\begin{equation}
	\begin{aligned}[t]
	&\frac{f(p_t/1-\varepsilon)}{f(p_t)} = \sum_{m \geq 0} p_t^m \frac{f^{(m)} (p_t)}{f(p_t)} \frac{1}{m!} \left( \frac{\varepsilon}{1-\varepsilon}\right)^m\\
  =& 1 +p_t \frac{f^{\prime}\left(p_t\right)}{f\left(p_t\right)} \varepsilon 
	+ \left[p_t \frac{f^{\prime}\left(p_t\right)}{f\left(p_t\right)}+\frac{1}{2} p_t^2 \frac{f^{\prime \prime}\left(p_t\right)}{f\left(p_t\right)}\right] \varepsilon^2 \\
	&+ \left[p_t \frac{f^{\prime}\left(p_t\right)}{f\left(p_t\right)}+\frac{2}{2 !} p_t^2 \frac{f^{\prime \prime}\left(p_t\right)}{f\left(p_t\right)}+\frac{1}{3 !} p_t^3 \frac{f^{\prime \prime \prime}\left(p_t\right)}{f\left(p_t\right)}\right] \varepsilon^3\\
	&+ \mathcal{O}(\epsilon^4) \\
	  =& 1 -n(p_T) \varepsilon + \left[-n(p_T)+\frac{1}{2} p_t^2 \frac{f^{\prime \prime}\left(p_t\right)}{f\left(p_t\right)}\right] \varepsilon^2 \\
	&+ \left[-n(p_T) + \frac{2}{2 !} p_t^2 \frac{f^{\prime \prime}\left(p_t\right)}{f\left(p_t\right)}+\frac{1}{3 !} p_t^3 \frac{f^{\prime \prime \prime}\left(p_t\right)}{f\left(p_t\right)}\right] \varepsilon^3\\
	&+ \mathcal{O}(\varepsilon^4),
\end{aligned}
\label{eqn:eqn-raa-taylor-expansion}
\end{equation}
where in the last line we have substituted $n(p_T) \equiv - p_T f'(p_T) / f(p_T)$ according to \cref{eqn:eqn-npt-derivative}. The equivalent procedure can be carried out for \cref{eqn:eqn-raa-power-law-simplifications} by expanding the factor $(1-\epsilon)^{n(p_T)}$,
\begin{align}
	&(1-\varepsilon)^{n(p_t)}=\nonumber\\
	&\sum_{m \geq 0} \frac{n(p_t)[n(p_t) - 1] \cdots [n(p_t)-m+1]}{m!} (-1)^m \epsilon^m\nonumber\\
 =&\; 1-n\left(p_t\right) \varepsilon+\frac{n(p t)\left[n\left(p_t\right)-1\right]}{2} \varepsilon^2\label{eqn:eqn-npt-expansion}\\
& -\frac{n\left(p_t\right)\left[n\left(p_t\right)-1\right]\left[n\left(p_t\right)-2\right]}{6} \varepsilon^3 +\mathcal{O}(\varepsilon^4).\nonumber
\end{align}

\begin{figure}[!htbp]
	\includegraphics[width=\linewidth]{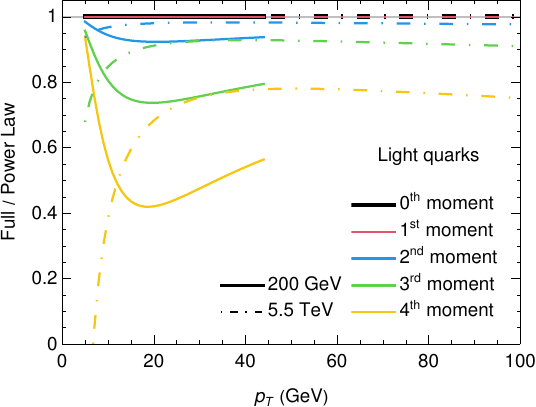}
	\caption{Order-by-order ratio of the $\mathcal{O}(\langle \epsilon^i \rangle)$ contribution to the $R_{AB}$ for the full result, to the same order contribution for the power law approximation. The order-by-order ratio is shown for the $0^{\text{th}}--4^{\text{th}}$ order contributions for light quarks produced at both $\sqrt{s_{NN}} = 200$ GeV and $\sqrt{s_{NN}} = 5.5$ TeV.
	}
	\label{fig:order_by_order_full_vs_npt_raa}
\end{figure}

It is immediately evident that \cref{eqn:eqn-raa-taylor-expansion} and \cref{eqn:eqn-npt-expansion} agree at zeroth and first order in $\epsilon$; however, the second and third order terms are not immediately comparable. \Cref{fig:order_by_order_full_vs_npt_raa} shows the contributions of various orders in $\epsilon$ to the $R_{AB}$ for the full $R_{AB}$ calculated with \cref{eqn:full_raa_spectrum_ratio} and the slowly-varying power law approximation to the $R_{AB}$ calculated with \cref{eqn:power-law-raa}. 

In the figure, we see that there are significant contributions from terms of order $\geq 2$ in $\epsilon$, for which the expansions in \cref{eqn:eqn-npt-expansion,eqn:eqn-raa-taylor-expansion} appear to be different.

We proceed by calculating the coefficients of the terms in the full $R_{AB}$ expansion in \cref{eqn:eqn-raa-taylor-expansion} in terms of the $n(p_T)$ function. Starting from $n(p_T) \equiv - p_T f'(p_T) / f(p_T)$ which implies $f'(p_T) = - f(p_T) n(p_T) / p_T$, we obtain
\begin{align}
p_t^2 \frac{f''(p_t)}{f(p_t)}  =& n\left(p_t\right)\left[n\left(p_t\right)+1\right] + p_T n'(p_T) \\
p_t^3 \frac{f'''p_t)}{f(p_t)}  =& - n(p_T) \left[n(p_T) + 1\right] \left[n(p_T) + 2\right] \nonumber\\
&+ p_T n'(p_T) [3 n(p_T) + 2] - p_T^2 n''(p_T)
\end{align}
	Now, we can read off the coefficient of the second order in $\epsilon$ as
	\begin{multline}
		p_t \frac{f'(p_t)}{f(p_t)} + \frac{1}{2} p_t^2 \frac{f''(p_t)}{f(p_t)}  = \frac{1}{2} n(p_T) [n(p_T) - 1] + \frac{1}{2} p_T n'(p_T)\\
	\end{multline}
	The coefficient of the third order in $\epsilon$ is
	\begin{multline}
		p_t \frac{f'(p_t)}{f(p_t)} + p_t^2 \frac{f''(p_t)}{f(p_t)}  + \frac{1}{3!} \frac{f'''(p_T)}{f(p_T)} =\\
		- \frac{1}{6} n(p_T) \left[n(p_T) - 1\right] \left[n(p_T) - 2\right] \\
		+ \frac{1}{2} p_T n'(p_T) \left( \frac{4}{3} + n(p_T)\right) - \frac{p_T^2 n''(p_T)}{6}.
		\label{eqn:third_order}
	\end{multline}

	The order by order difference between \cref{eqn:eqn-raa-taylor-expansion,eqn:eqn-npt-expansion} can then be written as
	\begin{multline}
		f(p_t/1-\varepsilon)/f(p_t) - (1-\varepsilon)^{n(p_T)} = \frac{1}{2} p_T n'(p_T) \epsilon^2 \\ + \left[\frac{1}{2} p_T n'(p_T) \left( \frac{4}{3} + n(p_T)\right) - \frac{p_T^2n''(p_T)}{6}\right] \epsilon^3 + \mathcal{O}(\epsilon^4).
		\label{eqn:difference_taylor_npt_spectrum}
	\end{multline}
We note the surprising cancellation that occurs at each order in $\epsilon$, leaving only terms proportional to $n'(p_T)$ or higher order derivatives of $n$.

The expansion of the $R_{AB}$ in terms of the moments of the energy loss probability distribution for the slowly-varying power law $R_{AB}$ and the full spectrum $R_{AB}$ is identical up to $\mathcal{O}(\epsilon^2)$. At higher orders in $\epsilon$ the slowly-varying power law $R_{AB}$ from \cref{eqn:power-law-raa} differs from the full $R_{AB}$ in \cref{eqn:full_raa_spectrum_ratio}.

From this analysis, it is unsurprising that we see a large relative difference in the slowly-varying power law $R_{AB}$ and the full $R_{AB}$ at low-$p_T$ in \coll{Pb}{Pb} and \coll{Au}{Au}, where there is a large amount of energy loss.

We conclude that the slowly-varying power law $R_{AB}$ expression in \cref{eqn:eqn-power-law} still useful for its explanatory power, and provides a good approximation for most of the phenomenologically relevant phase space. 
One should think of the slowly-varying power law $R_{AB}$ as an approximation in the limit of small fractional energy loss ($\epsilon \ll 1$) and slowly varying spectra ($p_T^m n^{(m)}(p_T) \ll 1 \forall m > 0$). Therefore, while the original calculation performed to derive \cref{eqn:power-law-raa} is not valid, the expression is still a well-defined approximation to the full $R_{AB}$, as motivated by the expansion of the $R_{AB}$ in moments of the energy loss probability distribution performed in this work. While initially, the agreement at the level of the $R_{AB}$ of the slowly-varying power law approximation to the full $R_{AB}$ is astounding, this analysis in terms of the moments makes it clear that the mistake is thinking of the function $n(p_T)$ as the power law of the production spectrum. Instead it should simply be defined as $n(p_T) \equiv p_T f'(p_T) / f(p_T)$, where $f(p_T)$ is the production spectrum.

Since the full expression for the $R_{AB}$ in \cref{eqn:full_raa_spectrum_ratio} is not significantly more numerically intensive, one should use this expression for all numerical results. 

\section{Conclusions}

In this article we presented first results for the nuclear modification factor of leading high-$p_T$ hadrons from a pQCD-based convolved radiative and elastic energy loss model which receives small system size corrections to \emph{both} the radiative \cite{Faraday:2023mmx, Kolbe:2015rvk, Kolbe:2015suq} \emph{and} elastic \cite{Wicks:2008zz} energy loss. 
Our model utilizes realistic production spectra, fragmentation functions, and a QGP medium geometry with fluctuating initial conditions generated by IP-Glasma \cite{Schenke:2020mbo,shen_private_communication,Schenke:2012hg, Schenke:2012wb}. In this work, we expanded our model \cite{Faraday:2023mmx} to include a more realistic elastic energy loss kernel \cite{Wicks:2008zz}, calculated within the HTL formalism \cite{Braaten:1989mz, Klimov:1982bv, Pisarski:1988vd, Weldon:1982aq, Weldon:1982bn}, and with reduced approximations in comparison to the elastic energy loss kernel \cite{Braaten:1991jj, Braaten:1991we} used previously \cite{Faraday:2023mmx}. This new elastic energy loss kernel allowed us to retain the distributional information of the elastic energy loss (i.e.\ $dN / dx$ and not just $\Delta E / E$), thereby removing the need to apply the central limit theorem \cite{Moore:2004tg} in the calculation, a common assumption in the literature \cite{Wicks:2005gt, Horowitz:2011gd, Zigic:2021rku}.

We varied the radiative energy loss kernel between the DGLV radiative energy loss kernel \cite{Djordjevic:2003zk} and the DGLV radiative energy loss kernel which receives a short pathlength correction \cite{Kolbe:2015rvk, Kolbe:2015suq}. We varied the elastic energy loss kernel between the Braaten and Thoma elastic energy loss kernel \cite{Braaten:1991jj, Braaten:1991we} (used in our previous work \cite{Faraday:2023mmx}), and an HTL-based elastic energy loss kernel \cite{Wicks:2008zz}. Additionally, we considered the full Poissonian distribution \cite{Gyulassy:2001nm}, which arises naturally from the HTL-based approach, as well as the Gaussian approximation \cite{Moore:2004tg} to this elastic energy loss distribution. 
This set of elastic energy loss kernels allowed us to interrogate the theoretical uncertainty in the elastic energy loss due to the crossover from HTL propagators to vacuum propagators \cite{Wicks:2008zz, Romatschke:2004au}, %
as well as the effect of the Gaussian approximation of the elastic energy loss distribution \cite{Moore:2004tg} according to the central limit theorem.

Model results for the nuclear modification factor were presented for light and heavy-flavor hadrons produced in central \coll{p}{Pb} and central and peripheral \coll{Pb}{Pb} collisions at LHC, and in central \collFour{p}{d}{He3}{Au} and central and peripheral \coll{Au}{Au} collisions at RHIC.

We found that for both heavy and light flavor mesons produced in central \coll{Pb}{Pb} and \coll{Au}{Au} collisions, results calculated with the Poisson HTL elastic energy loss differed from results calculated with the Gaussian HTL elastic energy loss by only $\mathcal{O}(5\text{--}15\%)$ for mid--high $p_T$. For light and heavy-flavor mesons produced in central \coll{p}{Pb} and \collFour{p}{d}{He3}{Au} collisions, we found that the difference between $R_{AB}$ results calculated with the Poisson HTL and Gaussian HTL elastic energy loss distributions was extremely small, $\mathcal{O}(2\text{--}5\%)$, for all $p_T \gtrsim 5$ GeV. This negligible difference between the Gaussian and Poisson results is especially surprising given the $\mathcal{O}(0\text{--}1)$ scatters that occur in small systems, making the central limit theorem inapplicable. 
This trend as a function of system size contradicts the expectation that the Gaussian approximation should \emph{improve} with increasing system size, according to the central limit theorem.
Paradoxically, the approximation is most accurate for small systems, where the central limit theorem is least applicable, and least accurate for large systems, where the central limit theorem is most applicable.

We demonstrated that the system size, $\sqrt{s_{NN}}$, and flavor dependence of the relative difference between $R_{AA}$ results calculated with Gaussian and Poisson elastic distributions can be explained by expanding the integrand of the $R_{AB}$ in the lost fractional energy $\epsilon$. Such an expansion leads to an expression for the $R_{AB}$ in terms of the moments of the underlying energy loss distributions.
We showed that when there is only a small degree of suppression, as there is in small systems, the $R_{AB}$ depends mostly on the first few moments, the zeroth and first of which are constrained to be identical for the Gaussian and Poisson elastic energy loss. Conversely, in large systems, the higher order moments are more important, exacerbating the difference between the $R_{AB}$ calculated with the Gaussian and the Poisson elastic energy loss distributions, respectively. The slope of the production spectra also impacts which moments are important in this expansion of the $R_{AA}$. At RHIC, the production spectra are steeper than for the equivalent partons at LHC, which leads to the higher-order moments being more important at RHIC vs at LHC. This can be understood simply by approximating $R_{AA} \sim \int d \epsilon \; (1-\epsilon)^{n-1} P_{\text{tot.}}(\epsilon)$ where $d \sigma / d p_T \propto p_T^{-n}$ is the production spectrum and $n$ is a constant \cite{Horowitz:2010dm}. At RHIC $n \sim 6\text{--}8$ while at LHC $n \sim 4\text{--}5$, from which it is evident that at RHIC, the higher order moments of a distribution are more important than at LHC. The $p_T$ dependence of the relative difference between results calculated with the Poisson HTL and Gaussian HTL elastic energy loss distributions is a result of the decreasing importance of the elastic energy loss compared to the radiative energy loss as a function of $p_T$.

We suggested previously \cite{Faraday:2023mmx} that if one assumes that medium-induced gluon emission and vacuum gluon emission are both independent, then a more realistic distribution for the \emph{additional} gluon emission in medium may be a Skellam distribution \cite{skellam:1946}. A further study of the dependence of the $R_{AB}$ on the underlying energy loss distribution, using the techniques introduced in this work, is necessary to understand how sensitive the $R_{AB}$ is to the modeling of the radiative energy loss distribution as Poissonian compared to, for instance, a Skellam distribution. 

In our previous work \cite{Faraday:2023mmx}, we found that elastic energy loss accounted for $\sim 90\%$ of the suppression in \coll{p}{Pb} collisions, which we suggested may be due to a breakdown of the application of the central limit theorem in these systems with $\mathcal{O}(0\text{--}1)$ elastic scatterings. Our results in this work imply that the large contribution of the elastic energy loss, in comparison to the radiative energy loss, in small systems is \textbf{not} due to the inapplicability of the Gaussian approximation, but rather a physical effect. The different length dependencies of the elastic and radiative energy loss---due to LPM interference the radiative energy loss scales as $L^2$, while the elastic energy loss scales as $L$---explains the large contribution of the elastic energy loss compared to the radiative energy loss in small systems.  
We showed that---over the range of phenomenologically interesting collision systems, final states, and momenta---the fractional contribution of the radiative vs elastic energy loss varies from $\sim \! 0.25 \text{--} 4$, demonstrating the importance of including both elastic and radiative energy loss in theoretical energy loss models.

We also made comparisons for all aforementioned systems between $R_{AB}$ results calculated with the Gaussian HTL \cite{Wicks:2008zz} and the Gaussian Braaten and Thoma (BT) \cite{Braaten:1991jj, Braaten:1991we} elastic energy loss distributions. These calculations both utilize HTL propagators \cite{Braaten:1989mz, Klimov:1982bv, Pisarski:1988vd, Weldon:1982aq, Weldon:1982bn} to calculate the elastic energy loss suffered by a high-$p_T$ parton moving through the QGP. They differ in that the BT result uses vacuum propagators at high momentum transfer, and HTL at low momentum transfer, while the HTL result \cite{Wicks:2008zz} uses HTL propagators for all momentum transfers. One may, therefore, consider the two results as an approximate measure of the theoretical uncertainty in the crossover region from HTL to vacuum propagators \cite{Wicks:2008zz, Romatschke:2004au}.  %
We found that the difference between $R_{AA}$ results calculated with the Poisson HTL and BT elastic energy loss kernels was $\mathcal{O}(30\text{--}50\%)$ for $p_T$ of $\mathcal{O}(10\text{--}50)$ GeV for both light and heavy flavor hadrons produced in central \coll{Pb}{Pb} and \coll{Au}{Au} collisions. For light and heavy flavor mesons produced in central \coll{p}{Pb} collisions, the difference between $R_{pA}$ results calculated with Poisson HTL and BT elastic energy loss kernels was $\mathcal{O}(5\text{--}10\%)$ for $p_T$ in the range $\mathcal{O}(5\text{--}50)$ GeV. For pions produced in central \collFour{p}{d}{He3}{Au} collisions we found a $\mathcal{O}(10\text{--}25\%)$ relative difference between the two calculations. In all cases, the relative difference between the $R_{AB}$ results calculated with the two different elastic energy loss kernels decreases as a function of $p_T$, due to the decreasing contribution of the elastic energy loss in comparison to the radiative energy loss. While the sensitivity seems decreased in small systems, this is an artifact of the small amount of energy loss in small systems. If one instead considers the difference at the level of the $1 - R_{AA}$, which more closely corresponds to energy loss, the difference in results calculated with HTL and BT propagators becomes $\mathcal{O}(50\text{--}100)\%$. 

We conclude that the uncertainty in the choice of elastic energy loss kernel is essential for making rigorous and quantitative phenomenological predictions for suppression in \emph{both} small and large systems. 
For models applied to systems of similar sizes, one may absorb this fundamental uncertainty into one's choice of the effective coupling constant, for instance. However, this may not be possible for models applied to a wide range of system sizes because the elastic energy loss scales like $\alpha_s^2$, while the radiative energy loss scales like $\alpha_s^3$, and various system sizes receive a vastly different proportion of their total energy loss from the elastic and radiative energy loss kernels; meaning that a global change of $\alpha_s$ will not globally affect the $R_{AA}$ in the same way as a change of the elastic energy loss kernel will. We propose that a system size scan will help to elucidate which prescription for changing between HTL and vacuum propagators is favored by suppression data, in lieu of an analytical approach to this problem.

In this work, we also investigated the impact of the short pathlength correction \cite{Kolbe:2015rvk, Kolbe:2015suq} to the DGLV radiative energy loss \cite{Djordjevic:2003zk}. As in our previous work \cite{Faraday:2023mmx}, we saw that the short pathlength correction is anomalously large for high-$p_T$ pions, which is likely due to the breakdown of the large formation time assumption, which we discussed in detail in \cite{Faraday:2023mmx}. We investigated the phenomenology of the short pathlength correction for the first time at RHIC energies,  where we found that the smaller momentum range, as well as the smaller fraction of gluons compared to light quarks, lead to the short pathlength correction being negligible in \emph{both} \coll{Au}{Au} and \collFour{p}{d}{He3}{Au} collisions. We also presented results for semi-central and peripheral \coll{Pb}{Pb} collisions for the first time, where we observed that the short pathlengths lead to anomalously large $R^{\pi}_{AA} \sim \mathcal{O}(1.1 \text{--}1.3)$ in the $p_T$ range $\mathcal{O}(50\text{--}400)$ GeV. We reiterate our conclusions from our previous work \cite{Faraday:2023mmx}: control over the large formation time approximation \cite{Faraday:2023mmx}---which breaks down dramatically for the short pathlength correction at high-$p_T$ and in small systems---will be crucial in making quantitative theoretical predictions. Future work \cite{Faraday:2024} will consider the effects of including a cut-off on the radiated transverse momentum, which ensures that no contributions from regions where the large formation time approximation is invalid are included in the calculation.

Finally, we investigated the validity of the slowly-varying power law approximation to the $R_{AA}$, which has previously been used \cite{Horowitz:2010dm, Wicks:2005gt, Horowitz:2011gd, Faraday:2023mmx}. The power law approximation to the $R_{AA}$ is given by $R_{AA} \simeq \int d \epsilon \; (1-\epsilon)^{n(p_T)-1}$, where $f(p_T) \equiv d \sigma / d p_T \propto p_T^{-n(p_T)}$ is the production spectrum and $n(p_T)$ is a slowly-varying function. We found that if one compares the spectra $f(p_T)$ and $ A p_T^{-n(p_T)}$, where $A$ is a fitted constant, the spectra differ by many orders of magnitude. However if one compares the power law $R_{AA}$ to the full $R_{AA}$, these quantities differ by $\mathcal{O}(5\text{--}15\%)$ only. We presented a new derivation of the power law $R_{AA} \simeq \int d \epsilon \; (1-\epsilon)^{n(p_T) - 1}$, based on an expansion of the $R_{AA}$ in terms of the moments of the energy loss distribution $\langle \epsilon^i \rangle$. This expansion showed that $n(p_T)$ is \emph{not} the power of the spectrum but rather $n(p_T) \equiv p_T f'(p_T) / f(p_T)$, and that the power law $R_{AA}$ is valid in the limit of small fractional energy loss ($\langle \epsilon \rangle \ll 1$) and slowly varying $n(p_T)$ ($p_T^m n^{(m)}(p_T) \ll 1$ for all $m \geq 1$).

Since this work aimed to understand various approximations and uncertainties in our energy loss model, we made no comparisons to experimental data. Future work in preparation \cite{Faraday:2024} will perform a global fit of the strong coupling $\alpha_s$ to available large system \coll{A}{A} data at LHC and RHIC. This large-system-constrained model may then be used to make predictions for small \collFour{p}{d}{He3}{A} systems, which will allow us to quantitatively infer the extent to which a single pQCD energy loss model can quantitatively describe both small and large system suppression data simultaneously.

\section*{Acknowledgments}

Computations were performed using facilities provided by the University of Cape Town’s ICTS High Performance Computing team: \href{http://hpc.uct.ac.za}{hpc.uct.ac.za}.
CF and WAH thank the National Research Foundation and the SA-CERN collaboration for their generous financial support during the course of this work. CF additionally thanks CERN for their hospitality during the course of part of this work. WAH thanks the EIC Theory Institute for their hospitality during the course of part of this work.

\newpage

\bibliographystyle{apsrev4-2} %
\bibliography{manual,small_system_elastic}

\begin{thebibliography}{115}%
\makeatletter
\providecommand \@ifxundefined [1]{%
 \@ifx{#1\undefined}
}%
\providecommand \@ifnum [1]{%
 \ifnum #1\expandafter \@firstoftwo
 \else \expandafter \@secondoftwo
 \fi
}%
\providecommand \@ifx [1]{%
 \ifx #1\expandafter \@firstoftwo
 \else \expandafter \@secondoftwo
 \fi
}%
\providecommand \natexlab [1]{#1}%
\providecommand \enquote  [1]{``#1''}%
\providecommand \bibnamefont  [1]{#1}%
\providecommand \bibfnamefont [1]{#1}%
\providecommand \citenamefont [1]{#1}%
\providecommand \href@noop [0]{\@secondoftwo}%
\providecommand \href [0]{\begingroup \@sanitize@url \@href}%
\providecommand \@href[1]{\@@startlink{#1}\@@href}%
\providecommand \@@href[1]{\endgroup#1\@@endlink}%
\providecommand \@sanitize@url [0]{\catcode `\\12\catcode `\$12\catcode `\&12\catcode `\#12\catcode `\^12\catcode `\_12\catcode `\%12\relax}%
\providecommand \@@startlink[1]{}%
\providecommand \@@endlink[0]{}%
\providecommand \url  [0]{\begingroup\@sanitize@url \@url }%
\providecommand \@url [1]{\endgroup\@href {#1}{\urlprefix }}%
\providecommand \urlprefix  [0]{URL }%
\providecommand \Eprint [0]{\href }%
\providecommand \doibase [0]{http://dx.doi.org/}%
\providecommand \selectlanguage [0]{\@gobble}%
\providecommand \bibinfo  [0]{\@secondoftwo}%
\providecommand \bibfield  [0]{\@secondoftwo}%
\providecommand \translation [1]{[#1]}%
\providecommand \BibitemOpen [0]{}%
\providecommand \bibitemStop [0]{}%
\providecommand \bibitemNoStop [0]{.\EOS\space}%
\providecommand \EOS [0]{\spacefactor3000\relax}%
\providecommand \BibitemShut  [1]{\csname bibitem#1\endcsname}%
\let\auto@bib@innerbib\@empty
\bibitem [{\citenamefont {Adcox}\ \emph {et~al.}(2002)\citenamefont {Adcox} \emph {et~al.}}]{PHENIX:2001hpc}%
  \BibitemOpen
  \bibfield  {author} {\bibinfo {author} {\bibfnamefont {K.}~\bibnamefont {Adcox}} \emph {et~al.} (\bibinfo {collaboration} {PHENIX}),\ }\href {\doibase 10.1103/PhysRevLett.88.022301} {\bibfield  {journal} {\bibinfo  {journal} {Phys. Rev. Lett.}\ }\textbf {\bibinfo {volume} {88}},\ \bibinfo {pages} {022301} (\bibinfo {year} {2002})},\ \Eprint {http://arxiv.org/abs/nucl-ex/0109003}{arXiv:nucl-ex/0109003}\BibitemShut {NoStop}%
\bibitem [{\citenamefont {Adler}\ \emph {et~al.}(2002)\citenamefont {Adler} \emph {et~al.}}]{STAR:2002ggv}%
  \BibitemOpen
  \bibfield  {author} {\bibinfo {author} {\bibfnamefont {C.}~\bibnamefont {Adler}} \emph {et~al.} (\bibinfo {collaboration} {STAR}),\ }\href {\doibase 10.1103/PhysRevLett.89.202301} {\bibfield  {journal} {\bibinfo  {journal} {Phys. Rev. Lett.}\ }\textbf {\bibinfo {volume} {89}},\ \bibinfo {pages} {202301} (\bibinfo {year} {2002})},\ \Eprint {http://arxiv.org/abs/nucl-ex/0206011}{arXiv:nucl-ex/0206011}\BibitemShut {NoStop}%
\bibitem [{\citenamefont {Adler}\ \emph {et~al.}(2005)\citenamefont {Adler} \emph {et~al.}}]{PHENIX:2005yls}%
  \BibitemOpen
  \bibfield  {author} {\bibinfo {author} {\bibfnamefont {S.~S.}\ \bibnamefont {Adler}} \emph {et~al.} (\bibinfo {collaboration} {PHENIX}),\ }\href {\doibase 10.1103/PhysRevLett.94.232301} {\bibfield  {journal} {\bibinfo  {journal} {Phys. Rev. Lett.}\ }\textbf {\bibinfo {volume} {94}},\ \bibinfo {pages} {232301} (\bibinfo {year} {2005})},\ \Eprint {http://arxiv.org/abs/nucl-ex/0503003}{arXiv:nucl-ex/0503003}\BibitemShut {NoStop}%
\bibitem [{\citenamefont {Adler}\ \emph {et~al.}(2003{\natexlab{a}})\citenamefont {Adler} \emph {et~al.}}]{PHENIX:2003qdw}%
  \BibitemOpen
  \bibfield  {author} {\bibinfo {author} {\bibfnamefont {S.~S.}\ \bibnamefont {Adler}} \emph {et~al.} (\bibinfo {collaboration} {PHENIX}),\ }\href {\doibase 10.1103/PhysRevLett.91.072303} {\bibfield  {journal} {\bibinfo  {journal} {Phys. Rev. Lett.}\ }\textbf {\bibinfo {volume} {91}},\ \bibinfo {pages} {072303} (\bibinfo {year} {2003}{\natexlab{a}})},\ \Eprint {http://arxiv.org/abs/nucl-ex/0306021}{arXiv:nucl-ex/0306021}\BibitemShut {NoStop}%
\bibitem [{\citenamefont {Adams}\ \emph {et~al.}(2003)\citenamefont {Adams} \emph {et~al.}}]{STAR:2003pjh}%
  \BibitemOpen
  \bibfield  {author} {\bibinfo {author} {\bibfnamefont {J.}~\bibnamefont {Adams}} \emph {et~al.} (\bibinfo {collaboration} {STAR}),\ }\href {\doibase 10.1103/PhysRevLett.91.072304} {\bibfield  {journal} {\bibinfo  {journal} {Phys. Rev. Lett.}\ }\textbf {\bibinfo {volume} {91}},\ \bibinfo {pages} {072304} (\bibinfo {year} {2003})},\ \Eprint {http://arxiv.org/abs/nucl-ex/0306024}{arXiv:nucl-ex/0306024}\BibitemShut {NoStop}%
\bibitem [{\citenamefont {Adler}\ \emph {et~al.}(2007)\citenamefont {Adler} \emph {et~al.}}]{PHENIX:2006mhb}%
  \BibitemOpen
  \bibfield  {author} {\bibinfo {author} {\bibfnamefont {S.~S.}\ \bibnamefont {Adler}} \emph {et~al.} (\bibinfo {collaboration} {PHENIX}),\ }\href {\doibase 10.1103/PhysRevLett.98.172302} {\bibfield  {journal} {\bibinfo  {journal} {Phys. Rev. Lett.}\ }\textbf {\bibinfo {volume} {98}},\ \bibinfo {pages} {172302} (\bibinfo {year} {2007})},\ \Eprint {http://arxiv.org/abs/nucl-ex/0610036}{arXiv:nucl-ex/0610036}\BibitemShut {NoStop}%
\bibitem [{\citenamefont {Dainese}\ \emph {et~al.}(2005)\citenamefont {Dainese}, \citenamefont {Loizides},\ and\ \citenamefont {Paic}}]{Dainese:2004te}%
  \BibitemOpen
  \bibfield  {author} {\bibinfo {author} {\bibfnamefont {A.}~\bibnamefont {Dainese}}, \bibinfo {author} {\bibfnamefont {C.}~\bibnamefont {Loizides}}, \ and\ \bibinfo {author} {\bibfnamefont {G.}~\bibnamefont {Paic}},\ }\href {\doibase 10.1140/epjc/s2004-02077-x} {\bibfield  {journal} {\bibinfo  {journal} {Eur. Phys. J. C}\ }\textbf {\bibinfo {volume} {38}},\ \bibinfo {pages} {461} (\bibinfo {year} {2005})},\ \Eprint {http://arxiv.org/abs/hep-ph/0406201}{arXiv:hep-ph/0406201}\BibitemShut {NoStop}%
\bibitem [{\citenamefont {Schenke}\ \emph {et~al.}(2009)\citenamefont {Schenke}, \citenamefont {Gale},\ and\ \citenamefont {Jeon}}]{Schenke:2009gb}%
  \BibitemOpen
  \bibfield  {author} {\bibinfo {author} {\bibfnamefont {B.}~\bibnamefont {Schenke}}, \bibinfo {author} {\bibfnamefont {C.}~\bibnamefont {Gale}}, \ and\ \bibinfo {author} {\bibfnamefont {S.}~\bibnamefont {Jeon}},\ }\href {\doibase 10.1103/PhysRevC.80.054913} {\bibfield  {journal} {\bibinfo  {journal} {Phys. Rev. C}\ }\textbf {\bibinfo {volume} {80}},\ \bibinfo {pages} {054913} (\bibinfo {year} {2009})},\ \Eprint {http://arxiv.org/abs/0909.2037}{arXiv:0909.2037 [hep-ph]}\BibitemShut {NoStop}%
\bibitem [{\citenamefont {Wicks}\ \emph {et~al.}(2007)\citenamefont {Wicks}, \citenamefont {Horowitz}, \citenamefont {Djordjevic},\ and\ \citenamefont {Gyulassy}}]{Wicks:2005gt}%
  \BibitemOpen
  \bibfield  {author} {\bibinfo {author} {\bibfnamefont {S.}~\bibnamefont {Wicks}}, \bibinfo {author} {\bibfnamefont {W.}~\bibnamefont {Horowitz}}, \bibinfo {author} {\bibfnamefont {M.}~\bibnamefont {Djordjevic}}, \ and\ \bibinfo {author} {\bibfnamefont {M.}~\bibnamefont {Gyulassy}},\ }\href {\doibase 10.1016/j.nuclphysa.2006.12.048} {\bibfield  {journal} {\bibinfo  {journal} {Nucl. Phys. A}\ }\textbf {\bibinfo {volume} {784}},\ \bibinfo {pages} {426} (\bibinfo {year} {2007})},\ \Eprint {http://arxiv.org/abs/nucl-th/0512076}{arXiv:nucl-th/0512076}\BibitemShut {NoStop}%
\bibitem [{\citenamefont {Horowitz}(2013)}]{Horowitz:2012cf}%
  \BibitemOpen
  \bibfield  {author} {\bibinfo {author} {\bibfnamefont {W.~A.}\ \bibnamefont {Horowitz}},\ }\href {\doibase 10.1016/j.nuclphysa.2013.01.061} {\bibfield  {journal} {\bibinfo  {journal} {Nucl. Phys. A}\ }\textbf {\bibinfo {volume} {904-905}},\ \bibinfo {pages} {186c} (\bibinfo {year} {2013})},\ \Eprint {http://arxiv.org/abs/1210.8330}{arXiv:1210.8330 [nucl-th]}\BibitemShut {NoStop}%
\bibitem [{\citenamefont {Aamodt}\ \emph {et~al.}(2011)\citenamefont {Aamodt} \emph {et~al.}}]{ALICE:2010mlf}%
  \BibitemOpen
  \bibfield  {author} {\bibinfo {author} {\bibfnamefont {K.}~\bibnamefont {Aamodt}} \emph {et~al.} (\bibinfo {collaboration} {ALICE}),\ }\href {\doibase 10.1103/PhysRevLett.106.032301} {\bibfield  {journal} {\bibinfo  {journal} {Phys. Rev. Lett.}\ }\textbf {\bibinfo {volume} {106}},\ \bibinfo {pages} {032301} (\bibinfo {year} {2011})},\ \Eprint {http://arxiv.org/abs/1012.1657}{arXiv:1012.1657 [nucl-ex]}\BibitemShut {NoStop}%
\bibitem [{\citenamefont {Chatrchyan}\ \emph {et~al.}(2012)\citenamefont {Chatrchyan} \emph {et~al.}}]{CMS:2012oiv}%
  \BibitemOpen
  \bibfield  {author} {\bibinfo {author} {\bibfnamefont {S.}~\bibnamefont {Chatrchyan}} \emph {et~al.} (\bibinfo {collaboration} {CMS}),\ }\href {\doibase 10.1016/j.physletb.2012.02.077} {\bibfield  {journal} {\bibinfo  {journal} {Phys. Lett. B}\ }\textbf {\bibinfo {volume} {710}},\ \bibinfo {pages} {256} (\bibinfo {year} {2012})},\ \Eprint {http://arxiv.org/abs/1201.3093}{arXiv:1201.3093 [nucl-ex]}\BibitemShut {NoStop}%
\bibitem [{\citenamefont {Chatrchyan}\ \emph {et~al.}(2011)\citenamefont {Chatrchyan} \emph {et~al.}}]{CMS:2011zfr}%
  \BibitemOpen
  \bibfield  {author} {\bibinfo {author} {\bibfnamefont {S.}~\bibnamefont {Chatrchyan}} \emph {et~al.} (\bibinfo {collaboration} {CMS}),\ }\href {\doibase 10.1103/PhysRevLett.106.212301} {\bibfield  {journal} {\bibinfo  {journal} {Phys. Rev. Lett.}\ }\textbf {\bibinfo {volume} {106}},\ \bibinfo {pages} {212301} (\bibinfo {year} {2011})},\ \Eprint {http://arxiv.org/abs/1102.5435}{arXiv:1102.5435 [nucl-ex]}\BibitemShut {NoStop}%
\bibitem [{\citenamefont {Abelev}\ \emph {et~al.}(2013)\citenamefont {Abelev} \emph {et~al.}}]{ALICE:2012mj}%
  \BibitemOpen
  \bibfield  {author} {\bibinfo {author} {\bibfnamefont {B.}~\bibnamefont {Abelev}} \emph {et~al.} (\bibinfo {collaboration} {ALICE}),\ }\href {\doibase 10.1103/PhysRevLett.110.082302} {\bibfield  {journal} {\bibinfo  {journal} {Phys. Rev. Lett.}\ }\textbf {\bibinfo {volume} {110}},\ \bibinfo {pages} {082302} (\bibinfo {year} {2013})},\ \Eprint {http://arxiv.org/abs/1210.4520}{arXiv:1210.4520 [nucl-ex]}\BibitemShut {NoStop}%
\bibitem [{\citenamefont {Abelev}\ \emph {et~al.}(2012)\citenamefont {Abelev} \emph {et~al.}}]{ALICE:2012ab}%
  \BibitemOpen
  \bibfield  {author} {\bibinfo {author} {\bibfnamefont {B.}~\bibnamefont {Abelev}} \emph {et~al.} (\bibinfo {collaboration} {ALICE}),\ }\href {\doibase 10.1007/JHEP09(2012)112} {\bibfield  {journal} {\bibinfo  {journal} {JHEP}\ }\textbf {\bibinfo {volume} {09}},\ \bibinfo {pages} {112} (\bibinfo {year} {2012})},\ \Eprint {http://arxiv.org/abs/1203.2160}{arXiv:1203.2160 [nucl-ex]}\BibitemShut {NoStop}%
\bibitem [{\citenamefont {Adam}\ \emph {et~al.}(2016{\natexlab{a}})\citenamefont {Adam} \emph {et~al.}}]{ALICE:2015vxz}%
  \BibitemOpen
  \bibfield  {author} {\bibinfo {author} {\bibfnamefont {J.}~\bibnamefont {Adam}} \emph {et~al.} (\bibinfo {collaboration} {ALICE}),\ }\href {\doibase 10.1007/JHEP03(2016)081} {\bibfield  {journal} {\bibinfo  {journal} {JHEP}\ }\textbf {\bibinfo {volume} {03}},\ \bibinfo {pages} {081} (\bibinfo {year} {2016}{\natexlab{a}})},\ \Eprint {http://arxiv.org/abs/1509.06888}{arXiv:1509.06888 [nucl-ex]}\BibitemShut {NoStop}%
\bibitem [{\citenamefont {Andronic}\ \emph {et~al.}(2016)\citenamefont {Andronic} \emph {et~al.}}]{Andronic:2015wma}%
  \BibitemOpen
  \bibfield  {author} {\bibinfo {author} {\bibfnamefont {A.}~\bibnamefont {Andronic}} \emph {et~al.},\ }\href {\doibase 10.1140/epjc/s10052-015-3819-5} {\bibfield  {journal} {\bibinfo  {journal} {Eur. Phys. J. C}\ }\textbf {\bibinfo {volume} {76}},\ \bibinfo {pages} {107} (\bibinfo {year} {2016})},\ \Eprint {http://arxiv.org/abs/1506.03981}{arXiv:1506.03981 [nucl-ex]}\BibitemShut {NoStop}%
\bibitem [{\citenamefont {Ackermann}\ \emph {et~al.}(2001)\citenamefont {Ackermann} \emph {et~al.}}]{STAR:2000ekf}%
  \BibitemOpen
  \bibfield  {author} {\bibinfo {author} {\bibfnamefont {K.~H.}\ \bibnamefont {Ackermann}} \emph {et~al.} (\bibinfo {collaboration} {STAR}),\ }\href {\doibase 10.1103/PhysRevLett.86.402} {\bibfield  {journal} {\bibinfo  {journal} {Phys. Rev. Lett.}\ }\textbf {\bibinfo {volume} {86}},\ \bibinfo {pages} {402} (\bibinfo {year} {2001})},\ \Eprint {http://arxiv.org/abs/nucl-ex/0009011}{arXiv:nucl-ex/0009011}\BibitemShut {NoStop}%
\bibitem [{\citenamefont {Adler}\ \emph {et~al.}(2003{\natexlab{b}})\citenamefont {Adler} \emph {et~al.}}]{PHENIX:2003qra}%
  \BibitemOpen
  \bibfield  {author} {\bibinfo {author} {\bibfnamefont {S.~S.}\ \bibnamefont {Adler}} \emph {et~al.} (\bibinfo {collaboration} {PHENIX}),\ }\href {\doibase 10.1103/PhysRevLett.91.182301} {\bibfield  {journal} {\bibinfo  {journal} {Phys. Rev. Lett.}\ }\textbf {\bibinfo {volume} {91}},\ \bibinfo {pages} {182301} (\bibinfo {year} {2003}{\natexlab{b}})},\ \Eprint {http://arxiv.org/abs/nucl-ex/0305013}{arXiv:nucl-ex/0305013}\BibitemShut {NoStop}%
\bibitem [{\citenamefont {Aamodt}\ \emph {et~al.}(2010)\citenamefont {Aamodt} \emph {et~al.}}]{ALICE:2010suc}%
  \BibitemOpen
  \bibfield  {author} {\bibinfo {author} {\bibfnamefont {K.}~\bibnamefont {Aamodt}} \emph {et~al.} (\bibinfo {collaboration} {ALICE}),\ }\href {\doibase 10.1103/PhysRevLett.105.252302} {\bibfield  {journal} {\bibinfo  {journal} {Phys. Rev. Lett.}\ }\textbf {\bibinfo {volume} {105}},\ \bibinfo {pages} {252302} (\bibinfo {year} {2010})},\ \Eprint {http://arxiv.org/abs/1011.3914}{arXiv:1011.3914 [nucl-ex]}\BibitemShut {NoStop}%
\bibitem [{\citenamefont {Romatschke}\ and\ \citenamefont {Romatschke}(2007)}]{Romatschke:2007mq}%
  \BibitemOpen
  \bibfield  {author} {\bibinfo {author} {\bibfnamefont {P.}~\bibnamefont {Romatschke}}\ and\ \bibinfo {author} {\bibfnamefont {U.}~\bibnamefont {Romatschke}},\ }\href {\doibase 10.1103/PhysRevLett.99.172301} {\bibfield  {journal} {\bibinfo  {journal} {Phys. Rev. Lett.}\ }\textbf {\bibinfo {volume} {99}},\ \bibinfo {pages} {172301} (\bibinfo {year} {2007})},\ \Eprint {http://arxiv.org/abs/0706.1522}{arXiv:0706.1522 [nucl-th]}\BibitemShut {NoStop}%
\bibitem [{\citenamefont {Song}\ and\ \citenamefont {Heinz}(2008)}]{Song:2007ux}%
  \BibitemOpen
  \bibfield  {author} {\bibinfo {author} {\bibfnamefont {H.}~\bibnamefont {Song}}\ and\ \bibinfo {author} {\bibfnamefont {U.~W.}\ \bibnamefont {Heinz}},\ }\href {\doibase 10.1103/PhysRevC.77.064901} {\bibfield  {journal} {\bibinfo  {journal} {Phys. Rev. C}\ }\textbf {\bibinfo {volume} {77}},\ \bibinfo {pages} {064901} (\bibinfo {year} {2008})},\ \Eprint {http://arxiv.org/abs/0712.3715}{arXiv:0712.3715 [nucl-th]}\BibitemShut {NoStop}%
\bibitem [{\citenamefont {Schenke}\ \emph {et~al.}(2010)\citenamefont {Schenke}, \citenamefont {Jeon},\ and\ \citenamefont {Gale}}]{Schenke:2010nt}%
  \BibitemOpen
  \bibfield  {author} {\bibinfo {author} {\bibfnamefont {B.}~\bibnamefont {Schenke}}, \bibinfo {author} {\bibfnamefont {S.}~\bibnamefont {Jeon}}, \ and\ \bibinfo {author} {\bibfnamefont {C.}~\bibnamefont {Gale}},\ }\href {\doibase 10.1103/PhysRevC.82.014903} {\bibfield  {journal} {\bibinfo  {journal} {Phys. Rev. C}\ }\textbf {\bibinfo {volume} {82}},\ \bibinfo {pages} {014903} (\bibinfo {year} {2010})},\ \Eprint {http://arxiv.org/abs/1004.1408}{arXiv:1004.1408 [hep-ph]}\BibitemShut {NoStop}%
\bibitem [{\citenamefont {Aad}\ \emph {et~al.}(2016)\citenamefont {Aad} \emph {et~al.}}]{ATLAS:2015hzw}%
  \BibitemOpen
  \bibfield  {author} {\bibinfo {author} {\bibfnamefont {G.}~\bibnamefont {Aad}} \emph {et~al.} (\bibinfo {collaboration} {ATLAS}),\ }\href {\doibase 10.1103/PhysRevLett.116.172301} {\bibfield  {journal} {\bibinfo  {journal} {Phys. Rev. Lett.}\ }\textbf {\bibinfo {volume} {116}},\ \bibinfo {pages} {172301} (\bibinfo {year} {2016})},\ \Eprint {http://arxiv.org/abs/1509.04776}{arXiv:1509.04776 [hep-ex]}\BibitemShut {NoStop}%
\bibitem [{\citenamefont {Acharya}\ \emph {et~al.}(2024)\citenamefont {Acharya} \emph {et~al.}}]{ALICE:2023ulm}%
  \BibitemOpen
  \bibfield  {author} {\bibinfo {author} {\bibfnamefont {S.}~\bibnamefont {Acharya}} \emph {et~al.} (\bibinfo {collaboration} {ALICE}),\ }\href {\doibase 10.1103/PhysRevLett.132.172302} {\bibfield  {journal} {\bibinfo  {journal} {Phys. Rev. Lett.}\ }\textbf {\bibinfo {volume} {132}},\ \bibinfo {pages} {172302} (\bibinfo {year} {2024})},\ \Eprint {http://arxiv.org/abs/2311.14357}{arXiv:2311.14357 [nucl-ex]}\BibitemShut {NoStop}%
\bibitem [{\citenamefont {Aad}\ \emph {et~al.}(2013)\citenamefont {Aad} \emph {et~al.}}]{ATLAS:2013jmi}%
  \BibitemOpen
  \bibfield  {author} {\bibinfo {author} {\bibfnamefont {G.}~\bibnamefont {Aad}} \emph {et~al.} (\bibinfo {collaboration} {ATLAS}),\ }\href {\doibase 10.1016/j.physletb.2013.06.057} {\bibfield  {journal} {\bibinfo  {journal} {Phys. Lett. B}\ }\textbf {\bibinfo {volume} {725}},\ \bibinfo {pages} {60} (\bibinfo {year} {2013})},\ \Eprint {http://arxiv.org/abs/1303.2084}{arXiv:1303.2084 [hep-ex]}\BibitemShut {NoStop}%
\bibitem [{\citenamefont {Abelev}\ \emph {et~al.}(2014{\natexlab{a}})\citenamefont {Abelev} \emph {et~al.}}]{ALICE:2014dwt}%
  \BibitemOpen
  \bibfield  {author} {\bibinfo {author} {\bibfnamefont {B.~B.}\ \bibnamefont {Abelev}} \emph {et~al.} (\bibinfo {collaboration} {ALICE}),\ }\href {\doibase 10.1103/PhysRevC.90.054901} {\bibfield  {journal} {\bibinfo  {journal} {Phys. Rev. C}\ }\textbf {\bibinfo {volume} {90}},\ \bibinfo {pages} {054901} (\bibinfo {year} {2014}{\natexlab{a}})},\ \Eprint {http://arxiv.org/abs/1406.2474}{arXiv:1406.2474 [nucl-ex]}\BibitemShut {NoStop}%
\bibitem [{\citenamefont {Khachatryan}\ \emph {et~al.}(2015)\citenamefont {Khachatryan} \emph {et~al.}}]{CMS:2015yux}%
  \BibitemOpen
  \bibfield  {author} {\bibinfo {author} {\bibfnamefont {V.}~\bibnamefont {Khachatryan}} \emph {et~al.} (\bibinfo {collaboration} {CMS}),\ }\href {\doibase 10.1103/PhysRevLett.115.012301} {\bibfield  {journal} {\bibinfo  {journal} {Phys. Rev. Lett.}\ }\textbf {\bibinfo {volume} {115}},\ \bibinfo {pages} {012301} (\bibinfo {year} {2015})},\ \Eprint {http://arxiv.org/abs/1502.05382}{arXiv:1502.05382 [nucl-ex]}\BibitemShut {NoStop}%
\bibitem [{\citenamefont {Adare}\ \emph {et~al.}(2013)\citenamefont {Adare} \emph {et~al.}}]{PHENIX:2013ktj}%
  \BibitemOpen
  \bibfield  {author} {\bibinfo {author} {\bibfnamefont {A.}~\bibnamefont {Adare}} \emph {et~al.} (\bibinfo {collaboration} {PHENIX}),\ }\href {\doibase 10.1103/PhysRevLett.111.212301} {\bibfield  {journal} {\bibinfo  {journal} {Phys. Rev. Lett.}\ }\textbf {\bibinfo {volume} {111}},\ \bibinfo {pages} {212301} (\bibinfo {year} {2013})},\ \Eprint {http://arxiv.org/abs/1303.1794}{arXiv:1303.1794 [nucl-ex]}\BibitemShut {NoStop}%
\bibitem [{\citenamefont {Adare}\ \emph {et~al.}(2015{\natexlab{a}})\citenamefont {Adare} \emph {et~al.}}]{PHENIX:2014fnc}%
  \BibitemOpen
  \bibfield  {author} {\bibinfo {author} {\bibfnamefont {A.}~\bibnamefont {Adare}} \emph {et~al.} (\bibinfo {collaboration} {PHENIX}),\ }\href {\doibase 10.1103/PhysRevLett.114.192301} {\bibfield  {journal} {\bibinfo  {journal} {Phys. Rev. Lett.}\ }\textbf {\bibinfo {volume} {114}},\ \bibinfo {pages} {192301} (\bibinfo {year} {2015}{\natexlab{a}})},\ \Eprint {http://arxiv.org/abs/1404.7461}{arXiv:1404.7461 [nucl-ex]}\BibitemShut {NoStop}%
\bibitem [{\citenamefont {Adare}\ \emph {et~al.}(2015{\natexlab{b}})\citenamefont {Adare} \emph {et~al.}}]{PHENIX:2015idk}%
  \BibitemOpen
  \bibfield  {author} {\bibinfo {author} {\bibfnamefont {A.}~\bibnamefont {Adare}} \emph {et~al.} (\bibinfo {collaboration} {PHENIX}),\ }\href {\doibase 10.1103/PhysRevLett.115.142301} {\bibfield  {journal} {\bibinfo  {journal} {Phys. Rev. Lett.}\ }\textbf {\bibinfo {volume} {115}},\ \bibinfo {pages} {142301} (\bibinfo {year} {2015}{\natexlab{b}})},\ \Eprint {http://arxiv.org/abs/1507.06273}{arXiv:1507.06273 [nucl-ex]}\BibitemShut {NoStop}%
\bibitem [{\citenamefont {Aidala}\ \emph {et~al.}(2017)\citenamefont {Aidala} \emph {et~al.}}]{PHENIX:2016cfs}%
  \BibitemOpen
  \bibfield  {author} {\bibinfo {author} {\bibfnamefont {C.}~\bibnamefont {Aidala}} \emph {et~al.} (\bibinfo {collaboration} {PHENIX}),\ }\href {\doibase 10.1103/PhysRevC.95.034910} {\bibfield  {journal} {\bibinfo  {journal} {Phys. Rev. C}\ }\textbf {\bibinfo {volume} {95}},\ \bibinfo {pages} {034910} (\bibinfo {year} {2017})},\ \Eprint {http://arxiv.org/abs/1609.02894}{arXiv:1609.02894 [nucl-ex]}\BibitemShut {NoStop}%
\bibitem [{\citenamefont {Aidala}\ \emph {et~al.}(2018)\citenamefont {Aidala} \emph {et~al.}}]{PHENIX:2017xrm}%
  \BibitemOpen
  \bibfield  {author} {\bibinfo {author} {\bibfnamefont {C.}~\bibnamefont {Aidala}} \emph {et~al.} (\bibinfo {collaboration} {PHENIX}),\ }\href {\doibase 10.1103/PhysRevLett.120.062302} {\bibfield  {journal} {\bibinfo  {journal} {Phys. Rev. Lett.}\ }\textbf {\bibinfo {volume} {120}},\ \bibinfo {pages} {062302} (\bibinfo {year} {2018})},\ \Eprint {http://arxiv.org/abs/1707.06108}{arXiv:1707.06108 [nucl-ex]}\BibitemShut {NoStop}%
\bibitem [{\citenamefont {Weller}\ and\ \citenamefont {Romatschke}(2017)}]{Weller:2017tsr}%
  \BibitemOpen
  \bibfield  {author} {\bibinfo {author} {\bibfnamefont {R.~D.}\ \bibnamefont {Weller}}\ and\ \bibinfo {author} {\bibfnamefont {P.}~\bibnamefont {Romatschke}},\ }\href {\doibase 10.1016/j.physletb.2017.09.077} {\bibfield  {journal} {\bibinfo  {journal} {Phys. Lett. B}\ }\textbf {\bibinfo {volume} {774}},\ \bibinfo {pages} {351} (\bibinfo {year} {2017})},\ \Eprint {http://arxiv.org/abs/1701.07145}{arXiv:1701.07145 [nucl-th]}\BibitemShut {NoStop}%
\bibitem [{\citenamefont {Schenke}\ \emph {et~al.}(2020)\citenamefont {Schenke}, \citenamefont {Shen},\ and\ \citenamefont {Tribedy}}]{Schenke:2020mbo}%
  \BibitemOpen
  \bibfield  {author} {\bibinfo {author} {\bibfnamefont {B.}~\bibnamefont {Schenke}}, \bibinfo {author} {\bibfnamefont {C.}~\bibnamefont {Shen}}, \ and\ \bibinfo {author} {\bibfnamefont {P.}~\bibnamefont {Tribedy}},\ }\href {\doibase 10.1103/PhysRevC.102.044905} {\bibfield  {journal} {\bibinfo  {journal} {Phys. Rev. C}\ }\textbf {\bibinfo {volume} {102}},\ \bibinfo {pages} {044905} (\bibinfo {year} {2020})},\ \Eprint {http://arxiv.org/abs/2005.14682}{arXiv:2005.14682 [nucl-th]}\BibitemShut {NoStop}%
\bibitem [{\citenamefont {Adam}\ \emph {et~al.}(2016{\natexlab{b}})\citenamefont {Adam} \emph {et~al.}}]{ALICE:2016sdt}%
  \BibitemOpen
  \bibfield  {author} {\bibinfo {author} {\bibfnamefont {J.}~\bibnamefont {Adam}} \emph {et~al.} (\bibinfo {collaboration} {ALICE}),\ }\href {\doibase 10.1007/JHEP06(2016)050} {\bibfield  {journal} {\bibinfo  {journal} {JHEP}\ }\textbf {\bibinfo {volume} {06}},\ \bibinfo {pages} {050} (\bibinfo {year} {2016}{\natexlab{b}})},\ \Eprint {http://arxiv.org/abs/1603.02816}{arXiv:1603.02816 [nucl-ex]}\BibitemShut {NoStop}%
\bibitem [{\citenamefont {Adam}\ \emph {et~al.}(2016{\natexlab{c}})\citenamefont {Adam} \emph {et~al.}}]{ALICE:2015mpp}%
  \BibitemOpen
  \bibfield  {author} {\bibinfo {author} {\bibfnamefont {J.}~\bibnamefont {Adam}} \emph {et~al.} (\bibinfo {collaboration} {ALICE}),\ }\href {\doibase 10.1016/j.physletb.2016.05.027} {\bibfield  {journal} {\bibinfo  {journal} {Phys. Lett. B}\ }\textbf {\bibinfo {volume} {758}},\ \bibinfo {pages} {389} (\bibinfo {year} {2016}{\natexlab{c}})},\ \Eprint {http://arxiv.org/abs/1512.07227}{arXiv:1512.07227 [nucl-ex]}\BibitemShut {NoStop}%
\bibitem [{\citenamefont {Abelev}\ \emph {et~al.}(2014{\natexlab{b}})\citenamefont {Abelev} \emph {et~al.}}]{ALICE:2013wgn}%
  \BibitemOpen
  \bibfield  {author} {\bibinfo {author} {\bibfnamefont {B.~B.}\ \bibnamefont {Abelev}} \emph {et~al.} (\bibinfo {collaboration} {ALICE}),\ }\href {\doibase 10.1016/j.physletb.2013.11.020} {\bibfield  {journal} {\bibinfo  {journal} {Phys. Lett. B}\ }\textbf {\bibinfo {volume} {728}},\ \bibinfo {pages} {25} (\bibinfo {year} {2014}{\natexlab{b}})},\ \Eprint {http://arxiv.org/abs/1307.6796}{arXiv:1307.6796 [nucl-ex]}\BibitemShut {NoStop}%
\bibitem [{\citenamefont {Adam}\ \emph {et~al.}(2016{\natexlab{d}})\citenamefont {Adam} \emph {et~al.}}]{ALICE:2016yta}%
  \BibitemOpen
  \bibfield  {author} {\bibinfo {author} {\bibfnamefont {J.}~\bibnamefont {Adam}} \emph {et~al.} (\bibinfo {collaboration} {ALICE}),\ }\href {\doibase 10.1103/PhysRevC.94.054908} {\bibfield  {journal} {\bibinfo  {journal} {Phys. Rev. C}\ }\textbf {\bibinfo {volume} {94}},\ \bibinfo {pages} {054908} (\bibinfo {year} {2016}{\natexlab{d}})},\ \Eprint {http://arxiv.org/abs/1605.07569}{arXiv:1605.07569 [nucl-ex]}\BibitemShut {NoStop}%
\bibitem [{\citenamefont {Aad}\ \emph {et~al.}(2023{\natexlab{a}})\citenamefont {Aad} \emph {et~al.}}]{ATLAS:2022kqu}%
  \BibitemOpen
  \bibfield  {author} {\bibinfo {author} {\bibfnamefont {G.}~\bibnamefont {Aad}} \emph {et~al.} (\bibinfo {collaboration} {ATLAS}),\ }\href {\doibase 10.1007/JHEP07(2023)074} {\bibfield  {journal} {\bibinfo  {journal} {JHEP}\ }\textbf {\bibinfo {volume} {07}},\ \bibinfo {pages} {074} (\bibinfo {year} {2023}{\natexlab{a}})},\ \Eprint {http://arxiv.org/abs/2211.15257}{arXiv:2211.15257 [hep-ex]}\BibitemShut {NoStop}%
\bibitem [{\citenamefont {Acharya}\ \emph {et~al.}(2022)\citenamefont {Acharya} \emph {et~al.}}]{PHENIX:2021dod}%
  \BibitemOpen
  \bibfield  {author} {\bibinfo {author} {\bibfnamefont {U.~A.}\ \bibnamefont {Acharya}} \emph {et~al.} (\bibinfo {collaboration} {PHENIX}),\ }\href {\doibase 10.1103/PhysRevC.105.064902} {\bibfield  {journal} {\bibinfo  {journal} {Phys. Rev. C}\ }\textbf {\bibinfo {volume} {105}},\ \bibinfo {pages} {064902} (\bibinfo {year} {2022})},\ \Eprint {http://arxiv.org/abs/2111.05756}{arXiv:2111.05756 [nucl-ex]}\BibitemShut {NoStop}%
\bibitem [{\citenamefont {Adam}\ \emph {et~al.}(2015)\citenamefont {Adam} \emph {et~al.}}]{ALICE:2014xsp}%
  \BibitemOpen
  \bibfield  {author} {\bibinfo {author} {\bibfnamefont {J.}~\bibnamefont {Adam}} \emph {et~al.} (\bibinfo {collaboration} {ALICE}),\ }\href {\doibase 10.1103/PhysRevC.91.064905} {\bibfield  {journal} {\bibinfo  {journal} {Phys. Rev. C}\ }\textbf {\bibinfo {volume} {91}},\ \bibinfo {pages} {064905} (\bibinfo {year} {2015})},\ \Eprint {http://arxiv.org/abs/1412.6828}{arXiv:1412.6828 [nucl-ex]}\BibitemShut {NoStop}%
\bibitem [{\citenamefont {Adare}\ \emph {et~al.}(2014)\citenamefont {Adare} \emph {et~al.}}]{PHENIX:2013jxf}%
  \BibitemOpen
  \bibfield  {author} {\bibinfo {author} {\bibfnamefont {A.}~\bibnamefont {Adare}} \emph {et~al.} (\bibinfo {collaboration} {PHENIX}),\ }\href {\doibase 10.1103/PhysRevC.90.034902} {\bibfield  {journal} {\bibinfo  {journal} {Phys. Rev. C}\ }\textbf {\bibinfo {volume} {90}},\ \bibinfo {pages} {034902} (\bibinfo {year} {2014})},\ \Eprint {http://arxiv.org/abs/1310.4793}{arXiv:1310.4793 [nucl-ex]}\BibitemShut {NoStop}%
\bibitem [{\citenamefont {Kordell}\ and\ \citenamefont {Majumder}(2018)}]{Kordell:2016njg}%
  \BibitemOpen
  \bibfield  {author} {\bibinfo {author} {\bibfnamefont {M.}~\bibnamefont {Kordell}}\ and\ \bibinfo {author} {\bibfnamefont {A.}~\bibnamefont {Majumder}},\ }\href {\doibase 10.1103/PhysRevC.97.054904} {\bibfield  {journal} {\bibinfo  {journal} {Phys. Rev. C}\ }\textbf {\bibinfo {volume} {97}},\ \bibinfo {pages} {054904} (\bibinfo {year} {2018})},\ \Eprint {http://arxiv.org/abs/1601.02595}{arXiv:1601.02595 [nucl-th]}\BibitemShut {NoStop}%
\bibitem [{\citenamefont {Abdulameer}\ \emph {et~al.}(2023)\citenamefont {Abdulameer} \emph {et~al.}}]{PHENIX:2023dxl}%
  \BibitemOpen
  \bibfield  {author} {\bibinfo {author} {\bibfnamefont {N.~J.}\ \bibnamefont {Abdulameer}} \emph {et~al.} (\bibinfo {collaboration} {PHENIX}),\ }\href@noop {} {\  (\bibinfo {year} {2023})},\ \Eprint {http://arxiv.org/abs/2303.12899}{arXiv:2303.12899 [nucl-ex]}\BibitemShut {NoStop}%
\bibitem [{\citenamefont {Bzdak}\ \emph {et~al.}(2016)\citenamefont {Bzdak}, \citenamefont {Skokov},\ and\ \citenamefont {Bathe}}]{Bzdak:2014rca}%
  \BibitemOpen
  \bibfield  {author} {\bibinfo {author} {\bibfnamefont {A.}~\bibnamefont {Bzdak}}, \bibinfo {author} {\bibfnamefont {V.}~\bibnamefont {Skokov}}, \ and\ \bibinfo {author} {\bibfnamefont {S.}~\bibnamefont {Bathe}},\ }\href {\doibase 10.1103/PhysRevC.93.044901} {\bibfield  {journal} {\bibinfo  {journal} {Phys. Rev. C}\ }\textbf {\bibinfo {volume} {93}},\ \bibinfo {pages} {044901} (\bibinfo {year} {2016})},\ \Eprint {http://arxiv.org/abs/1408.3156}{arXiv:1408.3156 [hep-ph]}\BibitemShut {NoStop}%
\bibitem [{\citenamefont {Glauber}\ and\ \citenamefont {Matthiae}(1970)}]{Glauber:1970jm}%
  \BibitemOpen
  \bibfield  {author} {\bibinfo {author} {\bibfnamefont {R.~J.}\ \bibnamefont {Glauber}}\ and\ \bibinfo {author} {\bibfnamefont {G.}~\bibnamefont {Matthiae}},\ }\href {\doibase 10.1016/0550-3213(70)90511-0} {\bibfield  {journal} {\bibinfo  {journal} {Nucl. Phys. B}\ }\textbf {\bibinfo {volume} {21}},\ \bibinfo {pages} {135} (\bibinfo {year} {1970})}\BibitemShut {NoStop}%
\bibitem [{\citenamefont {Miller}\ \emph {et~al.}(2007)\citenamefont {Miller}, \citenamefont {Reygers}, \citenamefont {Sanders},\ and\ \citenamefont {Steinberg}}]{Miller:2007ri}%
  \BibitemOpen
  \bibfield  {author} {\bibinfo {author} {\bibfnamefont {M.~L.}\ \bibnamefont {Miller}}, \bibinfo {author} {\bibfnamefont {K.}~\bibnamefont {Reygers}}, \bibinfo {author} {\bibfnamefont {S.~J.}\ \bibnamefont {Sanders}}, \ and\ \bibinfo {author} {\bibfnamefont {P.}~\bibnamefont {Steinberg}},\ }\href {\doibase 10.1146/annurev.nucl.57.090506.123020} {\bibfield  {journal} {\bibinfo  {journal} {Ann. Rev. Nucl. Part. Sci.}\ }\textbf {\bibinfo {volume} {57}},\ \bibinfo {pages} {205} (\bibinfo {year} {2007})},\ \Eprint {http://arxiv.org/abs/nucl-ex/0701025}{arXiv:nucl-ex/0701025}\BibitemShut {NoStop}%
\bibitem [{\citenamefont {Aad}\ \emph {et~al.}(2023{\natexlab{b}})\citenamefont {Aad} \emph {et~al.}}]{ATLAS:2022iyq}%
  \BibitemOpen
  \bibfield  {author} {\bibinfo {author} {\bibfnamefont {G.}~\bibnamefont {Aad}} \emph {et~al.} (\bibinfo {collaboration} {ATLAS}),\ }\href {\doibase 10.1103/PhysRevLett.131.072301} {\bibfield  {journal} {\bibinfo  {journal} {Phys. Rev. Lett.}\ }\textbf {\bibinfo {volume} {131}},\ \bibinfo {pages} {072301} (\bibinfo {year} {2023}{\natexlab{b}})},\ \Eprint {http://arxiv.org/abs/2206.01138}{arXiv:2206.01138 [nucl-ex]}\BibitemShut {NoStop}%
\bibitem [{\citenamefont {Aad}\ \emph {et~al.}(2020)\citenamefont {Aad} \emph {et~al.}}]{ATLAS:2019vcm}%
  \BibitemOpen
  \bibfield  {author} {\bibinfo {author} {\bibfnamefont {G.}~\bibnamefont {Aad}} \emph {et~al.} (\bibinfo {collaboration} {ATLAS}),\ }\href {\doibase 10.1140/epjc/s10052-020-7624-4} {\bibfield  {journal} {\bibinfo  {journal} {Eur. Phys. J. C}\ }\textbf {\bibinfo {volume} {80}},\ \bibinfo {pages} {73} (\bibinfo {year} {2020})},\ \Eprint {http://arxiv.org/abs/1910.13978}{arXiv:1910.13978 [nucl-ex]}\BibitemShut {NoStop}%
\bibitem [{\citenamefont {Huss}\ \emph {et~al.}(2021{\natexlab{a}})\citenamefont {Huss}, \citenamefont {Kurkela}, \citenamefont {Mazeliauskas}, \citenamefont {Paatelainen}, \citenamefont {van~der Schee},\ and\ \citenamefont {Wiedemann}}]{Huss:2020whe}%
  \BibitemOpen
  \bibfield  {author} {\bibinfo {author} {\bibfnamefont {A.}~\bibnamefont {Huss}}, \bibinfo {author} {\bibfnamefont {A.}~\bibnamefont {Kurkela}}, \bibinfo {author} {\bibfnamefont {A.}~\bibnamefont {Mazeliauskas}}, \bibinfo {author} {\bibfnamefont {R.}~\bibnamefont {Paatelainen}}, \bibinfo {author} {\bibfnamefont {W.}~\bibnamefont {van~der Schee}}, \ and\ \bibinfo {author} {\bibfnamefont {U.~A.}\ \bibnamefont {Wiedemann}},\ }\href {\doibase 10.1103/PhysRevC.103.054903} {\bibfield  {journal} {\bibinfo  {journal} {Phys. Rev. C}\ }\textbf {\bibinfo {volume} {103}},\ \bibinfo {pages} {054903} (\bibinfo {year} {2021}{\natexlab{a}})},\ \Eprint {http://arxiv.org/abs/2007.13758}{arXiv:2007.13758 [hep-ph]}\BibitemShut {NoStop}%
\bibitem [{\citenamefont {Huss}\ \emph {et~al.}(2021{\natexlab{b}})\citenamefont {Huss}, \citenamefont {Kurkela}, \citenamefont {Mazeliauskas}, \citenamefont {Paatelainen}, \citenamefont {van~der Schee},\ and\ \citenamefont {Wiedemann}}]{Huss:2020dwe}%
  \BibitemOpen
  \bibfield  {author} {\bibinfo {author} {\bibfnamefont {A.}~\bibnamefont {Huss}}, \bibinfo {author} {\bibfnamefont {A.}~\bibnamefont {Kurkela}}, \bibinfo {author} {\bibfnamefont {A.}~\bibnamefont {Mazeliauskas}}, \bibinfo {author} {\bibfnamefont {R.}~\bibnamefont {Paatelainen}}, \bibinfo {author} {\bibfnamefont {W.}~\bibnamefont {van~der Schee}}, \ and\ \bibinfo {author} {\bibfnamefont {U.~A.}\ \bibnamefont {Wiedemann}},\ }\href {\doibase 10.1103/PhysRevLett.126.192301} {\bibfield  {journal} {\bibinfo  {journal} {Phys. Rev. Lett.}\ }\textbf {\bibinfo {volume} {126}},\ \bibinfo {pages} {192301} (\bibinfo {year} {2021}{\natexlab{b}})},\ \Eprint {http://arxiv.org/abs/2007.13754}{arXiv:2007.13754 [hep-ph]}\BibitemShut {NoStop}%
\bibitem [{\citenamefont {Soudi}\ \emph {et~al.}(2024)\citenamefont {Soudi} \emph {et~al.}}]{JETSCAPE:2024dgu}%
  \BibitemOpen
  \bibfield  {author} {\bibinfo {author} {\bibfnamefont {I.}~\bibnamefont {Soudi}} \emph {et~al.} (\bibinfo {collaboration} {JETSCAPE}),\ }\href@noop {} {\  (\bibinfo {year} {2024})},\ \Eprint {http://arxiv.org/abs/2407.17443}{arXiv:2407.17443 [hep-ph]}\BibitemShut {NoStop}%
\bibitem [{\citenamefont {Faraday}\ \emph {et~al.}(2023)\citenamefont {Faraday}, \citenamefont {Grindrod},\ and\ \citenamefont {Horowitz}}]{Faraday:2023mmx}%
  \BibitemOpen
  \bibfield  {author} {\bibinfo {author} {\bibfnamefont {C.}~\bibnamefont {Faraday}}, \bibinfo {author} {\bibfnamefont {A.}~\bibnamefont {Grindrod}}, \ and\ \bibinfo {author} {\bibfnamefont {W.~A.}\ \bibnamefont {Horowitz}},\ }\href {\doibase 10.1140/epjc/s10052-023-12234-y} {\bibfield  {journal} {\bibinfo  {journal} {Eur. Phys. J. C}\ }\textbf {\bibinfo {volume} {83}},\ \bibinfo {pages} {1060} (\bibinfo {year} {2023})},\ \Eprint {http://arxiv.org/abs/2305.13182}{arXiv:2305.13182 [hep-ph]}\BibitemShut {NoStop}%
\bibitem [{\citenamefont {Gyulassy}\ \emph {et~al.}(2001)\citenamefont {Gyulassy}, \citenamefont {Levai},\ and\ \citenamefont {Vitev}}]{Gyulassy:2000er}%
  \BibitemOpen
  \bibfield  {author} {\bibinfo {author} {\bibfnamefont {M.}~\bibnamefont {Gyulassy}}, \bibinfo {author} {\bibfnamefont {P.}~\bibnamefont {Levai}}, \ and\ \bibinfo {author} {\bibfnamefont {I.}~\bibnamefont {Vitev}},\ }\href {\doibase 10.1016/S0550-3213(00)00652-0} {\bibfield  {journal} {\bibinfo  {journal} {Nucl. Phys. B}\ }\textbf {\bibinfo {volume} {594}},\ \bibinfo {pages} {371} (\bibinfo {year} {2001})},\ \Eprint {http://arxiv.org/abs/nucl-th/0006010}{arXiv:nucl-th/0006010}\BibitemShut {NoStop}%
\bibitem [{\citenamefont {Djordjevic}\ and\ \citenamefont {Gyulassy}(2004)}]{Djordjevic:2003zk}%
  \BibitemOpen
  \bibfield  {author} {\bibinfo {author} {\bibfnamefont {M.}~\bibnamefont {Djordjevic}}\ and\ \bibinfo {author} {\bibfnamefont {M.}~\bibnamefont {Gyulassy}},\ }\href {\doibase 10.1016/j.nuclphysa.2003.12.020} {\bibfield  {journal} {\bibinfo  {journal} {Nucl. Phys. A}\ }\textbf {\bibinfo {volume} {733}},\ \bibinfo {pages} {265} (\bibinfo {year} {2004})},\ \Eprint {http://arxiv.org/abs/nucl-th/0310076}{arXiv:nucl-th/0310076}\BibitemShut {NoStop}%
\bibitem [{\citenamefont {Zakharov}(1997)}]{Zakharov:1997uu}%
  \BibitemOpen
  \bibfield  {author} {\bibinfo {author} {\bibfnamefont {B.~G.}\ \bibnamefont {Zakharov}},\ }\href {\doibase 10.1134/1.567389} {\bibfield  {journal} {\bibinfo  {journal} {JETP Lett.}\ }\textbf {\bibinfo {volume} {65}},\ \bibinfo {pages} {615} (\bibinfo {year} {1997})},\ \Eprint {http://arxiv.org/abs/hep-ph/9704255}{arXiv:hep-ph/9704255}\BibitemShut {NoStop}%
\bibitem [{\citenamefont {Baier}\ \emph {et~al.}(1997{\natexlab{a}})\citenamefont {Baier}, \citenamefont {Dokshitzer}, \citenamefont {Mueller}, \citenamefont {Peigne},\ and\ \citenamefont {Schiff}}]{Baier:1996kr}%
  \BibitemOpen
  \bibfield  {author} {\bibinfo {author} {\bibfnamefont {R.}~\bibnamefont {Baier}}, \bibinfo {author} {\bibfnamefont {Y.~L.}\ \bibnamefont {Dokshitzer}}, \bibinfo {author} {\bibfnamefont {A.~H.}\ \bibnamefont {Mueller}}, \bibinfo {author} {\bibfnamefont {S.}~\bibnamefont {Peigne}}, \ and\ \bibinfo {author} {\bibfnamefont {D.}~\bibnamefont {Schiff}},\ }\href {\doibase 10.1016/S0550-3213(96)00553-6} {\bibfield  {journal} {\bibinfo  {journal} {Nucl. Phys. B}\ }\textbf {\bibinfo {volume} {483}},\ \bibinfo {pages} {291} (\bibinfo {year} {1997}{\natexlab{a}})},\ \Eprint {http://arxiv.org/abs/hep-ph/9607355}{arXiv:hep-ph/9607355}\BibitemShut {NoStop}%
\bibitem [{\citenamefont {Kolbe}\ and\ \citenamefont {Horowitz}(2019)}]{Kolbe:2015rvk}%
  \BibitemOpen
  \bibfield  {author} {\bibinfo {author} {\bibfnamefont {I.}~\bibnamefont {Kolbe}}\ and\ \bibinfo {author} {\bibfnamefont {W.~A.}\ \bibnamefont {Horowitz}},\ }\href {\doibase 10.1103/PhysRevC.100.024913} {\bibfield  {journal} {\bibinfo  {journal} {Phys. Rev. C}\ }\textbf {\bibinfo {volume} {100}},\ \bibinfo {pages} {024913} (\bibinfo {year} {2019})},\ \Eprint {http://arxiv.org/abs/1511.09313}{arXiv:1511.09313 [hep-ph]}\BibitemShut {NoStop}%
\bibitem [{\citenamefont {Kolbe}(2015)}]{Kolbe:2015suq}%
  \BibitemOpen
  \bibfield  {author} {\bibinfo {author} {\bibfnamefont {I.}~\bibnamefont {Kolbe}},\ }\emph {\bibinfo {title} {{Short path length pQCD corrections to energy loss in the quark gluon plasma}}},\ \href@noop {} {Master's thesis},\ \bibinfo  {school} {Cape Town U.} (\bibinfo {year} {2015}),\ \Eprint {http://arxiv.org/abs/1509.06122}{arXiv:1509.06122 [hep-ph]}\BibitemShut {NoStop}%
\bibitem [{\citenamefont {Baier}\ \emph {et~al.}(1997{\natexlab{b}})\citenamefont {Baier}, \citenamefont {Dokshitzer}, \citenamefont {Mueller}, \citenamefont {Peigne},\ and\ \citenamefont {Schiff}}]{Baier:1996sk}%
  \BibitemOpen
  \bibfield  {author} {\bibinfo {author} {\bibfnamefont {R.}~\bibnamefont {Baier}}, \bibinfo {author} {\bibfnamefont {Y.~L.}\ \bibnamefont {Dokshitzer}}, \bibinfo {author} {\bibfnamefont {A.~H.}\ \bibnamefont {Mueller}}, \bibinfo {author} {\bibfnamefont {S.}~\bibnamefont {Peigne}}, \ and\ \bibinfo {author} {\bibfnamefont {D.}~\bibnamefont {Schiff}},\ }\href {\doibase 10.1016/S0550-3213(96)00581-0} {\bibfield  {journal} {\bibinfo  {journal} {Nucl. Phys. B}\ }\textbf {\bibinfo {volume} {484}},\ \bibinfo {pages} {265} (\bibinfo {year} {1997}{\natexlab{b}})},\ \Eprint {http://arxiv.org/abs/hep-ph/9608322}{arXiv:hep-ph/9608322}\BibitemShut {NoStop}%
\bibitem [{\citenamefont {Baier}\ \emph {et~al.}(1996)\citenamefont {Baier}, \citenamefont {Dokshitzer}, \citenamefont {Mueller}, \citenamefont {Peigne},\ and\ \citenamefont {Schiff}}]{Baier:1996vi}%
  \BibitemOpen
  \bibfield  {author} {\bibinfo {author} {\bibfnamefont {R.}~\bibnamefont {Baier}}, \bibinfo {author} {\bibfnamefont {Y.~L.}\ \bibnamefont {Dokshitzer}}, \bibinfo {author} {\bibfnamefont {A.~H.}\ \bibnamefont {Mueller}}, \bibinfo {author} {\bibfnamefont {S.}~\bibnamefont {Peigne}}, \ and\ \bibinfo {author} {\bibfnamefont {D.}~\bibnamefont {Schiff}},\ }\href {\doibase 10.1016/0550-3213(96)00426-9} {\bibfield  {journal} {\bibinfo  {journal} {Nucl. Phys. B}\ }\textbf {\bibinfo {volume} {478}},\ \bibinfo {pages} {577} (\bibinfo {year} {1996})},\ \Eprint {http://arxiv.org/abs/hep-ph/9604327}{arXiv:hep-ph/9604327}\BibitemShut {NoStop}%
\bibitem [{\citenamefont {Baier}\ \emph {et~al.}(1998)\citenamefont {Baier}, \citenamefont {Dokshitzer}, \citenamefont {Mueller},\ and\ \citenamefont {Schiff}}]{Baier:1998kq}%
  \BibitemOpen
  \bibfield  {author} {\bibinfo {author} {\bibfnamefont {R.}~\bibnamefont {Baier}}, \bibinfo {author} {\bibfnamefont {Y.~L.}\ \bibnamefont {Dokshitzer}}, \bibinfo {author} {\bibfnamefont {A.~H.}\ \bibnamefont {Mueller}}, \ and\ \bibinfo {author} {\bibfnamefont {D.}~\bibnamefont {Schiff}},\ }\href {\doibase 10.1016/S0550-3213(98)00546-X} {\bibfield  {journal} {\bibinfo  {journal} {Nucl. Phys. B}\ }\textbf {\bibinfo {volume} {531}},\ \bibinfo {pages} {403} (\bibinfo {year} {1998})},\ \Eprint {http://arxiv.org/abs/hep-ph/9804212}{arXiv:hep-ph/9804212}\BibitemShut {NoStop}%
\bibitem [{\citenamefont {Zakharov}(1996)}]{Zakharov:1996fv}%
  \BibitemOpen
  \bibfield  {author} {\bibinfo {author} {\bibfnamefont {B.~G.}\ \bibnamefont {Zakharov}},\ }\href {\doibase 10.1134/1.567126} {\bibfield  {journal} {\bibinfo  {journal} {JETP Lett.}\ }\textbf {\bibinfo {volume} {63}},\ \bibinfo {pages} {952} (\bibinfo {year} {1996})},\ \Eprint {http://arxiv.org/abs/hep-ph/9607440}{arXiv:hep-ph/9607440}\BibitemShut {NoStop}%
\bibitem [{\citenamefont {Armesto}\ \emph {et~al.}(2012)\citenamefont {Armesto} \emph {et~al.}}]{Armesto:2011ht}%
  \BibitemOpen
  \bibfield  {author} {\bibinfo {author} {\bibfnamefont {N.}~\bibnamefont {Armesto}} \emph {et~al.},\ }\href {\doibase 10.1103/PhysRevC.86.064904} {\bibfield  {journal} {\bibinfo  {journal} {Phys. Rev. C}\ }\textbf {\bibinfo {volume} {86}},\ \bibinfo {pages} {064904} (\bibinfo {year} {2012})},\ \Eprint {http://arxiv.org/abs/1106.1106}{arXiv:1106.1106 [hep-ph]}\BibitemShut {NoStop}%
\bibitem [{\citenamefont {Moore}\ and\ \citenamefont {Teaney}(2005)}]{Moore:2004tg}%
  \BibitemOpen
  \bibfield  {author} {\bibinfo {author} {\bibfnamefont {G.~D.}\ \bibnamefont {Moore}}\ and\ \bibinfo {author} {\bibfnamefont {D.}~\bibnamefont {Teaney}},\ }\href {\doibase 10.1103/PhysRevC.71.064904} {\bibfield  {journal} {\bibinfo  {journal} {Phys. Rev. C}\ }\textbf {\bibinfo {volume} {71}},\ \bibinfo {pages} {064904} (\bibinfo {year} {2005})},\ \Eprint {http://arxiv.org/abs/hep-ph/0412346}{arXiv:hep-ph/0412346}\BibitemShut {NoStop}%
\bibitem [{\citenamefont {Zigic}\ \emph {et~al.}(2022)\citenamefont {Zigic}, \citenamefont {Salom}, \citenamefont {Auvinen}, \citenamefont {Huovinen},\ and\ \citenamefont {Djordjevic}}]{Zigic:2021rku}%
  \BibitemOpen
  \bibfield  {author} {\bibinfo {author} {\bibfnamefont {D.}~\bibnamefont {Zigic}}, \bibinfo {author} {\bibfnamefont {I.}~\bibnamefont {Salom}}, \bibinfo {author} {\bibfnamefont {J.}~\bibnamefont {Auvinen}}, \bibinfo {author} {\bibfnamefont {P.}~\bibnamefont {Huovinen}}, \ and\ \bibinfo {author} {\bibfnamefont {M.}~\bibnamefont {Djordjevic}},\ }\href {\doibase 10.3389/fphy.2022.957019} {\bibfield  {journal} {\bibinfo  {journal} {Front. in Phys.}\ }\textbf {\bibinfo {volume} {10}},\ \bibinfo {pages} {957019} (\bibinfo {year} {2022})},\ \Eprint {http://arxiv.org/abs/2110.01544}{arXiv:2110.01544 [nucl-th]}\BibitemShut {NoStop}%
\bibitem [{\citenamefont {Wicks}(2008)}]{Wicks:2008zz}%
  \BibitemOpen
  \bibfield  {author} {\bibinfo {author} {\bibfnamefont {S.}~\bibnamefont {Wicks}},\ }\emph {\bibinfo {title} {{Fluctuations with small numbers: Developing the perturbative paradigm for jet physics in the QGP at RHIC and LHC}}},\ \href@noop {} {\bibinfo {type} {{PhD} thesis}} (\bibinfo {year} {2008})\BibitemShut {NoStop}%
\bibitem [{\citenamefont {Braaten}\ and\ \citenamefont {Pisarski}(1990)}]{Braaten:1989mz}%
  \BibitemOpen
  \bibfield  {author} {\bibinfo {author} {\bibfnamefont {E.}~\bibnamefont {Braaten}}\ and\ \bibinfo {author} {\bibfnamefont {R.~D.}\ \bibnamefont {Pisarski}},\ }\href {\doibase 10.1016/0550-3213(90)90508-B} {\bibfield  {journal} {\bibinfo  {journal} {Nucl. Phys. B}\ }\textbf {\bibinfo {volume} {337}},\ \bibinfo {pages} {569} (\bibinfo {year} {1990})}\BibitemShut {NoStop}%
\bibitem [{\citenamefont {Klimov}(1982)}]{Klimov:1982bv}%
  \BibitemOpen
  \bibfield  {author} {\bibinfo {author} {\bibfnamefont {V.~V.}\ \bibnamefont {Klimov}},\ }\href@noop {} {\bibfield  {journal} {\bibinfo  {journal} {Sov. Phys. JETP}\ }\textbf {\bibinfo {volume} {55}},\ \bibinfo {pages} {199} (\bibinfo {year} {1982})}\BibitemShut {NoStop}%
\bibitem [{\citenamefont {Pisarski}(1989)}]{Pisarski:1988vd}%
  \BibitemOpen
  \bibfield  {author} {\bibinfo {author} {\bibfnamefont {R.~D.}\ \bibnamefont {Pisarski}},\ }\href {\doibase 10.1103/PhysRevLett.63.1129} {\bibfield  {journal} {\bibinfo  {journal} {Phys. Rev. Lett.}\ }\textbf {\bibinfo {volume} {63}},\ \bibinfo {pages} {1129} (\bibinfo {year} {1989})}\BibitemShut {NoStop}%
\bibitem [{\citenamefont {Weldon}(1982{\natexlab{a}})}]{Weldon:1982aq}%
  \BibitemOpen
  \bibfield  {author} {\bibinfo {author} {\bibfnamefont {H.~A.}\ \bibnamefont {Weldon}},\ }\href {\doibase 10.1103/PhysRevD.26.1394} {\bibfield  {journal} {\bibinfo  {journal} {Phys. Rev. D}\ }\textbf {\bibinfo {volume} {26}},\ \bibinfo {pages} {1394} (\bibinfo {year} {1982}{\natexlab{a}})}\BibitemShut {NoStop}%
\bibitem [{\citenamefont {Weldon}(1982{\natexlab{b}})}]{Weldon:1982bn}%
  \BibitemOpen
  \bibfield  {author} {\bibinfo {author} {\bibfnamefont {H.~A.}\ \bibnamefont {Weldon}},\ }\href {\doibase 10.1103/PhysRevD.26.2789} {\bibfield  {journal} {\bibinfo  {journal} {Phys. Rev. D}\ }\textbf {\bibinfo {volume} {26}},\ \bibinfo {pages} {2789} (\bibinfo {year} {1982}{\natexlab{b}})}\BibitemShut {NoStop}%
\bibitem [{\citenamefont {Gyulassy}\ \emph {et~al.}(2002)\citenamefont {Gyulassy}, \citenamefont {Levai},\ and\ \citenamefont {Vitev}}]{Gyulassy:2001nm}%
  \BibitemOpen
  \bibfield  {author} {\bibinfo {author} {\bibfnamefont {M.}~\bibnamefont {Gyulassy}}, \bibinfo {author} {\bibfnamefont {P.}~\bibnamefont {Levai}}, \ and\ \bibinfo {author} {\bibfnamefont {I.}~\bibnamefont {Vitev}},\ }\href {\doibase 10.1016/S0370-2693(02)01990-1} {\bibfield  {journal} {\bibinfo  {journal} {Phys. Lett. B}\ }\textbf {\bibinfo {volume} {538}},\ \bibinfo {pages} {282} (\bibinfo {year} {2002})},\ \Eprint {http://arxiv.org/abs/nucl-th/0112071}{arXiv:nucl-th/0112071}\BibitemShut {NoStop}%
\bibitem [{\citenamefont {Romatschke}\ and\ \citenamefont {Strickland}(2005)}]{Romatschke:2004au}%
  \BibitemOpen
  \bibfield  {author} {\bibinfo {author} {\bibfnamefont {P.}~\bibnamefont {Romatschke}}\ and\ \bibinfo {author} {\bibfnamefont {M.}~\bibnamefont {Strickland}},\ }\href {\doibase 10.1103/PhysRevD.71.125008} {\bibfield  {journal} {\bibinfo  {journal} {Phys. Rev. D}\ }\textbf {\bibinfo {volume} {71}},\ \bibinfo {pages} {125008} (\bibinfo {year} {2005})},\ \Eprint {http://arxiv.org/abs/hep-ph/0408275}{arXiv:hep-ph/0408275}\BibitemShut {NoStop}%
\bibitem [{\citenamefont {Gossiaux}\ and\ \citenamefont {Aichelin}(2008)}]{Gossiaux:2008jv}%
  \BibitemOpen
  \bibfield  {author} {\bibinfo {author} {\bibfnamefont {P.~B.}\ \bibnamefont {Gossiaux}}\ and\ \bibinfo {author} {\bibfnamefont {J.}~\bibnamefont {Aichelin}},\ }\href {\doibase 10.1103/PhysRevC.78.014904} {\bibfield  {journal} {\bibinfo  {journal} {Phys. Rev. C}\ }\textbf {\bibinfo {volume} {78}},\ \bibinfo {pages} {014904} (\bibinfo {year} {2008})},\ \Eprint {http://arxiv.org/abs/0802.2525}{arXiv:0802.2525 [hep-ph]}\BibitemShut {NoStop}%
\bibitem [{\citenamefont {Faraday}\ and\ \citenamefont {Horowitz}(2024)}]{Faraday:2024}%
  \BibitemOpen
  \bibfield  {author} {\bibinfo {author} {\bibfnamefont {C.}~\bibnamefont {Faraday}}\ and\ \bibinfo {author} {\bibfnamefont {W.~A.}\ \bibnamefont {Horowitz}},\ }\href@noop {} {} (\bibinfo {year} {2024}),\ \bibinfo {note} {work in preparation}\BibitemShut {NoStop}%
\bibitem [{\citenamefont {Gyulassy}\ \emph {et~al.}(2000)\citenamefont {Gyulassy}, \citenamefont {Levai},\ and\ \citenamefont {Vitev}}]{Gyulassy:1999zd}%
  \BibitemOpen
  \bibfield  {author} {\bibinfo {author} {\bibfnamefont {M.}~\bibnamefont {Gyulassy}}, \bibinfo {author} {\bibfnamefont {P.}~\bibnamefont {Levai}}, \ and\ \bibinfo {author} {\bibfnamefont {I.}~\bibnamefont {Vitev}},\ }\href {\doibase 10.1016/S0550-3213(99)00713-0} {\bibfield  {journal} {\bibinfo  {journal} {Nucl. Phys. B}\ }\textbf {\bibinfo {volume} {571}},\ \bibinfo {pages} {197} (\bibinfo {year} {2000})},\ \Eprint {http://arxiv.org/abs/hep-ph/9907461}{arXiv:hep-ph/9907461}\BibitemShut {NoStop}%
\bibitem [{\citenamefont {Vitev}\ and\ \citenamefont {Gyulassy}(2002)}]{Vitev:2002pf}%
  \BibitemOpen
  \bibfield  {author} {\bibinfo {author} {\bibfnamefont {I.}~\bibnamefont {Vitev}}\ and\ \bibinfo {author} {\bibfnamefont {M.}~\bibnamefont {Gyulassy}},\ }\href {\doibase 10.1103/PhysRevLett.89.252301} {\bibfield  {journal} {\bibinfo  {journal} {Phys. Rev. Lett.}\ }\textbf {\bibinfo {volume} {89}},\ \bibinfo {pages} {252301} (\bibinfo {year} {2002})},\ \Eprint {http://arxiv.org/abs/hep-ph/0209161}{arXiv:hep-ph/0209161}\BibitemShut {NoStop}%
\bibitem [{\citenamefont {Braaten}\ and\ \citenamefont {Thoma}(1991{\natexlab{a}})}]{Braaten:1991jj}%
  \BibitemOpen
  \bibfield  {author} {\bibinfo {author} {\bibfnamefont {E.}~\bibnamefont {Braaten}}\ and\ \bibinfo {author} {\bibfnamefont {M.~H.}\ \bibnamefont {Thoma}},\ }\href {\doibase 10.1103/PhysRevD.44.1298} {\bibfield  {journal} {\bibinfo  {journal} {Phys. Rev. D}\ }\textbf {\bibinfo {volume} {44}},\ \bibinfo {pages} {1298} (\bibinfo {year} {1991}{\natexlab{a}})}\BibitemShut {NoStop}%
\bibitem [{\citenamefont {Braaten}\ and\ \citenamefont {Thoma}(1991{\natexlab{b}})}]{Braaten:1991we}%
  \BibitemOpen
  \bibfield  {author} {\bibinfo {author} {\bibfnamefont {E.}~\bibnamefont {Braaten}}\ and\ \bibinfo {author} {\bibfnamefont {M.~H.}\ \bibnamefont {Thoma}},\ }\href {\doibase 10.1103/PhysRevD.44.R2625} {\bibfield  {journal} {\bibinfo  {journal} {Phys. Rev. D}\ }\textbf {\bibinfo {volume} {44}},\ \bibinfo {pages} {R2625} (\bibinfo {year} {1991}{\natexlab{b}})}\BibitemShut {NoStop}%
\bibitem [{\citenamefont {Horowitz}\ and\ \citenamefont {Gyulassy}(2011)}]{Horowitz:2011gd}%
  \BibitemOpen
  \bibfield  {author} {\bibinfo {author} {\bibfnamefont {W.~A.}\ \bibnamefont {Horowitz}}\ and\ \bibinfo {author} {\bibfnamefont {M.}~\bibnamefont {Gyulassy}},\ }\href {\doibase 10.1016/j.nuclphysa.2011.09.018} {\bibfield  {journal} {\bibinfo  {journal} {Nucl. Phys. A}\ }\textbf {\bibinfo {volume} {872}},\ \bibinfo {pages} {265} (\bibinfo {year} {2011})},\ \Eprint {http://arxiv.org/abs/1104.4958}{arXiv:1104.4958 [hep-ph]}\BibitemShut {NoStop}%
\bibitem [{\citenamefont {Djordjevic}(2006)}]{Djordjevic:2006tw}%
  \BibitemOpen
  \bibfield  {author} {\bibinfo {author} {\bibfnamefont {M.}~\bibnamefont {Djordjevic}},\ }\href {\doibase 10.1103/PhysRevC.74.064907} {\bibfield  {journal} {\bibinfo  {journal} {Phys. Rev. C}\ }\textbf {\bibinfo {volume} {74}},\ \bibinfo {pages} {064907} (\bibinfo {year} {2006})},\ \Eprint {http://arxiv.org/abs/nucl-th/0603066}{arXiv:nucl-th/0603066}\BibitemShut {NoStop}%
\bibitem [{\citenamefont {Blaizot}\ and\ \citenamefont {Iancu}(2002)}]{Blaizot:2001nr}%
  \BibitemOpen
  \bibfield  {author} {\bibinfo {author} {\bibfnamefont {J.-P.}\ \bibnamefont {Blaizot}}\ and\ \bibinfo {author} {\bibfnamefont {E.}~\bibnamefont {Iancu}},\ }\href {\doibase 10.1016/S0370-1573(01)00061-8} {\bibfield  {journal} {\bibinfo  {journal} {Phys. Rept.}\ }\textbf {\bibinfo {volume} {359}},\ \bibinfo {pages} {355} (\bibinfo {year} {2002})},\ \Eprint {http://arxiv.org/abs/hep-ph/0101103}{arXiv:hep-ph/0101103}\BibitemShut {NoStop}%
\bibitem [{\citenamefont {Bellac}(2011)}]{Bellac:2011kqa}%
  \BibitemOpen
  \bibfield  {author} {\bibinfo {author} {\bibfnamefont {M.~L.}\ \bibnamefont {Bellac}},\ }\href {\doibase 10.1017/CBO9780511721700} {\emph {\bibinfo {title} {{Thermal Field Theory}}}},\ Cambridge Monographs on Mathematical Physics\ (\bibinfo  {publisher} {Cambridge University Press},\ \bibinfo {year} {2011})\BibitemShut {NoStop}%
\bibitem [{\citenamefont {Nakamura}\ \emph {et~al.}(2004)\citenamefont {Nakamura}, \citenamefont {Saito},\ and\ \citenamefont {Sakai}}]{Nakamura:2003pu}%
  \BibitemOpen
  \bibfield  {author} {\bibinfo {author} {\bibfnamefont {A.}~\bibnamefont {Nakamura}}, \bibinfo {author} {\bibfnamefont {T.}~\bibnamefont {Saito}}, \ and\ \bibinfo {author} {\bibfnamefont {S.}~\bibnamefont {Sakai}},\ }\href {\doibase 10.1103/PhysRevD.69.014506} {\bibfield  {journal} {\bibinfo  {journal} {Phys. Rev. D}\ }\textbf {\bibinfo {volume} {69}},\ \bibinfo {pages} {014506} (\bibinfo {year} {2004})},\ \Eprint {http://arxiv.org/abs/hep-lat/0311024}{arXiv:hep-lat/0311024}\BibitemShut {NoStop}%
\bibitem [{\citenamefont {Hart}\ \emph {et~al.}(2000)\citenamefont {Hart}, \citenamefont {Laine},\ and\ \citenamefont {Philipsen}}]{Hart:2000ha}%
  \BibitemOpen
  \bibfield  {author} {\bibinfo {author} {\bibfnamefont {A.}~\bibnamefont {Hart}}, \bibinfo {author} {\bibfnamefont {M.}~\bibnamefont {Laine}}, \ and\ \bibinfo {author} {\bibfnamefont {O.}~\bibnamefont {Philipsen}},\ }\href {\doibase 10.1016/S0550-3213(00)00418-1} {\bibfield  {journal} {\bibinfo  {journal} {Nucl. Phys. B}\ }\textbf {\bibinfo {volume} {586}},\ \bibinfo {pages} {443} (\bibinfo {year} {2000})},\ \Eprint {http://arxiv.org/abs/hep-ph/0004060}{arXiv:hep-ph/0004060}\BibitemShut {NoStop}%
\bibitem [{\citenamefont {Bert}\ \emph {et~al.}(2024)\citenamefont {Bert}, \citenamefont {Faraday},\ and\ \citenamefont {Horowitz}}]{Bert:2024}%
  \BibitemOpen
  \bibfield  {author} {\bibinfo {author} {\bibfnamefont {B.}~\bibnamefont {Bert}}, \bibinfo {author} {\bibfnamefont {C.}~\bibnamefont {Faraday}}, \ and\ \bibinfo {author} {\bibfnamefont {W.~A.}\ \bibnamefont {Horowitz}},\ }\href@noop {} {} (\bibinfo {year} {2024}),\ \bibinfo {note} {work in preparation}\BibitemShut {NoStop}%
\bibitem [{\citenamefont {Xu}\ \emph {et~al.}(2014)\citenamefont {Xu}, \citenamefont {Buzzatti},\ and\ \citenamefont {Gyulassy}}]{Xu:2014ica}%
  \BibitemOpen
  \bibfield  {author} {\bibinfo {author} {\bibfnamefont {J.}~\bibnamefont {Xu}}, \bibinfo {author} {\bibfnamefont {A.}~\bibnamefont {Buzzatti}}, \ and\ \bibinfo {author} {\bibfnamefont {M.}~\bibnamefont {Gyulassy}},\ }\href {\doibase 10.1007/JHEP08(2014)063} {\bibfield  {journal} {\bibinfo  {journal} {JHEP}\ }\textbf {\bibinfo {volume} {08}},\ \bibinfo {pages} {063} (\bibinfo {year} {2014})},\ \Eprint {http://arxiv.org/abs/1402.2956}{arXiv:1402.2956 [hep-ph]}\BibitemShut {NoStop}%
\bibitem [{\citenamefont {Horowitz}(2010)}]{Horowitz:2010dm}%
  \BibitemOpen
  \bibfield  {author} {\bibinfo {author} {\bibfnamefont {W.~A.}\ \bibnamefont {Horowitz}},\ }\emph {\bibinfo {title} {{Probing the Frontiers of QCD}}},\ \href@noop {} {\bibinfo {type} {{PhD} thesis}} (\bibinfo {year} {2010}),\ \Eprint {http://arxiv.org/abs/1011.4316}{arXiv:1011.4316 [nucl-th]}\BibitemShut {NoStop}%
\bibitem [{\citenamefont {Cacciari}\ \emph {et~al.}(2001)\citenamefont {Cacciari}, \citenamefont {Frixione},\ and\ \citenamefont {Nason}}]{Cacciari:2001td}%
  \BibitemOpen
  \bibfield  {author} {\bibinfo {author} {\bibfnamefont {M.}~\bibnamefont {Cacciari}}, \bibinfo {author} {\bibfnamefont {S.}~\bibnamefont {Frixione}}, \ and\ \bibinfo {author} {\bibfnamefont {P.}~\bibnamefont {Nason}},\ }\href {\doibase 10.1088/1126-6708/2001/03/006} {\bibfield  {journal} {\bibinfo  {journal} {JHEP}\ }\textbf {\bibinfo {volume} {03}},\ \bibinfo {pages} {006} (\bibinfo {year} {2001})},\ \Eprint {http://arxiv.org/abs/hep-ph/0102134}{arXiv:hep-ph/0102134}\BibitemShut {NoStop}%
\bibitem [{\citenamefont {Wang}()}]{wang_private_communication}%
  \BibitemOpen
  \bibfield  {author} {\bibinfo {author} {\bibfnamefont {X.~N.}\ \bibnamefont {Wang}},\ }\href@noop {} {}\bibinfo {howpublished} {private communication}\BibitemShut {NoStop}%
\bibitem [{\citenamefont {Cacciari}\ \emph {et~al.}(2006)\citenamefont {Cacciari}, \citenamefont {Nason},\ and\ \citenamefont {Oleari}}]{Cacciari:2005uk}%
  \BibitemOpen
  \bibfield  {author} {\bibinfo {author} {\bibfnamefont {M.}~\bibnamefont {Cacciari}}, \bibinfo {author} {\bibfnamefont {P.}~\bibnamefont {Nason}}, \ and\ \bibinfo {author} {\bibfnamefont {C.}~\bibnamefont {Oleari}},\ }\href {\doibase 10.1088/1126-6708/2006/04/006} {\bibfield  {journal} {\bibinfo  {journal} {JHEP}\ }\textbf {\bibinfo {volume} {04}},\ \bibinfo {pages} {006} (\bibinfo {year} {2006})},\ \Eprint {http://arxiv.org/abs/hep-ph/0510032}{arXiv:hep-ph/0510032}\BibitemShut {NoStop}%
\bibitem [{\citenamefont {de~Florian}\ \emph {et~al.}(2007)\citenamefont {de~Florian}, \citenamefont {Sassot},\ and\ \citenamefont {Stratmann}}]{deFlorian:2007aj}%
  \BibitemOpen
  \bibfield  {author} {\bibinfo {author} {\bibfnamefont {D.}~\bibnamefont {de~Florian}}, \bibinfo {author} {\bibfnamefont {R.}~\bibnamefont {Sassot}}, \ and\ \bibinfo {author} {\bibfnamefont {M.}~\bibnamefont {Stratmann}},\ }\href {\doibase 10.1103/PhysRevD.75.114010} {\bibfield  {journal} {\bibinfo  {journal} {Phys. Rev. D}\ }\textbf {\bibinfo {volume} {75}},\ \bibinfo {pages} {114010} (\bibinfo {year} {2007})},\ \Eprint {http://arxiv.org/abs/hep-ph/0703242}{arXiv:hep-ph/0703242}\BibitemShut {NoStop}%
\bibitem [{\citenamefont {Djordjevic}\ and\ \citenamefont {Gyulassy}(2003)}]{Djordjevic:2003be}%
  \BibitemOpen
  \bibfield  {author} {\bibinfo {author} {\bibfnamefont {M.}~\bibnamefont {Djordjevic}}\ and\ \bibinfo {author} {\bibfnamefont {M.}~\bibnamefont {Gyulassy}},\ }\href {\doibase 10.1103/PhysRevC.68.034914} {\bibfield  {journal} {\bibinfo  {journal} {Phys. Rev. C}\ }\textbf {\bibinfo {volume} {68}},\ \bibinfo {pages} {034914} (\bibinfo {year} {2003})},\ \Eprint {http://arxiv.org/abs/nucl-th/0305062}{arXiv:nucl-th/0305062}\BibitemShut {NoStop}%
\bibitem [{\citenamefont {Faraday}\ and\ \citenamefont {Horowitz}(2023)}]{Faraday:2023uay}%
  \BibitemOpen
  \bibfield  {author} {\bibinfo {author} {\bibfnamefont {C.}~\bibnamefont {Faraday}}\ and\ \bibinfo {author} {\bibfnamefont {W.~A.}\ \bibnamefont {Horowitz}},\ }in\ \href@noop {} {\emph {\bibinfo {booktitle} {{67th Annual Conference of the South African Institute of Physics}}}}\ (\bibinfo {year} {2023})\ \Eprint {http://arxiv.org/abs/2309.06246}{arXiv:2309.06246 [hep-ph]}\BibitemShut {NoStop}%
\bibitem [{\citenamefont {Bjorken}(1983)}]{Bjorken:1982qr}%
  \BibitemOpen
  \bibfield  {author} {\bibinfo {author} {\bibfnamefont {J.~D.}\ \bibnamefont {Bjorken}},\ }\href {\doibase 10.1103/PhysRevD.27.140} {\bibfield  {journal} {\bibinfo  {journal} {Phys. Rev. D}\ }\textbf {\bibinfo {volume} {27}},\ \bibinfo {pages} {140} (\bibinfo {year} {1983})}\BibitemShut {NoStop}%
\bibitem [{\citenamefont {Acharya}\ \emph {et~al.}(2020)\citenamefont {Acharya} \emph {et~al.}}]{ALICE:2019hno}%
  \BibitemOpen
  \bibfield  {author} {\bibinfo {author} {\bibfnamefont {S.}~\bibnamefont {Acharya}} \emph {et~al.} (\bibinfo {collaboration} {ALICE}),\ }\href {\doibase 10.1103/PhysRevC.101.044907} {\bibfield  {journal} {\bibinfo  {journal} {Phys. Rev. C}\ }\textbf {\bibinfo {volume} {101}},\ \bibinfo {pages} {044907} (\bibinfo {year} {2020})},\ \Eprint {http://arxiv.org/abs/1910.07678}{arXiv:1910.07678 [nucl-ex]}\BibitemShut {NoStop}%
\bibitem [{\citenamefont {Adare}\ \emph {et~al.}(2007)\citenamefont {Adare} \emph {et~al.}}]{PHENIX:2007kqm}%
  \BibitemOpen
  \bibfield  {author} {\bibinfo {author} {\bibfnamefont {A.}~\bibnamefont {Adare}} \emph {et~al.} (\bibinfo {collaboration} {PHENIX}),\ }\href {\doibase 10.1103/PhysRevD.76.051106} {\bibfield  {journal} {\bibinfo  {journal} {Phys. Rev. D}\ }\textbf {\bibinfo {volume} {76}},\ \bibinfo {pages} {051106} (\bibinfo {year} {2007})},\ \Eprint {http://arxiv.org/abs/0704.3599}{arXiv:0704.3599 [hep-ex]}\BibitemShut {NoStop}%
\bibitem [{\citenamefont {Sirunyan}\ \emph {et~al.}(2017)\citenamefont {Sirunyan} \emph {et~al.}}]{CMS:2017uoy}%
  \BibitemOpen
  \bibfield  {author} {\bibinfo {author} {\bibfnamefont {A.~M.}\ \bibnamefont {Sirunyan}} \emph {et~al.} (\bibinfo {collaboration} {CMS}),\ }\href {\doibase 10.1103/PhysRevLett.119.152301} {\bibfield  {journal} {\bibinfo  {journal} {Phys. Rev. Lett.}\ }\textbf {\bibinfo {volume} {119}},\ \bibinfo {pages} {152301} (\bibinfo {year} {2017})},\ \Eprint {http://arxiv.org/abs/1705.04727}{arXiv:1705.04727 [hep-ex]}\BibitemShut {NoStop}%
\bibitem [{\citenamefont {Shen}()}]{shen_private_communication}%
  \BibitemOpen
  \bibfield  {author} {\bibinfo {author} {\bibfnamefont {C.}~\bibnamefont {Shen}},\ }\href@noop {} {}\bibinfo {howpublished} {private communication}\BibitemShut {NoStop}%
\bibitem [{\citenamefont {Schenke}\ \emph {et~al.}(2012{\natexlab{a}})\citenamefont {Schenke}, \citenamefont {Tribedy},\ and\ \citenamefont {Venugopalan}}]{Schenke:2012hg}%
  \BibitemOpen
  \bibfield  {author} {\bibinfo {author} {\bibfnamefont {B.}~\bibnamefont {Schenke}}, \bibinfo {author} {\bibfnamefont {P.}~\bibnamefont {Tribedy}}, \ and\ \bibinfo {author} {\bibfnamefont {R.}~\bibnamefont {Venugopalan}},\ }\href {\doibase 10.1103/PhysRevC.86.034908} {\bibfield  {journal} {\bibinfo  {journal} {Phys. Rev. C}\ }\textbf {\bibinfo {volume} {86}},\ \bibinfo {pages} {034908} (\bibinfo {year} {2012}{\natexlab{a}})},\ \Eprint {http://arxiv.org/abs/1206.6805}{arXiv:1206.6805 [hep-ph]}\BibitemShut {NoStop}%
\bibitem [{\citenamefont {Schenke}\ \emph {et~al.}(2012{\natexlab{b}})\citenamefont {Schenke}, \citenamefont {Tribedy},\ and\ \citenamefont {Venugopalan}}]{Schenke:2012wb}%
  \BibitemOpen
  \bibfield  {author} {\bibinfo {author} {\bibfnamefont {B.}~\bibnamefont {Schenke}}, \bibinfo {author} {\bibfnamefont {P.}~\bibnamefont {Tribedy}}, \ and\ \bibinfo {author} {\bibfnamefont {R.}~\bibnamefont {Venugopalan}},\ }\href {\doibase 10.1103/PhysRevLett.108.252301} {\bibfield  {journal} {\bibinfo  {journal} {Phys. Rev. Lett.}\ }\textbf {\bibinfo {volume} {108}},\ \bibinfo {pages} {252301} (\bibinfo {year} {2012}{\natexlab{b}})},\ \Eprint {http://arxiv.org/abs/1202.6646}{arXiv:1202.6646 [nucl-th]}\BibitemShut {NoStop}%
\bibitem [{\citenamefont {Schenke}\ \emph {et~al.}(2011)\citenamefont {Schenke}, \citenamefont {Jeon},\ and\ \citenamefont {Gale}}]{Schenke:2010rr}%
  \BibitemOpen
  \bibfield  {author} {\bibinfo {author} {\bibfnamefont {B.}~\bibnamefont {Schenke}}, \bibinfo {author} {\bibfnamefont {S.}~\bibnamefont {Jeon}}, \ and\ \bibinfo {author} {\bibfnamefont {C.}~\bibnamefont {Gale}},\ }\href {\doibase 10.1103/PhysRevLett.106.042301} {\bibfield  {journal} {\bibinfo  {journal} {Phys. Rev. Lett.}\ }\textbf {\bibinfo {volume} {106}},\ \bibinfo {pages} {042301} (\bibinfo {year} {2011})},\ \Eprint {http://arxiv.org/abs/1009.3244}{arXiv:1009.3244 [hep-ph]}\BibitemShut {NoStop}%
\bibitem [{\citenamefont {Schenke}\ \emph {et~al.}(2012{\natexlab{c}})\citenamefont {Schenke}, \citenamefont {Jeon},\ and\ \citenamefont {Gale}}]{Schenke:2011bn}%
  \BibitemOpen
  \bibfield  {author} {\bibinfo {author} {\bibfnamefont {B.}~\bibnamefont {Schenke}}, \bibinfo {author} {\bibfnamefont {S.}~\bibnamefont {Jeon}}, \ and\ \bibinfo {author} {\bibfnamefont {C.}~\bibnamefont {Gale}},\ }\href {\doibase 10.1103/PhysRevC.85.024901} {\bibfield  {journal} {\bibinfo  {journal} {Phys. Rev. C}\ }\textbf {\bibinfo {volume} {85}},\ \bibinfo {pages} {024901} (\bibinfo {year} {2012}{\natexlab{c}})},\ \Eprint {http://arxiv.org/abs/1109.6289}{arXiv:1109.6289 [hep-ph]}\BibitemShut {NoStop}%
\bibitem [{\citenamefont {Bass}\ \emph {et~al.}(1998)\citenamefont {Bass} \emph {et~al.}}]{Bass:1998ca}%
  \BibitemOpen
  \bibfield  {author} {\bibinfo {author} {\bibfnamefont {S.~A.}\ \bibnamefont {Bass}} \emph {et~al.},\ }\href {\doibase 10.1016/S0146-6410(98)00058-1} {\bibfield  {journal} {\bibinfo  {journal} {Prog. Part. Nucl. Phys.}\ }\textbf {\bibinfo {volume} {41}},\ \bibinfo {pages} {255} (\bibinfo {year} {1998})},\ \Eprint {http://arxiv.org/abs/nucl-th/9803035}{arXiv:nucl-th/9803035}\BibitemShut {NoStop}%
\bibitem [{\citenamefont {Bleicher}\ \emph {et~al.}(1999)\citenamefont {Bleicher} \emph {et~al.}}]{Bleicher:1999xi}%
  \BibitemOpen
  \bibfield  {author} {\bibinfo {author} {\bibfnamefont {M.}~\bibnamefont {Bleicher}} \emph {et~al.},\ }\href {\doibase 10.1088/0954-3899/25/9/308} {\bibfield  {journal} {\bibinfo  {journal} {J. Phys. G}\ }\textbf {\bibinfo {volume} {25}},\ \bibinfo {pages} {1859} (\bibinfo {year} {1999})},\ \Eprint {http://arxiv.org/abs/hep-ph/9909407}{arXiv:hep-ph/9909407}\BibitemShut {NoStop}%
\bibitem [{\citenamefont {Zhao}\ \emph {et~al.}(2022)\citenamefont {Zhao}, \citenamefont {Shen},\ and\ \citenamefont {Schenke}}]{Zhao:2022ayk}%
  \BibitemOpen
  \bibfield  {author} {\bibinfo {author} {\bibfnamefont {W.}~\bibnamefont {Zhao}}, \bibinfo {author} {\bibfnamefont {C.}~\bibnamefont {Shen}}, \ and\ \bibinfo {author} {\bibfnamefont {B.}~\bibnamefont {Schenke}},\ }\href {\doibase 10.1103/PhysRevLett.129.252302} {\bibfield  {journal} {\bibinfo  {journal} {Phys. Rev. Lett.}\ }\textbf {\bibinfo {volume} {129}},\ \bibinfo {pages} {252302} (\bibinfo {year} {2022})},\ \Eprint {http://arxiv.org/abs/2203.06094}{arXiv:2203.06094 [nucl-th]}\BibitemShut {NoStop}%
\bibitem [{\citenamefont {Zhao}\ \emph {et~al.}(2023)\citenamefont {Zhao}, \citenamefont {Ryu}, \citenamefont {Shen},\ and\ \citenamefont {Schenke}}]{Zhao:2022ugy}%
  \BibitemOpen
  \bibfield  {author} {\bibinfo {author} {\bibfnamefont {W.}~\bibnamefont {Zhao}}, \bibinfo {author} {\bibfnamefont {S.}~\bibnamefont {Ryu}}, \bibinfo {author} {\bibfnamefont {C.}~\bibnamefont {Shen}}, \ and\ \bibinfo {author} {\bibfnamefont {B.}~\bibnamefont {Schenke}},\ }\href {\doibase 10.1103/PhysRevC.107.014904} {\bibfield  {journal} {\bibinfo  {journal} {Phys. Rev. C}\ }\textbf {\bibinfo {volume} {107}},\ \bibinfo {pages} {014904} (\bibinfo {year} {2023})},\ \Eprint {http://arxiv.org/abs/2211.16376}{arXiv:2211.16376 [nucl-th]}\BibitemShut {NoStop}%
\bibitem [{\citenamefont {Djordjevic}\ \emph {et~al.}(2006)\citenamefont {Djordjevic}, \citenamefont {Gyulassy}, \citenamefont {Vogt},\ and\ \citenamefont {Wicks}}]{Djordjevic:2005db}%
  \BibitemOpen
  \bibfield  {author} {\bibinfo {author} {\bibfnamefont {M.}~\bibnamefont {Djordjevic}}, \bibinfo {author} {\bibfnamefont {M.}~\bibnamefont {Gyulassy}}, \bibinfo {author} {\bibfnamefont {R.}~\bibnamefont {Vogt}}, \ and\ \bibinfo {author} {\bibfnamefont {S.}~\bibnamefont {Wicks}},\ }\href {\doibase 10.1016/j.physletb.2005.09.087} {\bibfield  {journal} {\bibinfo  {journal} {Phys. Lett. B}\ }\textbf {\bibinfo {volume} {632}},\ \bibinfo {pages} {81} (\bibinfo {year} {2006})},\ \Eprint {http://arxiv.org/abs/nucl-th/0507019}{arXiv:nucl-th/0507019}\BibitemShut {NoStop}%
\bibitem [{\citenamefont {Djordjevic}\ \emph {et~al.}(2005)\citenamefont {Djordjevic}, \citenamefont {Gyulassy},\ and\ \citenamefont {Wicks}}]{Djordjevic:2004nq}%
  \BibitemOpen
  \bibfield  {author} {\bibinfo {author} {\bibfnamefont {M.}~\bibnamefont {Djordjevic}}, \bibinfo {author} {\bibfnamefont {M.}~\bibnamefont {Gyulassy}}, \ and\ \bibinfo {author} {\bibfnamefont {S.}~\bibnamefont {Wicks}},\ }\href {\doibase 10.1103/PhysRevLett.94.112301} {\bibfield  {journal} {\bibinfo  {journal} {Phys. Rev. Lett.}\ }\textbf {\bibinfo {volume} {94}},\ \bibinfo {pages} {112301} (\bibinfo {year} {2005})},\ \Eprint {http://arxiv.org/abs/hep-ph/0410372}{arXiv:hep-ph/0410372}\BibitemShut {NoStop}%
\bibitem [{\citenamefont {Borsanyi}\ \emph {et~al.}(2012)\citenamefont {Borsanyi}, \citenamefont {Fodor}, \citenamefont {Katz}, \citenamefont {Krieg}, \citenamefont {Ratti},\ and\ \citenamefont {Szabo}}]{Borsanyi:2011sw}%
  \BibitemOpen
  \bibfield  {author} {\bibinfo {author} {\bibfnamefont {S.}~\bibnamefont {Borsanyi}}, \bibinfo {author} {\bibfnamefont {Z.}~\bibnamefont {Fodor}}, \bibinfo {author} {\bibfnamefont {S.~D.}\ \bibnamefont {Katz}}, \bibinfo {author} {\bibfnamefont {S.}~\bibnamefont {Krieg}}, \bibinfo {author} {\bibfnamefont {C.}~\bibnamefont {Ratti}}, \ and\ \bibinfo {author} {\bibfnamefont {K.}~\bibnamefont {Szabo}},\ }\href {\doibase 10.1007/JHEP01(2012)138} {\bibfield  {journal} {\bibinfo  {journal} {JHEP}\ }\textbf {\bibinfo {volume} {01}},\ \bibinfo {pages} {138} (\bibinfo {year} {2012})},\ \Eprint {http://arxiv.org/abs/1112.4416}{arXiv:1112.4416 [hep-lat]}\BibitemShut {NoStop}%
\bibitem [{\citenamefont {Skellam}(1946)}]{skellam:1946}%
  \BibitemOpen
  \bibfield  {author} {\bibinfo {author} {\bibfnamefont {J.~G.}\ \bibnamefont {Skellam}},\ }\href {http://www.jstor.org/stable/2981372} {\bibfield  {journal} {\bibinfo  {journal} {Journal of the Royal Statistical Society}\ }\textbf {\bibinfo {volume} {109}},\ \bibinfo {pages} {296} (\bibinfo {year} {1946})}\BibitemShut {NoStop}%
\bibitem [{\citenamefont {Arleo}(2017)}]{Arleo:2017ntr}%
  \BibitemOpen
  \bibfield  {author} {\bibinfo {author} {\bibfnamefont {F.}~\bibnamefont {Arleo}},\ }\href {\doibase 10.1103/PhysRevLett.119.062302} {\bibfield  {journal} {\bibinfo  {journal} {Phys. Rev. Lett.}\ }\textbf {\bibinfo {volume} {119}},\ \bibinfo {pages} {062302} (\bibinfo {year} {2017})},\ \Eprint {http://arxiv.org/abs/1703.10852}{arXiv:1703.10852 [hep-ph]}\BibitemShut {NoStop}%
\bibitem [{\citenamefont {Baier}(2003)}]{Baier:2002tc}%
  \BibitemOpen
  \bibfield  {author} {\bibinfo {author} {\bibfnamefont {R.}~\bibnamefont {Baier}},\ }\href {\doibase 10.1016/S0375-9474(02)01429-X} {\bibfield  {journal} {\bibinfo  {journal} {Nucl. Phys. A}\ }\textbf {\bibinfo {volume} {715}},\ \bibinfo {pages} {209} (\bibinfo {year} {2003})},\ \Eprint {http://arxiv.org/abs/hep-ph/0209038}{arXiv:hep-ph/0209038}\BibitemShut {NoStop}%
\end{thebibliography}%

\end{document}